\documentclass[aps,prx,reprint,superscriptaddress,nofootinbib]{revtex4-2}
\usepackage{blindtext}
\usepackage{lipsum}
\usepackage{graphics}
\usepackage{amsmath}
\usepackage{graphicx}
\usepackage{graphics}
\usepackage{amssymb}
\usepackage{verbatim}
\usepackage{physics}
\usepackage{verbatim}
\usepackage{float}
\usepackage{dsfont}
\usepackage[dvipsnames]{xcolor}
\usepackage{framed}
\definecolor{shadecolor}{RGB}{224,224,224}
\usepackage{amsthm}
\usepackage{amsfonts}
\usepackage{soul}

\makeatletter
\def\l@subsubsection#1#2{}
\makeatother

\usepackage[dvipsnames]{xcolor}
\usepackage[normalem]{ulem}
\usepackage{wasysym} 
\usepackage[export]{adjustbox}

\makeatletter
\DeclareFontFamily{OMX}{MnSymbolE}{}
\DeclareSymbolFont{MnLargeSymbols}{OMX}{MnSymbolE}{m}{n}
\SetSymbolFont{MnLargeSymbols}{bold}{OMX}{MnSymbolE}{b}{n}
\DeclareFontShape{OMX}{MnSymbolE}{m}{n}{
    <-6>  MnSymbolE5
   <6-7>  MnSymbolE6
   <7-8>  MnSymbolE7
   <8-9>  MnSymbolE8
   <9-10> MnSymbolE9
  <10-12> MnSymbolE10
  <12->   MnSymbolE12
}{}
\DeclareFontShape{OMX}{MnSymbolE}{b}{n}{
    <-6>  MnSymbolE-Bold5
   <6-7>  MnSymbolE-Bold6
   <7-8>  MnSymbolE-Bold7
   <8-9>  MnSymbolE-Bold8
   <9-10> MnSymbolE-Bold9
  <10-12> MnSymbolE-Bold10
  <12->   MnSymbolE-Bold12
}{}

\let\llangle\@undefined
\let\rrangle\@undefined
\DeclareMathDelimiter{\llangle}{\mathopen}%
                     {MnLargeSymbols}{'164}{MnLargeSymbols}{'164}
\DeclareMathDelimiter{\rrangle}{\mathclose}%
                     {MnLargeSymbols}{'171}{MnLargeSymbols}{'171}
\makeatother

\usepackage{tikz}
\usepackage{tikz-cd}
\usetikzlibrary{arrows}
\usetikzlibrary{snakes}
\usetikzlibrary{intersections}
\usetikzlibrary{shapes.geometric}
\usetikzlibrary{decorations.pathmorphing, patterns,shapes}
\usetikzlibrary{decorations.markings}

\tikzset{
	partial ellipse/.style args={#1:#2:#3}{
		insert path={+ (#1:#3) arc (#1:#2:#3)}
	}
}

\tikzset{
	mid arrow/.style={postaction={decorate,decoration={
				markings,
				mark=at position .575 with {\arrow[#1]{stealth}}
	}}},
	near arrow/.style={postaction={decorate,decoration={
				markings,
				mark=at position .275 with {\arrow[#1]{stealth}}
	}}},
	far arrow/.style={postaction={decorate,decoration={
				markings,
				mark=at position .800 with {\arrow[#1]{stealth}}
	}}},
}

\pgfdeclarepatternformonly{south west lines}{\pgfqpoint{-0pt}{-0pt}}{\pgfqpoint{3pt}{3pt}}{\pgfqpoint{3pt}{3pt}}{
	\pgfsetlinewidth{0.4pt}
	\pgfpathmoveto{\pgfqpoint{0pt}{0pt}}
	\pgfpathlineto{\pgfqpoint{3pt}{3pt}}
	\pgfpathmoveto{\pgfqpoint{2.8pt}{-.2pt}}
	\pgfpathlineto{\pgfqpoint{3.2pt}{.2pt}}
	\pgfpathmoveto{\pgfqpoint{-.2pt}{2.8pt}}
	\pgfpathlineto{\pgfqpoint{.2pt}{3.2pt}}
	\pgfusepath{stroke}}

\definecolor{orange(ryb)}{HTML}{FFA500}
\definecolor{lightorange(ryb)}{HTML}{FFB300}
\definecolor{dodgerblue}{HTML}{1E90FF}
\definecolor{lightdodgerblue}{HTML}{4dbff7}
\definecolor{crimson}{HTML}{FF4C4C}
\definecolor{pinkerton}{HTML}{e03185}
\definecolor{forest}{HTML}{6DD189}
\definecolor{lightishgray}{HTML}{DFDFDF}
\definecolor{error-red}{HTML}{EFB2B6}

\usepackage[colorlinks=true,citecolor=dodgerblue,linkcolor=dodgerblue,urlcolor=dodgerblue,pdftitle={Funky Ising}]{hyperref}

\def \beq {\begin{equation}}
\def \eeq {\end{equation}}
\def \beqa {\begin{eqnarray}}
\def \eeqa {\end{eqnarray}}
\def \bseq {\begin{subequations}}
\def \eseq {\end{subequations}}
\newcommand{\phii}{\varphi}

\begin{document}

\title{Enforced Gaplessness from States with Exponentially Decaying Correlations}

\author{Rahul Sahay}
\thanks{These authors contributed equally to this work.}
\affiliation{Department of Physics, Harvard University, Cambridge, MA 02138, USA}
\author{Curt von Keyserlingk}
\thanks{These authors contributed equally to this work.}
\affiliation{Department of Physics, King’s College London, United Kingdom}
\author{Ruben Verresen}
\affiliation{Pritzker School of Molecular Engineering, University of Chicago, Chicago, IL 60637, USA}
\author{Carolyn Zhang}
\affiliation{Department of Physics, Harvard University, Cambridge, MA 02138, USA}

\begin{abstract}
It is well known that an exponentially localized Hamiltonian must be gapless if its ground state has algebraic correlations.
We show that even certain exponentially decaying correlations can imply gaplessness.
This is exemplified by the deformed toric code $\propto \exp(\beta \sum_{\ell} Z_{\ell}) \ket{\mathsf{TC}}$, where $|\mathsf{TC}\rangle$ is a fixed-point toric code wavefunction. 
Although it has a confined regime for $\beta > \beta_c$, recent work has drawn attention to its perimeter law loop correlations.
Here, we show that these unusual loop correlations---namely, perimeter law coexisting with a 1-form symmetry whose disorder operator has long-range order---imply that any local parent Hamiltonian must either be gapless or have a degeneracy scaling with system size.
Moreover, we construct a variational low-energy state for arbitrary local frustration-free Hamiltonians, upper bounding the finite-size gap by $\mathcal{O}(1/L^3)$ on periodic boundary conditions.
Strikingly, these variational states look like loop waves---non-quasiparticle analogs of spin waves---generated from the ground state by non-local loop operators.
Our findings have implications for identifying the subset of Hilbert space to which gapped ground states belong, and the techniques have wide applicability. For instance, a corollary of our first result is that Glauber dynamics for the ordered phase of the two-dimensional classical Ising model on the torus must have a gapless Markov transition matrix, with our second result bounding its gap.
\end{abstract}

\maketitle

\setcounter{tocdepth}{1} 

{
  \tableofcontents
}
\section{Introduction}

Ground states of gapped, local\footnote{In this work we will take `local' to mean `exponentially localized'. See footnote \ref{footnote:local} for a precise definition.} Hamiltonians have been proven to exhibit exponentially decaying correlations~\cite{LSM2004Hastings, hastings2004locality,nachtergaele2006,hastings2006} and often feature an area law scaling in their entanglement entropy, which can been rigorously proven in some cases~\cite{hastings2007,arad2013,anshu2022}.
The latter helps justify the utility of tensor network states \cite{verstraete2004renormalizationalgorithmsquantummanybody, Verstraete_2004, Verstraete_2006, Cirac2021MPSReview} as good classical descriptions of gapped ground states.
A natural question is then: \textit{What about the converse?}
%
If we are given a state with all these properties, is there a gapped Hamiltonian for which it is the ground state?

In this work, we provide an example of a family of many-body ground states in two-spatial dimensions (2D) that have exponentially decaying correlations, area law entanglement, and even exact low-rank tensor network representations but nevertheless do not admit a gapped local parent Hamiltonian.
The specific wavefunctions naturally arise in many contexts, appearing in literature under names such as the ``deformed toric code'' \cite{Papanikolaou07,castelnovo2008} or ``Rokhsar-Kivelson'' wavefunctions \cite{rokhsar1988, Castelnovo_2005, henley2004classical}.
For a system of qubits on the links of a 2D lattice, the wavefunctions are
\begin{equation} \label{eq-deformed_TC} 
    \ket{\psi(\beta)} \propto \exp\left( \frac{\beta}{2} \sum_{\ell} Z_\ell\right) \ket{\mathsf{TC}},
\end{equation}
where $\beta$ is a real parameter henceforth assumed to be finite, $\ket{\mathsf{TC}}$ is the toric code wavefunction [precisely defined in Eq.~\eqref{eq-TCWF}] \cite{Kitaev_2003}, and $X_{\ell}, Y_{\ell}, Z_{\ell}$ are the Pauli matrices at link $\ell$.
Here, $\ket{\psi(\beta)}$ interpolates between the toric code state and the product state $\ket{\uparrow}^{\otimes N}$. The state provably admits a gapped parent Hamiltonian in the toric code phase  for $\beta < \beta_c$ (e.g., $\beta_c=  \frac{\ln(1 + \sqrt{2})}{2}$ on the square lattice) \cite{gappedHam}, while for $\beta = \beta_c$ the state has algebraic correlations and hence cannot admit a gapped parent Hamiltonian ~\cite{ardonne2004,LSM2004Hastings, hastings2004locality,hastings2010locality,Isakov11}.

\begin{figure}[!t]
    \centering
    \includegraphics[width=247pt]{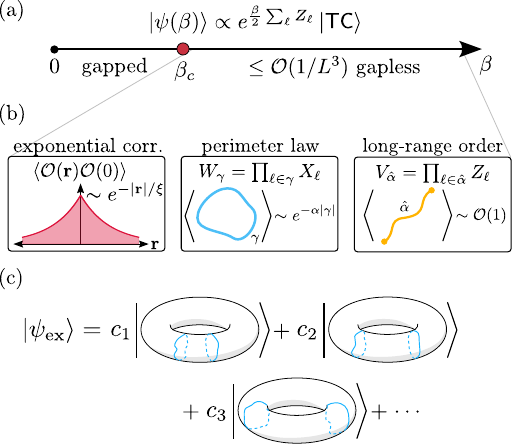}
    \caption{\textbf{Enforced Gaplessness from Exponentially Decaying Correlations.} (a) We study a family of wavefunctions [see Eq.~\eqref{eq-deformed_TC}] interpolating between a gapped toric code phase $\beta < \beta_c$ and a regime that limits to a product state $\beta > \beta_c$. While the toric code phase is known to be gapped and the point at $\beta = \beta_c$ has algebraic correlations and is manifestly gapless, we show in this work that any parent Hamiltonian for the wavefunctions with finite $\beta > \beta_c$ must be gapless (with a finite size gap $\leq \mathcal{O}(1/L^3)$ for frustration free parent Hamiltonians). (b) Crucially, this is despite the fact that the state has exponentially decaying correlations (left). This gaplessness is enforced due to an incompatiblity between the perimeter law scaling of Wilson loop operator in the state (middle) and the long-range order of the open 't Hooft string (right). (c) We conclude our work by showing, via a variational argument, that the low-energy states look like ``loop waves'' above the ground state wavefunction. }
    \label{fig:overview}
\end{figure}

For $\beta > \beta_c$, the state has exponentially decaying correlations, area law entanglement, and is not topologically ordered.
Furthermore, $\lim_{\beta \to \infty} \ket{\psi(\beta)} = \ket{\uparrow}^{N}$ and approaching this limit does not cause a diverging correlation length.
From this one might reasonably assume that in this regime, $\ket{\psi(\beta)}$ is a ground state of a gapped Hamiltonian in a trivial phase.
Here, we show that this is not the case: any local parent Hamiltonian of $\ket{\psi(\beta)}$ must be either gapless or infinitely degenerate for finite $\beta > \beta_c$. We illustrate the phase diagram in Fig.~\ref{fig:overview}(a).
This highlights that the regime $\beta > \beta_c$ is fundamentally different than the limit $\beta = \infty$, where the state is simply a product state and thus clearly admits a gapped parent Hamiltonian.

Specifically, we have two complementary results. First, we show the following (see Sec.~\ref{sec:summary} for the precise formulation):
\begin{shaded}
   \noindent \textbf{Result 1 (Informal).} There are subtle loop correlations present in $\ket{\psi(\beta > \beta_c)}$ [see Fig.~\ref{fig:overview}(b)] that
    imply that it does not admit a gapped local parent Hamiltonian on a torus.
\end{shaded}
\noindent

Next, to elucidate the nature of the gapless modes above $\ket{\psi(\beta)}$, we restrict to the case of ``frustration-free'' parent Hamiltonians and argue variationally that:
\begin{shaded}
    \noindent \textbf{Result 2 (Informal).}  Any local frustration-free parent Hamiltonian of $\ket{\psi(\beta > \beta_c)}$ has gapless modes with a gap bounded above by $c/L^{3}$, where $L$ is the linear extent of the system and $c$ is some constant.
    These modes are heuristically ``loop waves'' [Fig.~\ref{fig:overview}(c)]---analogs of spin waves in isotropic ferromagnets generated from non-local loop operators.
\end{shaded}
\noindent
In the above theorem the ``frustration-free'' condition on the Hamiltonian can be replaced with a less  restrictive (but less well-known) condition which we term ``locally annihilating'', discussed in App.~\ref{appendixvariational}.
Strictly speaking, this second result is only proven for a class of wavefunctions dual to $\ket{\psi(\beta)}$ under exact lattice dualities, which means the precise statement refers to certain symmetry properties (see next section).
Moreover, we remark that, intriguingly, the loop wave states above evade a quasi-particle description on account of their non-local nature.
We conclude by discussing how our arguments generalize to higher dimensions and outlining some directions for future work.

Our work challenges commonly held intuitions for when a given ground state admits a gapped local parent Hamiltonian, and has implications for understanding what properties distinguish gapped ground states from generic states in the Hilbert space.
We further remark that ideas similar to those presented in this work can be applied to mixed states; in a companion work~\cite{companion}, we use them to show that the toric code under decoherence is, strictly speaking, long-range entangled beyond its error threshold, albeit in a rather subtle way.
Finally, while most of our paper is about quantum systems, our work naturally provides tighter bounds
on the gap on the transition matrices that generate Markovian dynamics of the classical Ising model than have previously been known for $\beta > \beta_c$.

\section{Unusual Correlations and Summary of Results \label{sec:summary}}

Before presenting rigorous arguments, we highlight peculiar properties of the states in Eq.~\eqref{eq-deformed_TC} that are ultimately responsible for gaplessness when $\beta > \beta_c$.

\subsection{The Deformed Toric Code Wavefunctions}

The peculiar properties of the wavefunctions of Eq.~\eqref{eq-deformed_TC} can be understood by recasting them in the language of gauge theory.
In particular, if we treat the local states of the qubits as the states of a $\mathbb{Z}_2$-valued electric field $E$, visually $\left\{\ket{ \begin{tikzpicture}[scale = 0.5, baseline = {([yshift=-.5ex]current bounding box.center)}]
\draw[gray] (0, 0) -- (1, 0);
\node at (0.5, 0) {\normalsize $\downarrow$};
\end{tikzpicture}}= \ket{\frac{}{}\begin{tikzpicture}[scale = 0.5, baseline = {([yshift=-.5ex]current bounding box.center)}]
\draw[lightdodgerblue,line width=0.5mm] (0.0, 0) -- (1, 0);
\end{tikzpicture}}, \quad 
\ket{ \begin{tikzpicture}[scale = 0.5, baseline = {([yshift=-.5ex]current bounding box.center)}]
\draw[gray] (0, 0) -- (1, 0);
\node at (0.5, 0) {\normalsize $\uparrow$};
\end{tikzpicture}} = \ket{\frac{}{} \begin{tikzpicture}[scale = 0.5, baseline = {([yshift=-.5ex]current bounding box.center)}]
\draw[gray] (0, 0) -- (1, 0);
\end{tikzpicture}}\right\}$, then  $\ket{\mathsf{TC}}$ is an equal amplitude sum over all contractible\footnote{The reason to omit non-contractible loops on the torus is such that in the $\beta \to \infty$ limit of Eq.~\eqref{eq-deformed_TC} we tend to the product state $\ket{\uparrow}^{\otimes N}$.} loop configurations of $E$: 
\begin{equation}\label{eq-TCWF}
    \ket{\mathsf{TC}} \propto \ket{\frac{}{} \quad } + \ket{\frac{}{} \hspace{-0.5mm} \begin{tikzpicture}[scale = 0.5, baseline = {([yshift=-.5ex]current bounding box.center)}]
    \draw[lightdodgerblue,line width=0.5mm] (-0.5, -0.5) .. controls (-0.4, -0.7) and (-0.2, -0.9) .. (0.0, -0.5) -- (0.0, -0.5) .. controls (0.1, -0.425) and (0.5, -0.2) .. (0.25, 0.1) -- (0.25, 0.1) .. controls (0, 0.25) and (-0.3, 0.4) .. (-0.5, 0.1) -- (-0.5, 0.1) .. controls (-0.7, -0.2) and (-0.7, -0.4) .. (-0.5, -0.5);
    \draw[lightdodgerblue,line width=0.5mm]
    (0.3, 0.5) ellipse (0.3 and 0.15) -- cycle;
    \end{tikzpicture}
    } + \ket{\hspace{-2mm}\frac{}{}  \begin{tikzpicture}[scale = 0.5, baseline = {([yshift=-.5ex]current bounding box.center)}]
    \draw[lightdodgerblue,line width=0.5mm] (-0.5, -0.5) .. controls (-1, -0.2) and (-0.1, 0.2) .. (-0.5, 0.5) -- (-0.5, 0.5) .. controls (-0.9, 1) and (0.4, 1) .. (0.5, 0.5) -- (0.5, 0.5) .. controls (0.6, 0) and (0.9, -0.3) .. (0.5, -0.5) -- (0.5, -0.5) .. controls (0.1, -0.7) and (-0.1, -0.3) .. (-0.5, -0.5) -- cycle;
    \end{tikzpicture}\hspace{-1mm}}  +  \cdots,
\end{equation}
which manifestly satisfies the no-charge constraint $G_v \equiv \prod_{\ell \in v} Z_\ell = +1$, where the product is over links neighboring a vertex $v$ of the lattice.
In this language, the deformed toric code wavefunctions host two phases.
When $\beta < \beta_c$, $\psi(\beta)$ is close to the canonical toric code wavefunction which is in the topologically ordered \textit{deconfined} phase of the $\mathbb{Z}_2$ gauge theory. For $\beta > \beta_c$, the $e^{\frac{\beta}{2} \sum_{\ell} Z_{\ell}}$ appearing in Eq.~\eqref{eq-deformed_TC} gives a string tension which suppresses large electric field configurations, driving it into a \textit{confined} regime  limiting to an `empty' product state at $\beta = \infty$.

These wavefunction phases can be translated into phases of associated \textit{parent Hamiltonians}, which host them as ground states.
A family of such Hamiltonians was introduced by Castelnovo and Chamon\footnote{Eq.~\ref{eq-CastelnovoChamon} is written slightly differently than in Ref.~\onlinecite{castelnovo2008} to view the transition as a confinement transition rather than a Higgs transition.} \cite{castelnovo2008}:
\begin{equation} \label{eq-CastelnovoChamon}
    H(\beta) = \sum_{p} \left( e^{-\beta \sum_{\ell \in \partial p} Z_\ell} - \begin{tikzpicture}[scale = 0.5, baseline = {([yshift=-.5ex]current bounding box.center)}]
        \draw[color = gray] (-1, -1) -- (-1, 1) -- (1, 1) -- (1, -1) -- cycle;
        \node at (-1, 0) {\small $X$};
        \node at (1, 0) {\small $X$};
        \node at (0, -1) {\small $X$};
        \node at (0, 1) {\small $X$};
        \node at (0, 0) {\small $\textcolor{gray}{p}$};
    \end{tikzpicture}\ - \begin{tikzpicture}[scale = 0.5, baseline = {([yshift=-.5ex]current bounding box.center)}]
    \draw[gray] (1.5,0) -- (-1.5, 0);
    \draw[gray] (0,1.5) -- (0,-1.5);
    \node at (0.75, 0) {\normalsize $Z$};
    \node at (-0.75, 0) {\normalsize $Z$};
    \node at (0, 0.75) {\normalsize $Z$};
    \node at (0, -0.75) {\normalsize $Z$};
    \node at (-0.8, -0.85) {\small $\textcolor{gray}{p}$};
\end{tikzpicture}  \right),
\end{equation}
parameterized by $\beta$.
As we now explain, the confined regime ($\beta>\beta_c$) in the deformed toric code wavefunctions and their parent Hamiltonians is different to the usual confined phase of the $\mathbb{Z}_2$ gauge theory~\cite{Wegner71, Kogut_Review, Kogut_Susskind,FradkinShenker,Wilson74}.

The first piece of evidence for the peculiarity of the confined regime of $\ket{\psi(\beta)}$ is found by examining the correlations of the wavefunctions themselves.
In particular, the confinement transition (for a ``pure $\mathbb{Z}_2$ gauge theory'' where $G_v = +1$ for all $v$) is detected via the Wilson loop operator  $W_{\gamma} = \prod_{\ell \in \gamma} X_{\ell}$, where $\gamma$ is any closed loop on the lattice.
In particular, one expects that in the deconfined phase, the Wilson loop scales with a \textit{perimeter law} $\langle W_{\gamma} \rangle_{\psi} \sim e^{-\alpha |\gamma|}$ whereas in the confined phase it scales with an \textit{area law} $\langle W_{\gamma} \rangle_{\psi} \sim e^{-\alpha \text{Area}(\gamma)}$\footnote{The terminology of area/perimeter law for the Wilson loop is different than what is used in the scaling of the entanglement where dependence on the 2D area would correspond to a ``volume'' law and perimeter dependence would be an ``area'' law.}, with $|\gamma|$ being the length of the closed curve $\gamma$ and $\text{Area}(\gamma)$ being the smallest area it encloses\footnote{In modern parlance, the perimeter law can be interpreted as long-range order detecting that the deconfined phase spontaneously breaks a 1-form symmetry \cite{Gaiotto2015,McGreevy23}.}.

However, for the wavefunctions of Eq.~\eqref{eq-deformed_TC}, the expectation value of the Wilson loop obeys a perimeter law for \emph{all} values of $\beta$~\cite{huxford2023, fan2024diagnostics, bao2023mixed}, including those in the confined regime.
Explicitly,
\begin{equation} \label{eq-perimeterscaling}
    |\langle W_{\gamma} \rangle_{\psi(\beta)}| = \left|\left \langle \begin{tikzpicture}[scale = 0.9, baseline = {([yshift=-0.5ex]current bounding box.center)}]
    \draw[lightdodgerblue, line width =0.5mm] (0,0) circle (15pt);
    \node at (16 pt, -16 pt) {\small $\gamma$};
    \node at (16 pt, 16 pt) {\small $\ $};
    \node at (-16 pt, -16 pt) {\small $\ $};
    \node at (-16 pt, 16 pt) {\small $\ $};
    \end{tikzpicture}\right \rangle \right| \geq we^{- \alpha |\gamma|},
\end{equation}
where $\gamma$ is homologically trivial and sufficiently large compared to the correlation length, and  $w > 0$ is some positive constant (see Appendix~\ref{app-propertiesofdtc} for a derivation).

Indeed, the only distinction between the confined regime of $\ket{\psi(\beta)}$ and its deconfined phase is the behavior of the dual ```t Hooft'' operator $V_{\hat{\gamma}} = \prod_{\ell \in \hat{\gamma}} Z_{\ell}$.
For all closed loops on the dual lattice, contractible or non-contractible, $V_{\hat{\gamma}} \ket{\psi(\beta)} = \ket{\psi(\beta)}$ for all $\beta$. 
In other words, $V_{\hat{\gamma}}$ is a \textit{$1$-form symmetry} of the state.
However, the open 't Hooft string\footnote{In modern parlance, we say that the disorder operator of the 1-form symmetry has long-range order.} has long range order for $\beta > \beta_c$:
\begin{equation}\label{eq-fluxprolif}
    \lim_{|v - w| \to \infty}\langle V_{\hat{\alpha}} \rangle_{\psi(\beta > \beta_c)} = \left \langle\ \begin{tikzpicture}[scale = 0.8, baseline = {([yshift=-.5ex]current bounding box.center)}]
        \draw[color = lightorange(ryb), line width=0.45mm] (-0.2, -0.4) .. controls (-0.25, 0.0) and (-0.25, 0.1) .. (0.3, 0.2) .. controls (1.0, 0.3) and (1.0, 0.4) .. (1.0, 0.9);
        \filldraw[color = lightorange(ryb)] (-0.2, -0.4) circle (2 pt);
        \filldraw[color = lightorange(ryb)] (1.0, 0.9) circle (2 pt);
        \node at (0.35, 0.55) {\small $\hat{\alpha}$};
    \end{tikzpicture}\  \right \rangle_{\psi(\beta > \beta_c)} \neq 0
\end{equation}
where $v, w$ are the endpoints of the open curve $\hat{\alpha}$, but is short-ranged (i.e. decays to zero exponentially in the distance $|v-w|$) for $\beta < \beta_c$\footnote{We remark that there is ambiguity in how to truncate the closed loop operators $V_{\hat{\gamma}}$ to an open string, but the decay with respect to $|v-w|$ does not depend on the choice of truncation as long as $[V_{\hat{\alpha}},V_{\hat{\gamma}}]=0$ for all closed strings $\hat{\gamma}$. }.
This behavior occurs in the usual confined phase of $\mathbb{Z}_2$, and reflects the proliferation of $\mathbb{Z}_2$ fluxes therein.

This unusual state of affairs was recently emphasized by Huxford \textit{et al.} in Ref.~\onlinecite{huxford2023}, where it was linked to a ground state degeneracy of \eqref{eq-CastelnovoChamon} for \emph{any} $\beta$.
Here we emphasize that it in fact implies gaplessness (or degeneracy that \textit{scales with system size}).

\subsection{Deformed Ising Wavefunctions}

The perimeter law scaling described above is perhaps more surprising if we apply a combination of $em$-duality and the Kramers-Wannier duality to the wavefunctions of Eq.~\eqref{eq-deformed_TC}.
The resulting wavefunctions are defined on the vertices $v$ of the 2D lattice, taking the form: 
\begin{equation} \label{eq-deformedIsing}
    \ket{\phii(\beta)} \propto \exp\left(\frac{\beta}{2} \sum_{\langle v, w\rangle} Z_v Z_{w}\right) \ket{+}^{\otimes N}.
\end{equation}
These wavefunctions interpolate between the paramagnetic state $\ket{+}^{\otimes N}$ and an Ising ferromagnetic state $\ket{\mathsf{GHZ}} \propto \ket{\uparrow}^{\otimes N} + \ket{\downarrow}^{\otimes N}$.
Under the duality, the perimeter law of the Wilson loop (Eq.~\eqref{eq-perimeterscaling}) becomes a perimeter law of the disorder parameter:
\begin{equation}
\left|\left\langle \prod_{v \in \mathcal{R}} X_v \right \rangle_{\hspace{-1.5 mm}\phii(\beta)} \right| \geq w \; e^{-\alpha|\partial \mathcal{R}|}
\end{equation}
where $\mathcal{R}$ is a sufficiently large disk anywhere on the lattice and $\partial \mathcal{R}$ is its boundary. Moreover, the long-range order of the 't Hooft loop (Eq.~\eqref{eq-fluxprolif}) becomes long-range order of the order parameter $Z_v$ of the Ising ferromagnet:
\begin{equation}
\lim_{|v - w|\to \infty}\hspace{-0.5mm}\langle Z_v Z_w \rangle_{\phii(\beta > \beta_c)} \neq 0 .
\end{equation}
The coexistence of these two correlation functions is contrary to the typical behavior in an Ising ferromagnet, where one expects (following, say, a mean-field treatment) an area law of the disorder operator since the probability of a spin configuration appearing in the ground state wavefunction decreases multiplicatively in the number of flipped spins present. Indeed, the area law was recently proven rigorously in the vicinity of the ferromagnetic fixed point in Ref.~\onlinecite{mcdonough2025}.

\subsection{Overview of Key Results} \label{subsec-summ}

In this work, we show that the peculiar correlations present in $\ket{\psi(\beta > \beta_c)}$ and its dual $\ket{\phii(\beta > \beta_c)}$ cannot appear in the ground state subspace of any gapped local parent Hamiltonian.
Specifically, our first result extends beyond the particular form of these wavefunctions: it shows that \textit{any state} with the perimeter law scaling of Eq.~\eqref{eq-perimeterscaling} and long-range order of Eq.~\eqref{eq-fluxprolif} is incompatible with being a gapped ground state with finite degeneracy\footnote{Note that if we give up on finite degeneracy, the zero Hamiltonian $H=0$ is a parent Hamiltonian for any state.}.

However, to prove this more general result, we introduce a condition (obeyed by our deformed wavefunctions) that serves to exclude states that achieve such correlations in a trivial way---i.e., by forming a superposition of a state exhibiting a perimeter law scaling of the Wilson loop and long-range order of the 't Hooft string\footnote{We thank Chong Wang for an insightful discussion on this point.}.
An example of such a state is
\begin{equation}
    \ket{\psi} = \frac{1}{\sqrt{2}}(\ket{\uparrow}^{\otimes N} + \ket{\mathsf{TC}}),   
\end{equation}
which can naturally occur as one of a number of degenerate but otherwise gapped ground states at the first-order transitions between the toric code and confined state (see, e.g., Ref.~\onlinecite{OBrien18} for a one-dimensional analogue).
To exclude this `cat state' possibility, we demand that the connected correlations between two open strings $\hat{\alpha}_1$ and $\hat{\alpha}_2$ decays exponentially as:
\begin{equation} \label{eq-notcat}
    |\langle V_{\hat{\alpha}_1} V_{\hat{\alpha}_2} \rangle  - \langle V_{\hat{\alpha}_1} \rangle \langle V_{\hat{\alpha}_2}\rangle |\leq c e^{-\mu \ell_{12}}
\end{equation}
where $\ell_{12}$ is the minimum distance between the endpoints of $\hat{\alpha}_1$ and $\hat{\alpha}_2$, $\mu$ is a positive constant that is independent of system size,\footnote{Henceforth, when we say a constant is ``independent of system size'' we mean to say that it is independent of $L_x$ and $L_y$.} and $c$ depends only polynomially on the distance between the endpoints of both strings.
Indeed, for $\ket{\psi} \propto \ket{0}^{\otimes N} + \ket{\mathsf{TC}}$, the left hand side of the above quantity saturates to $1/4$, violating the inequality and revealing its cat state nature.
With this diagnostic in mind, we are able to state our first key result:
\begin{shaded}
    \noindent \textbf{Theorem 1.} (\textit{Gaplessness from Correlations)} Let $\ket{\psi}$ be a wavefunction defined on a 2D lattice of finite dimensional qudits with the global topology of a torus\footnotemark.
    Suppose further that there exist $\mathbb{Z}_n$ unitary string operators $W_{\gamma}$ and $V_{\hat{\gamma}}$ that braid non-trivially (i.e. $W_{\gamma} V_{\hat{\gamma}} = e^{i \theta} V_{\hat{\gamma}} W_{\gamma}$ for $\theta \neq 0$ and $\gamma$ and $\hat{\gamma}$ intersecting once).
    If the state $\ket{\psi}$
    exhibits the following properties:
    \begin{enumerate}    
        \item[(i)] a $\mathbb{Z}_n$ $1$-form symmetry given by closed $V_{\hat{\gamma}}$ loops,

        \item[(ii)] a perimeter law for closed $W_{\gamma}$ loops [Eq.~\eqref{eq-perimeterscaling}] and long-range order of open $V_{\hat \alpha}$ strings [Eq.~\eqref{eq-fluxprolif}], 

        \item[(iii)] cluster decomposition of open $V_{\hat \alpha}$ strings [Eq.~\eqref{eq-notcat}], 
        \end{enumerate}
        \noindent
        then $\ket{\psi}$ cannot be the  ground state of any gapped local Hamiltonian with finite ground state degeneracy.
\end{shaded}
\footnotetext[\value{footnote}]{In fact, the result holds for lattices whose global topology corresponds to any closed manifold with genus $g \geq 1$.}

\noindent

In the above, we define a local Hamiltonian $H$ to be one that can be written as a sum of terms $H_X$ supported on regions of the lattice $X$ whose operator norm decays at least exponentially quickly in the diameter of their support.
Notice that all of the above conditions are obeyed by $\ket{\psi(\beta)}$, implying that its parent Hamiltonians (e.g., the Castelnovo-Chamon Hamiltonian of Eq.~\eqref{eq-CastelnovoChamon}) must be gapless.
We highlight that we do \emph{not} require the Hamiltonian to have the 1-form symmetry; in fact, all our conditions are on the state $\ket \psi$, not on the Hamiltonian.

The above theorem does rely on \textbf{two plausible assumptions about gapped Hamiltonians}.
First, our theorem is strictly proven for an aspect ratio $a = L_x/L_y$ beyond a finite\footnote{It is important to keep the aspect ratio finite, otherwise the effective dimensionality of the problem can change, which can have drastic consequences.} threshold value which is independent of system size (where $L_{x, y}$ are the linear extent of the system in the $x$ and $y$ directions).
To extend our result to all finite aspect ratios, we assume the following conjecture about gapped states of matter in 2D: \textit{if a family of quantum states defined on increasing number of qudits $L_x L_y$ remains gapped in the thermodynamic limit on an open range of finite aspect ratios, it remains gapped for any finite aspect ratio.}
Such a conjecture can be proven for a wide class of topological orders (e.g. those with string-net fixed points, chiral topological orders in Kitaev's 16-fold way classification, etc.)~and is closely related to the concept of entanglement renormalization and the ``generalized $s$-source'' conjecture of Ref.~\onlinecite{swingle2016ssource}.
Relatedly, in our proof we assume that after taking the thermodynamic limit for any fixed nonzero aspect ratio, the gap and the number of degenerate ground states are bounded independent of aspect ratio.

While this first assumption is sufficient to prove our theorem for Hamiltonians with an exact ground state degeneracy, we can use a further assumption to extend beyond this.
In particular, since our proof for the case with exact degeneracy also works for Hamiltonians with exponentially suppressed long-range tails, it seems reasonable that even if one started with a Hamiltonian with finite-size splitting, one could utilize this freedom of adding long-range terms to reduce to the case with exact degeneracy.
To justify this, let us recall that in gapped quantum systems, the origin of finite-size splitting is often attributed to extensive operators $A_{L_x, L_y}$ generated in perturbation theory that distinguish between or connect different ground states (e.g. see the discussion in Ref.~\cite{Kitaev_2003}). 
Consequently, we assume that the finite-size splitting between degenerate ground states of gapped local Hamiltonians is exponentially small in the support of such operators $\sim e^{-\kappa \cdot \text{support}(A_{L_x, L_y})}$ and that such operators can be added to the Hamiltonian to then exactly cancel this degeneracy (see Appendix~\ref{app-assumption-discussion} for a further discussion on this point).


We remark that an immediate corollary of the above theorem is that any $\mathbb{Z}_2$ symmetric parent Hamiltonian of the deformed Ising wavefunction $\ket{\phii(\beta)}$ is gapless for $\beta>\beta_c$.
Indeed, for any such parent Hamiltonian of $\ket{\phii(\beta)}$ [Eq.~\eqref{eq-deformedIsing}], a combination of Kramers-Wannier and $em$-duality can be used to map the Hamiltonian to a parent Hamiltonian of $\ket{\psi(\beta)}$, which the above implies is gapless.

While the theorem above is powerful enough to imply gaplessness of $\ket{\psi(\beta)}$, it does not give us insight into the \emph{nature} of the low-energy excitations above $\ket{\psi(\beta)}$.
To get some intuition about these states, we restrict to the case of \textit{frustration-free} parent Hamiltonians---Hamiltonians that can be written as a sum of local, positive semi-definite\footnote{Theorem 2  holds even when we drop this positive semi-definiteness requirement on the local Hamiltonian terms; see Appendix~\ref{appendixvariational}.} terms that each annihilate the ground state, an example of which is given by the Castelnovo-Chamon Hamiltonian.
We work with the dual Ising wavefunctions \eqref{eq-deformedIsing} since we are then guaranteed that loops (being domain walls) cannot be broken\footnote{Alternatively, this result can be applied to $1$-form symmetric parent Hamiltonians of $\ket{\psi(\beta)}$. In the gauge theory language, these Hamiltonians are ``pure gauge theories,'' i.e. have no matter present.}.
For this setting we present a variational argument that applies in all dimensions greater than one to prove:
\begin{shaded}
    \noindent \textbf{(Physicist's) Theorem 2.} Let $\ket{\phii(\beta)}$ be the deformed Ising wavefunction on the $D$-dimensional torus.
    Then, if $\beta > \beta_c$, the many-body gap $\Delta E$ of any frustration-free parent Hamiltonian of $\ket{\phii(\beta)}$ is bounded as $\Delta E \leq c/L^{D + 1}$ where $L$ is the linear extent of the system and $c$ is an $\mathcal{O}(1)$ constant.
\end{shaded}
\noindent
 We call the above theorem a ``physicist's" theorem because the energy bound uses some approximations based on properties of the classical Ising model in the ferromagnetic phase, and is therefore not fully rigorous.

The variational argument used yields insight into the low-energy states of frustration-free parent Hamiltonians of $\ket{\phii(\beta)}$.
Indeed, we find that the low-energy states look like ``loop waves'', or analogs of spin waves in Heisenberg ferromagnets except generated from non-local loop operators (see Fig.~\ref{fig:overview}(c)).
Crucially, these excitations do not have a quasi-particle description.

Finally, we restrict to the case where these frustration-free Hamiltonians are sign-problem free.
In this case, we review a connection between such Hamiltonians and generators of Markovian dynamics that have as a steady state the Gibbs distribution of the classical Ising model at inverse temperature $\beta$.
Given this connection, we remark upon a striking corollary of Theorem 1 (we discuss the connection to existing literature in Sec.~\ref{ssurvey}):
\begin{shaded}
    \noindent \textbf{Corollary 1.} Suppose that $\mathcal{L}(\beta)$ is a $\mathbb{Z}_2$-symmetric and local generator of Markovian dynamics that is defined for Ising spins living on a torus, has as a steady state the Gibbs distribution of the classical Ising model at inverse temperature $\beta$, and whose local terms individually satisfy detailed balance\footnotemark.
    Then, for finite $\beta > \beta_c$, $\mathcal{L}(\beta)$ is either gapless or infinitely degenerate.
\end{shaded}
\footnotetext[\value{footnote}]{\label{footnote:detailedbalance}For the Ising model, let $P = e^{\beta \sum_{\langle v, w \rangle} Z_v Z_w}$. By detailed balance, we mean that the Markov generator $\mathcal{L}$ can be expressed as $\mathcal{L} = \sum_v \mathcal{L}_v(\beta)$ where $\mathcal{L}_v$ is finitely supported near site $v$ such that:
$$ P \mathcal{L}_v = \mathcal{L}_v^{\mathsf{T}} P$$
where the $\mathsf{T}$ indicates the transpose operation.
}
\noindent
In particular, this proves that the continuous time glauber dynamics of the classical Ising model on a torus has a gapless Markov generator in its ordered phase under the plausible assumptions about gapped Hamiltonians we remarked upon earlier in this section.
We conclude by reviewing existing statistical mechanics arguments pioneered by Huse and Fisher \cite{HuseFisher, miyashita1985,henkel2011}, which can be used to argue that such Markov generators and Hamiltonians are gapless in 2D.
Moreover, the results of Theorem 2 give new, tighter bounds for the gap of the Markov generator on a torus that crucially apply to all $\beta > \beta_c$.

We relegate several of the finer details of the proofs of Theorem 1 and 2 to the appendices, in which we also prove a number of interesting intermediate results (e.g. Theorem S.1. of Appendix~\ref{app-locality} and the lemmas of Appendix~\ref{app-twolemmas}) that may be of independent interest.

\begin{figure}
    \centering
    \includegraphics[width=200pt]{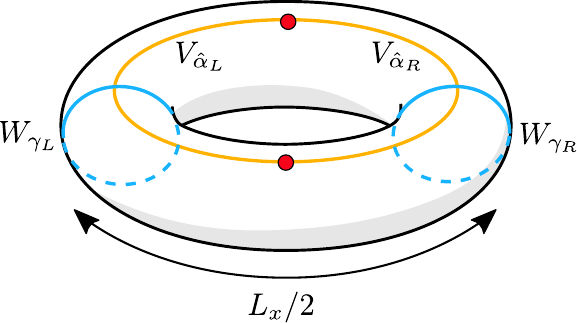}
    \caption{\textbf{Loop Configurations on the Torus.} In the proof of Theorem 1, we consider the above arrangement of string operators on the torus. 
    Notably, the closed string operators $W_{\gamma_{L/R}}$ are supported along non-contractible loops of length $L_y$ and are separated from one another by $L_x/2$.
    The open string operators $V_{\hat{\alpha}_{L/R}}$ each have a support of length $L_x/2$ and product to closed string operator $V_{\hat{\alpha}_L} V_{\hat{\alpha}_R}$ with support along a non-contractible loop.
    }
    \label{fig:loop_configs_on_torus}
\end{figure}

\section{Gaplessness from Correlations}

We now provide the proof of Theorem 1.
We start by supposing, for sake of deriving a contradiction, that $\Omega \ni \ket{\psi}$ is the ground state subspace of a gapped local Hamiltonian.\footnote{ \label{footnote:local} To be precise, by local we mean that $H = \sum_X H_X$ where $H_X$ is supported on sets $X$, such that for any lattice site $v$, there exists constants $\mu$ and $s$ independent of system size (in our case, independent of $L_x$ and $L_y$) such that:
\begin{equation}
    \sum_{X\ni v}\|H_X\||X|e^{\mu\mathrm{diam}(X)}\leq s.
\end{equation}}
Generically, there is a small splitting between degenerate ground states in a many-body system.
However, given our assumptions on gapped phases of matter discussed below Theorem 1, we can take this splitting to zero, i.e., make the degeneracy is exact.
This is because, as we discuss in Appendix~\ref{app-assumption-discussion}, if the finite-size splitting is exponentially small and our second assumption holds, we can construct a gapped and exponentilaly localized Hamiltonian with the same ground state subspace but with an exact degeneracy.

Correlations in gapped ground states must satisfy strong bounds. 
In particular, if $A_X$ and $B_Y$ are operators supported on sets $X$ and $Y$ respectively and the ground state degeneracy is exact, then correlations in $\Omega$ must obey the following ``clustering'' property
\cite{LSM2004Hastings, hastings2004locality,nachtergaele2006,hastings2010locality}:%
\begin{equation} \label{eq-clusteringHastings}
        |\langle A_X B_Y \rangle_{\psi} -\langle A_X \mathbb{P}_{\Omega} B_Y\rangle_{\psi}| \leq C_{X, Y} \|A_X\| \|B_Y\| e^{-\mu d(X, Y)},
\end{equation}
where $\mathbb{P}_{\Omega}$ is the projector onto $\Omega$: $\mathbb{P}_{\Omega}=\sum_{n\in\Omega}|n\rangle\langle n|$, $\| \mathcal{O} \| = \text{max}_{\ket{\psi}, \braket{\psi} = 1} \sqrt{\langle \mathcal{O}^{\dagger} \mathcal{O} \rangle_{\psi}}$ is the operator norm, $d(X, Y)=\min_{x\in X,y\in Y}\text{dist}(x,y)$ is the minimum distance between sets $X$ and $Y$, $\mu$ is an order one constant, and $C_{X, Y}$ is proportional to the 
size of $X$ and $Y$. 
Assuming this property, we will compute the quantity:
\begin{equation} \label{eq-contradictionquantity}
    \langle W^{\dagger}_{\gamma_L} V_{\hat{\alpha}_L} W_{\gamma_R} \rangle_{\psi},
\end{equation}
for curves $\gamma_L, \gamma_R,$ and $\hat{\alpha}_L$ that are defined in Fig.~\ref{fig:loop_configs_on_torus}.
By evaluating this expression using two different methods, we will arrive at a contradiction.
We note that Refs.~\onlinecite{hastings2011finiteT,Aharonov18,Li2024} also used incompatibility of loop correlations to show that certain states (e.g. the toric code) must be long-range entangled. However, in those cases the properties were still consistent with a gapped phase of matter.
Our case is markedly different, since we prove that states with the correlations specified in the statement of Theorem 1 are not only long-range entangled but are incompatible with any gapped, local Hamiltonian.

\subsection{First Approach from Gapped Assumption} \label{sec-method1}
Our first way of computing Eq.~\eqref{eq-contradictionquantity} involves clustering the operators on the left of the torus ($W_{\gamma_L}^{\dagger} V_{\hat{\alpha}_L}$) away from those on the right ($W_{\gamma_R}$).
In particular, we know that since $W_{\gamma_L}^\dagger V_{\hat{\alpha}_L}$ is supported a distance $L_x/4$ from $W_{\gamma_R}$, so
\begin{align}
    \langle W_{\gamma_L}^{\dagger} V_{\hat{\alpha}_L}  W_{\gamma_R} \rangle_{\psi} &=\langle W^{\dagger}_{\gamma_L} V_{\hat{\alpha}_L} \mathbb{P}_{\Omega} W_{\gamma_R} \rangle_{\psi} + \varepsilon^{(1)}_{L_x, L_y},
\end{align}
with the error bound $|\varepsilon^{(1)}_{L_x, L_y}| \leq \mathsf{poly}(L_x, L_y) e^{- \mu L_x/4}$. 
Moreover, using the fact that $V_{\hat{\alpha}_L}$ and $W_{\hat{\gamma}_L}$ braid non-trivially with one another and the fact that $V_{\hat{\alpha}_L} V_{\hat{\alpha}_R}$ is a 1-form symmetry of the state [condition (i) of the theorem] we can easily derive that
\begin{equation}
    \langle W_{\gamma_L}^{\dagger} V_{\hat{\alpha}_L} \mathbb{P}_{\Omega} W_{\gamma_R} \rangle_{\psi} = e^{-i \theta} \langle V_{\hat{\alpha}_R}^{\dagger} W_{\gamma_L}^{\dagger} \mathbb{P}_{\Omega} W_{\gamma_R} \rangle_{\psi}.
\end{equation}

We now apply a form of clustering once more on the right-hand side.
In particular, in Theorem S.1. of Appendix~\ref{app-locality}, we prove a clustering theorem [closely related to \eqref{eq-clusteringHastings}] that implies that
\begin{equation}
 \langle V_{\hat{\alpha}_R}^{\dagger} W^{\dagger}_{\gamma_L} \mathbb{P}_{\Omega} W_{\gamma_R} \rangle_{\psi} =  \langle V_{\hat{\alpha}_R}^{\dagger} \mathbb{P}_{\Omega} W^{\dagger}_{\gamma_L} \mathbb{P}_{\Omega} W_{\gamma_R} \rangle_{\psi} + \varepsilon^{(2)}_{L_x, L_y} ,  
\end{equation}
where once again $\varepsilon^{(2)}$ decays exponentially in $L_x$ (with a coefficient proportional to $L_{x, y}$).

At this point, we use an important lemma (Lemma S.2), which we prove in Appendix~\ref{app-twolemmas}
This lemma states that if condition (iii) in the statement of Theorem 1 holds and the ground state degeneracy is finite, then there exists an open string $\hat{\alpha}_L$ (which we could have chosen) that wraps half the torus and satisfies: 
\begin{equation}
    \mathbb{P}_{\Omega} V_{\hat{\alpha}_L} \ket{\psi} =  \langle V_{\hat{\alpha}_L} \rangle_{\psi} \ket{\psi} + \varepsilon^{(3)}_{L_x} \ket{\phii}
\end{equation}
where $\ket{\phii} \in \Omega$ is a normalized ground state that is orthogonal to $\ket{\psi}$, $\langle V_{\hat{\alpha}_L} \rangle_{\psi} \sim \mathcal{O}(1)$ by assumption (ii) of our Theorem,  and $|\varepsilon^{(3)}_{L_x}| \leq \mathsf{poly}(L_x) e^{-b \mu L_x}$ for some constant $b$ that is independent of system size.
Intuitively, the above equation says that $V_{\hat{\alpha}_L}$ does not drive transitions out of $\ket{\psi}$ and into other ground states.
With this in mind, we can conclude that:
\begin{equation}\label{eq-way1}
\langle W^{\dagger}_{\gamma_L} V_{\hat{\alpha}_L}  W_{\gamma_R} \rangle_{\psi} = e^{-i \theta} \langle V_{\hat{\alpha}_R}^{\dagger} \rangle_{\psi} \langle W^{\dagger}_{\gamma_L}\mathbb{P}_{\Omega} W_{\gamma_R} \rangle_{\psi} + \varepsilon^{(4)}_{L_x, L_y}  ,  
\end{equation}
where once again $\varepsilon^{(4)}_{L_x, L_y} \leq  \mathsf{poly}(L_x, L_y) e^{- \mu' L_x/2}$ for $\mu'$ independent of system size.

\subsection{Second Approach from Gapped Assumption}

Our second method of computing Eq.~\eqref{eq-contradictionquantity} involves \textit{first} using the fact that $V_{\hat{\alpha}_L} V_{\hat{\alpha}_R}$ is a $1$-form symmetry of the state [condition (i)].
In particular, we can re-write the quantity of interest as
\begin{align}
\langle W^{\dagger}_{\gamma_L} V_{\hat{\alpha}_L} W_{\gamma_R} \rangle_{\psi} &= \langle W^{\dagger}_{\gamma_L} W_{\gamma_R} V_{\hat{\alpha}_R}^{\dagger} \rangle_{\psi} \\
&=\langle W^{\dagger}_{\gamma_L}  \mathbb{P}_{\Omega} W_{\gamma_R}  V_{\hat{\alpha}_R}^{\dagger} \rangle_{\psi} + \delta^{(1)}_{L_x, L_y},
\end{align}
where the second line follows from clustering and $\delta^{(1)}$ is the associated error term bounded similarly as $\varepsilon^{(1)}$ earlier.
Then, once again using the $1$-form symmetry property of condition (i), we can re-write $V_{\hat{\alpha}_R}^{\dagger}\ket{\psi} = V_{\hat{\alpha}_L} \ket{\psi}$ and use Theorem S.1. of Appendix~\ref{app-locality} to cluster $V_{\hat{\alpha}_L}$ away from $W_{\gamma_R}$ to arrive at 
\begin{equation} \label{eq-way2}
    \langle W^{\dagger}_{\gamma_L} V_{\hat{\alpha}_L} W_{\gamma_R}  \rangle_{\psi} = \langle W^{\dagger}_{\gamma_L} \mathbb{P}_{\Omega} W_{\gamma_R} \rangle_{\psi}  \langle V_{\hat{\alpha}_R}^{\dagger} \rangle_{\psi} + \delta^{(4)}_{L_x, L_y}.
\end{equation}
In the above, we once again used condition (iii) of Theorem 1 and Lemma S.2. of Appendix~\ref{app-twolemmas} to split off the expectation value of $V_{\hat{\alpha}_L}$ (see discussion in Sec.~\ref{sec-method1}). We also the $1$-form symmetry [condition (i)] to write $\langle V_{\hat{\alpha}_L} \rangle_{\psi} = \langle V_{\hat{\alpha}_R}^{\dagger} \rangle_{\psi}$.
The error term $\delta^{(4)}_{L_x, L_y}$ is bounded as $|\delta^{(4)}_{L_x, L_y}| \leq  \mathsf{poly}(L_x, L_y) e^{- \mu' L_x/2}$.

\subsection{Arriving at a Contradiction}

At this point, we compare Eq.~\eqref{eq-way1}~and~\eqref{eq-way2}.
Indeed, if we equate the two expressions, we arrive at 
\begin{equation}
    (1 - e^{-i \theta})\langle W^{\dagger}_{\gamma_L} \mathbb{P}_{\Omega} W_{\gamma_R} \rangle_{\psi}  \langle V_{\hat{\alpha}_R}^{\dagger} \rangle_{\psi} = \varepsilon^{(4)}_{L_x, L_y} - \delta^{(4)}_{L_x, L_y}.
\end{equation}
At this point, we take the modulus of both sides.
We remark that using soley the perimeter law of  $W_{\gamma}$ for homologically trivial loops [assumed in condition (ii)], Lemma~S.1. (proven in Appendix~\ref{app-twolemmas}) shows that
\begin{equation}
 |\langle W^{\dagger}_{\gamma_L} \mathbb{P}_{\Omega} W_{\gamma_R} \rangle_{\psi}| \geq w'e^{-2\alpha' L_y} - \mathsf{poly}(L_x, L_y) e^{-\mu L_x/2},
\end{equation}
for some $w'$ and $\alpha'$.
Consequently, the result of taking the modulus yields: 
\begin{align}
    2 \sin(\theta/2) |\langle V_{\alpha_R}^{\dagger} \rangle_{\psi}| &\left(w'e^{-2\alpha' L_y} -  \mathsf{poly}(L_x, L_y) e^{-\mu L_x/2}\right) \nonumber \\
    &\leq \mathsf{poly}(L_x, L_y) e^{-\mu L_x/2}, \label{eq:inequality}
\end{align}
where in the second line we bounded the error term $\varepsilon^{(4)}_{L_x,L_y}-\delta^{(4)}_{L_x,L_y}$. 
Crucially, since $|\langle V_{\alpha_R}^{\dagger} \rangle_{\psi}|$ is order $1$ by condition (ii) and since $\theta \neq 0$, we learn that (by moving all the $L_x$-dependent exponentials in Eq.~\eqref{eq:inequality} to the right):
\begin{equation}
e^{-2\alpha' L_y} \leq \mathsf{poly}(L_x, L_y) e^{-\mu L_x/2}.
\end{equation}
In the limit of large $L_x,L_y$, this imposes a condition
\begin{equation}
2 \alpha' L_y \geq \mu L_x/2,
\end{equation}
which can be violated by choosing an aspect ratio $a = \frac{L_x}{L_y} > \frac{4 \alpha'}{\mu}$.
%
%
This provides the desired contradiction and demonstrates that $\Omega$ cannot be the ground state subspace of any gapped, local Hamiltonian. 

\hspace{0.45\textwidth} $\blacksquare$

\section{Gapless Mode Structure and Connection to Markovian Dynamics}

Having established a rigorous, correlation-based proof of gaplessness in parent Hamiltonians of $\ket{\psi(\beta)}$, we now turn to understanding the structure of the gapless modes above this state.
To do so, it will convenient to work with the dual Ising wavefunctions $\ket{\phii(\beta)}$ of Eq.~\eqref{eq-deformedIsing} and restrict to the case of frustration-free parent Hamiltonians; in Appendix~\ref{appendixvariational} we establish the same result for the more general class of `locally annihilating' Hamiltonians.

First, we present a variational argument proving the gaplessness of any frustration-free parent Hamiltonian in any dimension that further elucidates the structure of the gapless modes above the state (Theorem 2 of Sec.~\ref{subsec-summ}).
Subsequently, in the restricted case where these Hamiltonians are $\mathbb{Z}_2$ symmetric and sign-problem free, we review a known connection between $\ket{\phii(\beta)}$ and the Markovian dynamics of the classical Ising model.
We then leverage prior results on the Markovian dynamics of the classical Ising to argue for the gaplessness of such Hamiltonians and use our theorems to prove and bound the gaplessness of the Markov generator of these dynamics.

\subsection{Variational Class of States}\label{sgaplessstates}

We will now show via a variational argument that any frustration-free parent Hamiltonian has a gap that scales like an inverse power of the system size.
While we restrict the discussion below to two-dimensions, the generalizations to higher dimensions is evident from the argument presented. 
It is worth mentioning that any gapped local Hamiltonian can be written as a gapped, quasilocal Hamiltonian with superpolynomially decaying interactions \cite{kitaev2006}\footnote{While the construction in \cite{kitaev2006} assumes a unique ground state, it also applies to Hamiltonians with degenerate ground states with small splitting as long as the Hamiltonian does not have matrix elements between these ground states. This is true if the Hamiltonian is symmetric and the ground states lie in different symmetry sectors. We discuss this in more detail in Appendix~\ref{appendixff}.} (see Appendix~\ref{appendixff}). While we will assume for simplicity that the Hamiltonian is finite-range, it is not hard to see that the bound also holds for superpolynomially decaying Hamiltonians. Therefore, the states described in this section preclude gapped (in the Ising symmetric sector) Hamiltonians more generally.

To start, we introduce a class of variational states that will be convenient for defining our low-energy states.
In particular, suppose that $H = \sum_{v} H_v$ is a frustration-free parent Hamiltonian of $\ket{\phii(\beta)}$ for $\beta > \beta_c$, with each local Hamiltonian term $H_v$ being positive semi-definite (without loss of generality) and satisfying $H_v \ket{\phii(\beta)} = 0$.
To find the variational low-energy states of $H$, it will be useful to consider a ``domain wall'' representation of both $H$ and $\ket{\phii(\beta)}$.
In particular, note that the Hilbert space of both can be viewed as the span of all domain wall configuration $\text{span}\{\mathcal{C}\}$ above the Ising ferromagnetic background.
In this basis, we express the ground state wavefunction as: 
\begin{equation}
    \ket{\phii}  =  \sum_{\mathcal{C}} \phii_{\mathcal{C}} \ket{\mathcal{C}} \equiv   \sum_{\mathcal{C}} e^{-\frac{\beta}{2} |\mathcal{C}|} \ket{\mathcal{C}},
\end{equation}
up to a normalization factor, where $|\mathcal{C}|$ is the total length of the domain walls present in $\mathcal{C}$ and the $\beta$ dependence of $\ket{\phii}$ is suppressed for brevity.
As a consequence of locality, we remark that each Hamiltonian term $H_v$ can only connect domain wall configurations $\mathcal{C}$ that are close in Hamming distance; we will assume (without loss of generality) that $\mathcal{C}, \mathcal{C}'$ must be a distance of $1$ apart.
Moreover, the wavefunction amplitude $\phii_{\mathcal{C}}$ varies uniformly with respect to Hamming distance---i.e. if two domain wall configurations $\mathcal{C}, \mathcal{C}'$ are connected by the Hamiltonian, $|\phii_{\mathcal{C}}| \leq \alpha |\phii_{\mathcal{C}'}|$ for some $\alpha$.

In this domain wall representation, we define our (unnormalized) variational class of states as
\begin{equation} \label{eq-variationalclass}
    \ket{\phii[\theta]} = \sum_{\mathcal{C}} \theta_{\mathcal{C}}\, \phii_{\mathcal{C}} \ket{\mathcal{C}},
\end{equation}
where $\theta$ is a function on domain wall configurations that we must determine.
Variational class in hand, our goal now will be to find a $\theta$ that creates states with low energy which have little overlap with $\ket{\phii(\beta)}$.

\subsubsection{Gapless Mode Structure}\label{vargapless}

\begin{figure}[!t]
    \centering
    \includegraphics[width=247pt]{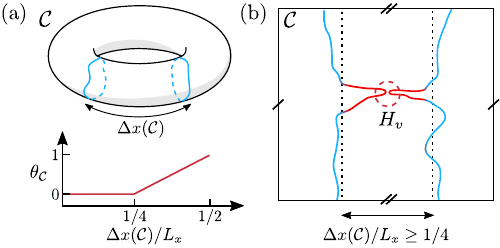}
    \caption{\textbf{Variational States.} We introduce a class of variational states whose energy is bounded above by $\sim 1/L_x^2 L_y$.
   (a) Our variational states look like ``domain wall waves.''
    Specifically, they take the form of the variational ansatz of Eq.~\eqref{eq-variationalclass}~and~\eqref{eq-trialtheta}, which projects the ground state to the space with two non-contractible domain walls (top) and deforms it with a function $\theta_{\mathcal{C}}$ depending on the separation between the domain walls' center of mass (bottom).
    The resulting states have an energy  $\lesssim 1/L_x^2 L_y$ because the Hamiltonian terms $H_v$ typically only connect configurations where $\theta$ is only slightly changed.
    (d) There are cases where $\theta$ can drastically change---e.g. for the meeting point shown, $H_v$ can change $\mathcal{C}$ to a configuration $\mathcal{C}'$ with no non-contractible loops and hence $(\theta_{C} - \theta_{\mathcal{C}'}) \sim \mathcal{O}(1)$.
    However, such configurations occur with low probability due to the large center of mass departures required (shown in red), bounding their contribution to the energy.
    }
    \label{fig:variational_states}
\end{figure}

To develop our gapless modes, we choose $\theta$ as follows.
Let $\mathbb{W}$ be the space of states consisting of two (non-contractible) domain walls that wrap around a system in the $y$-direction and any number of contractible domain walls [see Fig.~\ref{fig:variational_states}(a)].
Moreover, let $\Delta x(\mathcal{C})$ denote that difference in the center of mass of between these two domain walls, which can be at most $L_x/2$ on a torus.
Then $\theta_{\mathcal{C}}$ is defined as [Fig.~\ref{fig:variational_states}(a)]
\begin{equation} \label{eq-trialtheta}
    \theta_{\mathcal{C}} = \begin{cases}
        \text{min} \left( \frac{4\Delta x(\mathcal{C})}{L_x} - 1, 1 \right) & \mathcal{C} \in \mathbb{W}\, \&\, \frac{\Delta x}{L_x} > \frac{1}{4}\\ 
        0 & \text{otherwise}
    \end{cases},
\end{equation}
which has support $\mathbb{S} \subset \mathbb{W}$ on configurations where the the center of mass of the domain walls $\Delta x(\mathcal{C})$ is greater than a distance $L_x/4$ and smoothly interpolates from $0$ to $1$ as the domain walls get further apart.

At this point, we aim to bound the energy of the state generated by $\ket{\phii[\theta]}$ and then use the variational principle to thereby bound the gap of $H$.
The variational principle can be used because the state  $\ket{\phii[\theta]}$ becomes orthogonal to $\ket{\phii(\beta)}$ in the thermodynamic limit. 
In particular, note that $|\braket{\phii(\beta)}{\phii[\theta]}|^2$ is bounded from above by the probability of a configuration being within $\mathbb{S} \subset \mathbb{W}$.
Such a probability is equivalent to the probability of the same configuration appearing within the $D$-dimensional classical Ising model.
Since domain wall configurations are suppressed exponentially in their length for $\beta > \beta_c$, this means that $|\braket{\phii(\beta)}{\phii[\theta]}|^2$ decays exponentially in $L_y$.

With this in mind, we now show that the energy of such states $E = \sum_v \bra{\phii[\theta]}H_v \ket{\phii[\theta]}/\braket{\phii[\theta]} \leq c/(L_x^2 L_y)$. 
While the full derivation of this result is relegated to Appendix~\ref{appendixvariational}, here we sketch the high-level ideas.
In particular, one can show that $H_v |\varphi(\beta)\rangle=0$ implies, that the expectation value in the numerator is bounded by:
\begin{equation} \label{eq-variationalenergyboundmain}
    |\bra{\phii[\theta]} H_v \ket{\phii[\theta]}| \leq  \frac{\alpha \|H_v \|}{2} \sum_{\mathcal{C}, \mathcal{C}'} (\mathcal{\theta}_{\mathcal{C}} - \mathcal{\theta}_{\mathcal{C}'})^2 |\phii_{\mathcal{C}}|^2 \delta_{\mathcal{C}, \mathcal{C}'}^{v},
\end{equation}
where $\delta^v$ is one when $\mathcal{C}, \mathcal{C}'$ are connected by $H_v$ and zero otherwise.
The above makes clear that low-energy states must have a $\theta$ that mostly ``varies slowly'' in $\mathcal{C}$ and where it doesn't, the probability $|\phii_{\mathcal{C}}|^2$ should be small.
We now show that this is the case for our trial $\theta$ in Eq.~\eqref{eq-trialtheta}.

To see this, we break down the sum over $\mathcal{C}, \mathcal{C}'$ into three cases.
In the first case, where $\mathcal{C}, \mathcal{C}' \notin \mathbb{S}$, note that $\theta_{\mathcal{C}}$ is not varying with $\mathcal{C}$ and hence the contribution to Eq.~\eqref{eq-variationalenergyboundmain} is zero.
Therefore, the only cases of relevance are the second two: either when both $\mathcal{C}, \mathcal{C}' \in \mathbb{S}$  or when $\mathcal{C} \in \mathbb{S}$ and $\mathcal{C}' \notin \mathbb{S}$ (or vice versa).
We will find that in the former, $\theta_{\mathcal{C}}$ is varying slowly leading to a small energy contribution in Eq.~\eqref{eq-variationalenergyboundmain}; in the latter case, $\theta$ can vary quickly but the probability of such configurations are small similarly leading to small contributions.

In the former case, we note that $\theta_{\mathcal{C}} = 4\Delta x(\mathcal{C})/(L_x)$. 
In the most likely scenario, $\mathcal{C}$ is changed into $\mathcal{C}'$ by the action of a local Hamiltonian term $H_v$ on the domain walls of $\mathcal{C}$.
This shifts the center of mass  the center of mass of a non-contractible loop by $1/L_{y}$.
Consequently, in such cases, 
\begin{equation}\label{eq:thetadiff}
    (\theta_{\mathcal{C}} - \theta_{\mathcal{C}'})^2 =  \frac{16(\Delta x(\mathcal{C}) - \Delta x(\mathcal{C}'))^2}{L_x^2} \leq \frac{16}{L_x^2 L_y^2}.    
\end{equation}
We remark that the center of mass could change by more than $1/L_y$ if, say, a contractible domain wall in $\mathcal{C}$ is ``fused'' into one of the non-contractible domain walls.
Nevertheless, such configurations are exponentially suppressed relative to the scenario earlier and can be argued to be safely ignored (see Appendix~\ref{appendixvariational}).
Combining the above results, and relegating  the details to an appendix, the energy contribution from these configurations ($E_{\mathcal{C}, \mathcal{C}' \in \mathbb{S}}$) is bounded by:
\begin{equation}
    E_{\mathcal{C}, \mathcal{C}' \in \mathbb{S}} \leq \frac{c'}{L_x^2 L_y^2}\sum_v \frac{\alpha \|H_v \|}{2} \left(\frac{\sum_{\mathcal{C}, \mathcal{C}' \in \mathbb{S}} |\phii_{\mathcal{C}}|^2 \delta_{\mathcal{C}, \mathcal{C}'}^v \delta_{v \in \mathrm{ld}(\mathcal{C})}}{\sum_{\mathcal{C}} |\theta_{\mathcal{C}}|^2|\phii_{\mathcal{C}}|^2}\right),
\end{equation}
where $c'$ is a constant independent of system size and $\delta_{v \in \mathrm{ld}(\mathcal{C})}$ enforces that $v$ lies along one of the long winding domain walls in $\mathcal{C}$, which ultimately comes from the fact that the LHS of Eq.~\eqref{eq:thetadiff}, and hence the contribution to the energy, vanishes unless $H_v$ shifts the center of mass difference $\Delta x(\mathcal{C})$. Note that by summing over $v$, we would naively get an extra factor of $L_x L_y$. Due to this delta function, for each configuration $\mathcal{C}$, only the $v$ that lie in the vicinity of the domain walls can change $\theta(\mathcal{C})$.
Consequently, this sum simply gives yields an additional  factor of $L_y$.
Finally, it is possible to argue that the term in the parentheses is order $1$ (independent of system size) by re-expressing it as a probabilisitic computation in the classical Ising model (see Appendix~\ref{appendixff}).
The result is that $E_{\mathcal{C}, \mathcal{C}' \in \mathbb{S}} \leq c/(L_x^2 L_y)$ for some constant $c$ independent of system size.

In the latter case,  $H_v$ connects $\mathcal{C} \in \mathbb{S}$ and $\mathcal{C}' \notin \mathbb{S}$,  could potentially raise the energy of these variational states.
The situation that raises the most concern is where $\mathcal{C}$ contains a ``meeting'' point as shown in Fig.~\ref{fig:variational_states}(c), where $H_v$ is able to take a configuration with $\theta_{\mathcal{C}} \sim \mathcal{O}(1)$ to $\mathcal{C}' \in \mathbb{S}$, for which $\theta_{\mathcal{C}'} = 0$.
Consequently, $(\theta_{\mathcal{C}} - \theta_{\mathcal{C}'}) \sim \mathcal{O}(1)$, which could naively produce a large energy penalty in Eq.~\eqref{eq-variationalenergyboundmain}.
However, let us note that, by a mapping to the classical Ising model, one can show that the variance of the center of mass of a domain wall in $\mathcal{C}$ scales like $\sqrt{L_x}$.
Conversely, the meeting point requires a departure from the center of mass of order $L_x$ [shown in red in Fig.~\ref{fig:variational_states}(b)].
Hence, one can show that these meeting configurations are suppressed exponentially in $L_x$ (see Appendix~\ref{appendixvariational}).
As such, the energy contributions from this case ($E_{\mathcal{C} \in \mathbb{S}, \mathcal{C}' \notin \mathbb{S}}$) are exponentially suppressed in $L_x$.

As a consequence of the above calculations, we find that the energy of the states $\ket{\phii[\theta]}$ are bounded from above by $E = E_{\mathcal{C},\mathcal{C}' \in \mathbb{S}} + E_{\mathcal{C} \in \mathbb{S}, \mathcal{C}' \notin \mathbb{S}} \leq c/(L_x^2 L_y)$ as desired.\footnote{More precisely, the tightest bound is $E\leq \frac{c}{L_xL_y\max(L_x,L_y)}$, but we assume that the aspect ratio $L_x/L_y$ is $\mathcal{O}(1)$ and can be absorbed into $c$.}
We remark that in Appendix~\ref{app-numerics}, we provide preliminary numerical evidence that this energy bound is tight by performing density matrix renormalization group simulations on a paradigmatic frustration-free Hamiltonian for $\ket{\phii(\beta)}$ and obtaining its first excited state.
Intriguingly, we find that the excited states share qualitative similarities to the states the domain wall waves described in Eq.~\eqref{eq-trialtheta}.

We note that our variational wavefunctions share similarities to those developed in the context of $d$-isotopy loop models \cite{freedman2003non, Freedman04, Freedman05a, Freedman05b, freedman2008, dai2020}.
Indeed, both our variational wavefunctions and those constructed in these seminal works have gapless modes since the space of configurations containing large domain walls is well separated from the configurations only containing small domain walls.
However, in the works described above, the gapless modes are unstable to the addition of terms which do not change the ground state wavefunction (so-called \textit{Jones-Wenzl projectors}), which join large domain wall configurations together [e.g. the meeting points of Fig.~\ref{fig:variational_states}(b)].
In contrast, our states remain gapless even if the Hamiltonian is allowed to contain such terms. 

\subsection{Connection to Markovian Dynamics and Prior Work}
We conclude by reviewing connections between $\ket{\phii(\beta)}$ and prior work on the Markovian dynamics of the classical Ising model \cite{henley2004classical, Castelnovo_2005, Verstraete_2006}, highlighting how this links to gaplessness.
Subsequently, we will survey previous work in literature arguing for the gaplessness of certain parent Hamiltonians of $\ket{\phii(\beta)}$ (or limiting cases of these wavefunctions) based on this connection.

\subsubsection{Connections to Markovian Dynamics}

We start by considering the case of  $\mathbb{Z}_2$-symmetric frustration-free Hamiltonians that are ``sign-problem free''---i.e. their off-diagonal elements are real and non-positive\footnote{This implies that the thermal partition function of such a Hamiltonian is equivalent to the partition function of a classical statistical mechanics model in one higher spatial dimension.}, a natural example of which is the Hamiltonian
\begin{equation} \label{eq-dualCCHam}
  \widetilde{H}(\beta) = \sum_v \left( e^{-\beta \sum_{\langle v, w \rangle} Z_v Z_w} - X_v \right),   
\end{equation}
which is dual to the Castelnovo-Chamon Hamiltonian of Eq.~\eqref{eq-CastelnovoChamon} and has $\ket{\phii(\beta)}$ as its exact ground state.
In such a case, we will now review the well-known connection between such parent Hamiltonians and the Markovian dynamics (e.g., Glauber, Metropolis, etc.) of the classical Ising model and use this to provide another argument for gaplessness.

To demonstrate this, it is convenient to work with the wavefunctions of Eq.~\eqref{eq-deformedIsing}---dual to the deformed toric code wavefunctions---and express them as $\ket{\phii(\beta)} = P^{1/2} \ket{+}^{\otimes N}$, where $P = \exp\left(\beta \sum_{\langle v, w \rangle} Z_v Z_w\right)$.
Then, if $H(\beta)$ is a sign-problem-free and frustration-free Hamiltonian for $\ket{\phii(\beta)}$, it has been shown in prior works~\cite{henley2004classical, Castelnovo_2005,Verstraete_2006} that $H$ is related via a similarity transformation to a local generator of Markovian dynamics:
\begin{equation}\label{eq-Markov}
     \mathcal{L}(\beta) =  -P^{-1/2} H(\beta) P^{1/2},
\end{equation}
whose local terms will manifestly obey detailed balance (see footnote~\ref{footnote:detailedbalance}).
As a concrete example, for the Hamiltonian of Eq.~\eqref{eq-dualCCHam}, the local Markov generator is  given by
\begin{equation}
    \mathcal{L}(\beta) = -\sum_v  e^{-\beta \sum_{\langle v, w \rangle} Z_v Z_w} \left(1  - X_v \right).
\end{equation}
This generator describes the dynamics of probability distributions defined over the space of classical spin configurations $\boldsymbol{\sigma}$.
In particular, if we specify the probability distribution at time $t$, $\pi(\boldsymbol{\sigma}, t)$, with a row vector $\bra{\boldsymbol{\pi}(t)} = \sum_{\{\sigma\}} \pi(\boldsymbol{\sigma}, t) \bra{\boldsymbol{\sigma}}$, then the distribution vector evolves according to
\begin{equation}
    \frac{d}{dt}\bra{\boldsymbol{\pi}(t)} = \bra{\boldsymbol{\pi}(t)} \mathcal{L}(\beta),
\end{equation}
 and the steady state of the above process corresponds to the thermal ensemble of the classical Ising model at inverse temperature $\beta$: $\bra{\boldsymbol{\pi}_{\beta}} \propto Z^{-1}(\beta)\sum_{\boldsymbol{\sigma}} e^{\beta \sum_{\langle v, w \rangle} \sigma_v \sigma_{w}} \bra{\boldsymbol{\sigma}}$.

Connection to Markov dynamics in hand, we remark that the gaplessness of $H(\beta)$ can be phenomenologically understood by examining the dynamical spin-spin autocorrelation of the Markovian dynamics:
\begin{equation}
    C_{\beta}(t) = \langle Z_v(t) Z_v(0) \rangle_{\beta}  - \langle Z_v \rangle_{\beta}^2,
\end{equation}
as a function of time $t$, which is crucially equivalent to the imaginary-time autocorrelation function of $H(\beta)$ in the state $\ket{\phii(\beta)}$.
With some care, the decay of the above correlator can be connected to the second largest eigenvalue of $\mathcal{L}(\beta)$, whose spectrum is bounded above by $0$\footnote{
Specifically, note that via the spectral decomposition: $\langle Z(t)  Z(0) \rangle_{\beta} = \sum_{n} |\bra{\pi(\beta)} Z \ket{n}|^2 e^{-(E_n - E_0)t}$. Since the $|\bra{\pi(\beta)} Z \ket{n}|^2$ appearing the sum are all positive and their sum is equal to $1$, we know that if there is a gap from the largest to second largest eigenvalue $\gamma$: $\langle Z(t)  Z(0) \rangle_{\beta} \leq e^{-t \gamma}$.}.
In particular, if the second largest eigenvalue of $\mathcal{L}(\beta)$ is $-\gamma$ , then the autocorrelation function decays as $C_{\beta}(t) \sim e^{-t \gamma}$ at long times.
For the case of the 2D classical Ising model in its ordered phase ($\beta > \beta_c$), it was first argued in a seminal work by Huse and Fisher \cite{HuseFisher} and further evidenced numerically \cite{HF_extended_discussion, Ogielski_HF, martinelli1994two} that the ``curvature driven dynamics'' of domain wall fluctuations imply that the dynamical autocorrelation function decays with a \textit{stretched exponential} form:
\begin{equation} \label{eq:stretched}
    C_{\beta>\beta_c}(t) \sim e^{-(\gamma t)^{\alpha}},
\end{equation}
where $\alpha \neq 1$ was initially predicted to be $1/2$.
This deviation from exponential decay naturally implies that the eigenvalues of $\mathcal{L}$ must cluster around $0$, with gaps between these eigenvalues decaying to zero in the thermodynamic limit---a fact that can be made more precise by extracting the many-body spectral function from $C_{\beta}(t)$.
%
Since the spectrum of the Markov generator is $\text{spec}(\mathcal{L}(\beta)) = -\text{spec}(H(\beta))$ from Eq.~\eqref{eq-Markov}, the above naturally implies that $H(\beta)$ is gapless. 
We remark that this plausibility argument is not a proof, since the stretched exponential behavior remains a conjecture to date.

Our work places this gaplessness on rigorous footing.
In particular, let us note that, as stated in Sec.~\ref{subsec-summ}, an immediate corollary of our first theorem is that any local $\mathbb{Z}_2$-symmetric parent Hamiltonian of the deformed Ising wavefunction $\ket{\phii(\beta)}$ is gapless (as such a parent Hamiltonian maps to one for the deformed toric code under lattice dualities).
With this in mind, suppose that $\mathcal{L}(\beta)$ is a $\mathbb{Z}_2$-symmetric and local generator of Markovian dynamics that is defined for Ising spins living on a torus, has as a steady state the Gibbs distribution of the classical Ising model at inverse temperature $\beta$, and whose local terms satisfy detailed balance.
Then, by Eq.~\eqref{eq-Markov}, it is related via a similarity transformation to a $\mathbb{Z}_2$ symmetric local parent Hamiltonian for $\ket{\phii(\beta)}$.
Since such a Hamiltonian is gapless if $\beta > \beta_c$ by Theorem 1, the local Markov generator is gapless for $\beta>\beta_c$.
This proves Corollary 1 of Sec.~\ref{subsec-summ}.
As a further remark, our second result (Sec.~\ref{subsec-summ}) can be used to bound the gap of this Markov generator by $\mathcal{O}(1/L^3)$ since the Hamiltonian from Eq.~\eqref{eq-Markov} will be frustration-free.

\subsubsection{Survey of Prior Work}\label{ssurvey}

We conclude by mentioning that there is a rich literature on the dynamics of the Markovian (or ``kinetic") Ising model that have alluded to, conjectured, or argued for its gaplessness.
The aforementioned work by Huse and Fisher \cite{HuseFisher} on $\beta>\beta_c$ builds on work from Lifshitz, Allen and Cahn~\cite{lifshitz1962,allen1979}, identifying a $T\sim L^2$ timescale due to the curvature-driven relaxation of (contractible) large domain walls. 
This timescale was confirmed in Monte Carlo simulations in numerous works including Refs.~\cite{sahni1981,ohta1982,sahni1983,gunton1984}. 
In Ref.~\onlinecite{miyashita1985}, this $L^2$ timescale were linked to a suggested $1/L^2$ gap in the spectrum of $\mathcal{L}(\beta>\beta_c)$. 
Ref.~\onlinecite{masaoka2024} mentioned the above works about potential gaplessness for $\beta > \beta_c$ as a consistency check of their conjecture that if a frustration-free Hamiltonian has non-exact ground state degeneracy it must be gapless.

Furthermore, Ref.~\onlinecite{miyashita1985} conjectured that $\mathcal{L}(\beta > \beta_c)$ might be $1/L^3$ gapless when there are boundary conditions or a torus topology allowing for ``band-like" flat domain walls, due to the presence of a $L^3$ relaxation timescale for these flat domain walls. Note that a long timescale does not always imply a corresponding small gap~\cite{haga2021,rakovszky2024}, so even if one has reasonable confidence in the $L^3$ timescale, this does not immediately imply a $1/L^3$ gap.
This conjecture is corroborated by our second (physicist's) theorem and placed on a more rigorous footing. 
Moreover, our bound does not rely on the Hamiltonian being sign-problem-free (which is necessary for the mapping to a Markov process), indicating that the $1/L^3$ gap appears more generally, beyond the intuition given by the Markov process timescale.

More recently,  Ref.~\onlinecite{masaoka2024} provided exact $1/L^2$ gapless eigenstates for the Hamiltonian of Eq.~\eqref{eq-dualCCHam} at $\beta=\infty$, using the exact solvability of the 1D version of \eqref{eq-dualCCHam}. At $\beta=\infty$, the 1D version of \eqref{eq-dualCCHam} have $1/L^2$ gapless states consisting of superpositions of domain wall states with complex weights.
In 2D, at $\beta=\infty$, configurations consisting of straight domain walls along a cycle of the torus become exact ground states. Along these straight domain walls, the problem reduces to the 1D model, and one can decorate these straight domain walls with the 1D gapless modes to obtain $1/L^2$ gapless states in 2D.
If one interpolates the gapless modes that we develop in Sec.~\ref{vargapless} for $\beta > \beta_c$ to the $\beta \to \infty$ limit, the result would correspond to a superposition of straight domain wall configurations, which would have zero energy as $\beta \to \infty$ (specifically, for $\beta\geq\mathcal{O}(\log L)$).

One nice feature of the states in Ref.~\onlinecite{masaoka2024} for $\beta\to\infty$ is that they are \emph{locally gapless} above the straight domain wall ground states in that limit.
In particular, we can obtain $\mathcal{O}(1/L)$ overlap with these states by acting on the straight domain wall states with sums of operators up to size $\mathcal{O}(\log L)$.
On the other hand, the states in Sec.~\ref{vargapless} do not seem locally gapless, because the ground state is dominated by states without large domains wrapping the torus and we cannot get even $1/\mathsf{poly}(L)$ overlap with the states $|\varphi[\theta]\rangle$ with sums of local operators.
However, as we discuss more in Sec.~\ref{sdiscussion}, the stretched exponential autocorrelation function implies that there are locally gapless states, likely at slightly higher (though still $1/\mathsf{poly}(L)$) energies than those of Sec.~\ref{vargapless}, that are locally generated. It would be intriguing to see if such locally gapless states at finite $\beta$ are related to the $1/L^2$ gapless states in Ref.~\onlinecite{masaoka2024}.

\section{Discussion}\label{sdiscussion}

In this work, we showed that certain patterns of exponentially decaying correlations appearing within a quantum ground state can imply that any local parent Hamiltonian of the state must be gapless.
To do so, our work primarily centered around the peculiar loop correlations that appear in the deformed toric code wavefunctions of Eq.~\eqref{eq-deformed_TC} and their duals [Eq.~\eqref{eq-deformedIsing}].
In particular, we first proved on the torus that whenever a perimeter law for a loop operator charged under a 1-form symmetry [Eq.~\eqref{eq-perimeterscaling}] coexists with long-range order of the 1-form's disorder operator [Eq.~\eqref{eq-fluxprolif}], then any local parent Hamiltonian of the wavefunction cannot be gapped with finite ground state degeneracy.

Subsequently, in the context of the dual deformed Ising wavefunctions, we constructed variational gapless modes on the torus [Eq.~\eqref{eq-variationalclass}~and~\eqref{eq-trialtheta}] for frustration-free (more generally, locally annihilating) parent Hamiltonians whose energy decreased with system size as $\sim 1/L^{D+1}$, where $D$ is the spatial dimension of the wavefunctions.
These variational states revealed that the gapless modes above these wavefunctions are not pointlike quasi-particles, instead looking like waves of extended loops or domain walls. 

Overall, our work highlights that algebraically-decaying correlation functions are not the only static correlation indicators which disallow a gapped parent Hamiltonian.
Indeed, the states we considered had only exponentially decaying (or long-range-ordered) correlation functions. 
Consequently, it naturally invites further exploration into the question of what patterns of entanglement and correlations distinguish gapped ground states over other states in the Hilbert space.

\subsection{Lore about Higher-Form Symmetry Breaking and Topological Order}

Let us comment on how our findings can be relevant to further developing the concept of topological order \cite{Wen_90, Wilczek82, Wen89, Read89, Kivelson89, Wenbook, einarsson_1990, sachdev_book_2023, simon_topological_2023} and related phenomena. 
One definition of \textit{spontaneous 1-form symmetry breaking} (1-form SSB) commonly used in the literature is the existence of an exact $1$-form symmetry (here $V_{\hat{\gamma}}$) together with a perimeter law loop (here of the loop operator $W_{\gamma}$) with which it braids nontrivially \cite{Gaiotto2015,zhao2021}. For the case at hand, we would say there is a spontaneously broken $\mathbb{Z}_2$ 1-form symmetry.
It has been argued\footnote{This follows if there exist ground states for which the perimeter law also applies to non-contractible loops in a certain basis, which plausibly follows from the perimeter law for contractible loops---at least in gapped phases of matter.} \cite{McGreevy23} that such correlations imply that on the torus there exist ground states that are \textit{not invariant} under the 1-form generator on non-trivial cycles. 
It is often assumed that one can always dress the perimeter-law loop $W_{\gamma}$ to produce another unitary 1-form symmetry that braids nontrivially with $V_{\hat{\gamma}}$, so there is actually $\mathbb{Z}_2\times\mathbb{Z}_2$ 1-form SSB.
If so, then perimeter-law $W_{\gamma}$ together with exact $V_{\hat{\gamma}}$ would imply nontrivial topological order.

However, a recent work by Huxford et al.~\cite{huxford2023} has shown that the above definition of 1-form SSB need not imply topological order; we review their argument here.
Indeed, the Castelnovo-Chamon Hamiltonian of Eq.~\eqref{eq-CastelnovoChamon} has a perimeter law and a corresponding non-trivial ground state degeneracy \cite{matsumoto2020} for $\beta > \beta_c$, but it seems to not be in the topologically ordered toric code phase. In particular, it has zero topological entanglement entropy \cite{castelnovo2008}, violating the lower bound required for nontrivial topological order \cite{kim2023}.
To highlight in what way the above definition is weaker, consider the following properties that we would like for two ground states $\ket{\psi_1}$ and $\ket{\psi_2}$ to have in a topologically ordered phase, both of which we refer to as \emph{local indistinguishability} \cite{bravyi_2006_topological, Bravyi_2010,Bonderson13,huxford2023}:
\begin{itemize}
    \item[(i)] they should have the same expectation values for any local operator:
    \begin{equation}
        \bra{\psi_1} \mathcal O \ket{\psi_1} \approx \bra{\psi_2} \mathcal O \ket{\psi_2} ,\label{eq:ind1}
    \end{equation}
    where the error term is typically required to be superpolynomially small in system size;
    \item[(ii)] a local operator cannot connect the states:
    \begin{equation}
        \bra{\psi_1} \mathcal O \ket{\psi_2} \approx 0, \label{eq:ind2}
    \end{equation}
    again with an error superpolynomially small in system size.
\end{itemize}
Notice that the two properties are related by a basis transformation $\ket{\psi_1} \to \ket{\psi_1} + \ket{\psi_2}$ and $\ket{\psi_2} \to \ket{\psi_1} - \ket{\psi_2}$. Hence, if \emph{one} of these two properties holds for \emph{all} choices of ground states, then \emph{both} properties automatically hold. The reason to be interested in these properties is that they suggest that local perturbations of the Hamiltonian cannot split these degenerate states (although see Ref.~\onlinecite{Bravyi_2010} for subtleties); such robust degeneracy is a defining property of topological order. 

The perimeter-law decay of $W_{{\gamma}}$ and exact $V_{\hat{\gamma}}$ is consistent with (i) and (ii) holding for particular ground states, but they may not hold for \emph{all} ground states. For the Castelnovo-Chamon model \eqref{eq-CastelnovoChamon}, it was observed in Ref.~\onlinecite{matsumoto2020} that the local magnetization $\langle Z_{\ell} \rangle$ differs by $\mathcal{O}(1/L)$ rather than $\mathcal{O}(L^{-\infty})$ in different ground states that are eigenstates of $V_{\hat{\gamma}}$ along the two non-contractible cycles on a torus---this violates the superpolynomially small error term in Eq.~\eqref{eq:ind1}. If we label these four states $|(x=\pm1,y=\pm 1)\rangle$,\footnote{There may be additional ground states. These do not spoil the argument here.} then superpositions $\frac{1}{\sqrt{2}}|(1,1)\rangle\pm|(-1,-1)\rangle,\frac{1}{\sqrt{2}}|(1,-1)\rangle\pm|(-1,1)\rangle$ form states that are not invariant under the 1-form symmetry generators $V_{\hat{\gamma}}$ along noncontractible cycles but violate (ii) \cite{huxford2023}.
In Refs.~\onlinecite{huxford2023,McGreevy23} this was pointed out as an example of how this definition of higher-form symmetry breaking is weaker than topological order (conditions (i) and (ii), i.e., Eqs.~\eqref{eq:ind1} and \eqref{eq:ind2}).

The Castelnovo-Chamon model demonstrates that the perimeter-law decay of $W_{\gamma}$ together with exact $V_{\hat{\gamma}}$ does not imply that there exists a quasilocal dressed $W_{\gamma}$ that stabilizes the ground states with $V_{\hat{\gamma}}$. This is because the existence of such a dressed $W_{\gamma}$ would give a nonzero topological entanglement entropy. Therefore, we are only guaranteed SSB of a single $\mathbb{Z}_2$ 1-form symmetry with trivial self-statistics. In the Castelnovo-Chamon model for $\beta>\beta_c$, there is a topological line operator that acts nontrivially in the ground state subspace, and there does not appear to be any other quasilocal topological line that braids nontrivially with it. This violates braiding non-degeneracy (``modularity"); such states are believed to only be ground states of local Hamiltoniants if they are gapless \cite{kitaev2006}. This is consistent with our result that any local parent Hamiltonian of $|\psi(\beta)\rangle$ is gapless (or has infinite degeneracy in the thermodynamic limit).

Our work raises the question whether one can prove that for \emph{gapped} phases of matter the above definition of 1-form SSB implies topological order. 
We leave this interesting open question to future work.
It is also worth pointing out that even though (i) and (ii) do not hold for the usual error term $\mathcal O(L^{-\infty})$, they \emph{do} hold for polynomial error terms. In particular, the error terms are $\mathcal{O}(1/L)$, so they still decay to zero in the thermodynamic limit. Therefore, ground states cannot distinguished by a \emph{local} measurement, even if they are distinguished by quantities like \emph{global} magnetization. Such indistinguishability for ground states for gapless phases have been discussed before \cite{Rasmussen18} and is reminiscent of the distinct topological sectors in the the $(3+1)D$ $U(1)$ spin liquid \cite{Hermele04} where Eqs.~\eqref{eq:ind1} and \eqref{eq:ind2} are satisfied with error terms $\mathcal O(1/L^2)$. It would be interesting to explore the implications of such a (weaker) form of indistinguishability.

\subsection{Future Directions}
Our work opens a number of worthwhile research directions.
First, there are many notable aspects of the proof of Theorem 1 of our work that invite a closer examination.
In particular, our proof relied on placing these wavefunctions on a torus and assuming an aspect ratio of the system $a = L_x/L_y$ above some critical aspect ratio $a_c$ independent of system size.
While it is natural to conjecture that the aspect ratio of a quantum system should not influence its status of being gapped (see closely related conjectures in Ref.~\onlinecite{swingle2016ssource} and the discussion below Theorem 1 in Ref.~\ref{subsec-summ}), it would be interesting to see if it is possible to develop a proof that avoids a discussion of aspect ratio and works on general manifolds (including the sphere).
Our variational $1/L^3$ gapless states also seem to rely on the geometry and topology of a torus, crucially using the fact that a large domain wall wrapping a cycle has roughly equal probability of appearing at all locations along the direction perpendicular to the cycle. Since the gapless modes are highly non-local, they may depend on the topology of the manifold and the boundary conditions. 
It would be interesting to obtain a more careful analysis of this dependence.
Moreover, to extend our proof of Theorem 1 to the case of finite ground state splitting, we assumed that this splitting was of order $\sim e^{-\kappa \cdot \text{support}(A_{L_x, L_y})}$, where $A_{L_x, L_y}$ is an extensively supported operator that can be added to the Hamiltonian to exactly cancel this splitting (see discussion below Theorem 1 in Ref.~\ref{subsec-summ}).
While this assumption can be justified following intuition from perturbation theory (see Appendix~\ref{app-assumption-discussion}), it would be interesting to see if it can be rigorously justified in future work.

Furthermore, as stated in the main text, our variational states demonstrate that for frustration-free Hamiltonians in any dimension, there are low-energy states above the deformed Ising ground state that look like domain wall waves.
While strictly speaking we only obtained one such state, it is not hard to see that we can construct many other orthogonal states in a similar way, for example by using higher domain wall winding sectors and giving the coefficients complex phases.
This has intriguing implications for the Markovian dynamics of the classical Ising model.
In particular, while the stretched exponential behavior of the autocorrelation function $\langle Z_v(t) Z_v(0) \rangle$ of the classical Ising model has been argued by Huse and Fisher \cite{HuseFisher} to exist in two-dimensions, such arguments do not extend to higher dimensions.
This is not in contradiction with our gapless modes.
Since our modes are non-local they may not be detected via a local autocorrelation function.
However, since there are gapless signatures in this autocorrelation function in $D = 2$, this suggests that there could exist local gapless modes---i.e. states obtained from the ground state by acting with a sum of local operators---beyond those constructed in our work (see Sec.~\ref{ssurvey} for more discussion about this point).
We leave the understanding of this tension to future work.

Beyond our variational states, we note that the wavefunctions of Eqs.~\eqref{eq-deformed_TC}~and~\eqref{eq-deformedIsing} lie within a broader class of \textit{deformed wavefunctions}---fixed point wavefunctions of various orders acted upon by some positive deformation (e.g., imaginary time-evolution).
Such wavefunctions have a long history in the literature for the study of quantum phases and their transitions \cite{Freedman2003, Freedman04, Freedman05a, Freedman05b, castelnovo2008, Fendley_2008, Fidkowski2009, Fendley13, haegeman_2015_gauging, zhu2019gapless, schotte2019tensornetwork, xu2020tensor, xu2021characterization, sala2024decoherencewavefunctiondeformationd4,sala2024stabilityloopmodelsdecohering} and have further appeared within the context of measurement-based and dynamical state preparation \cite{Sahay_2023, sahay2024finitedepthpreparationtensornetwork, sahay2025classifying, zhu2023nishimori}, and various studies of quantum order under decoherence (e.g., see Refs.~\onlinecite{bao2023mixed, fan2024diagnostics, Lee2023quantum, lee2024exactcalculationscoherentinformation, su2024tapestry, lyons2024understandingstabilizercodeslocal, li2024replicatopologicalorderquantum,
chen_2024_unconventional, chen_2024_separability, sala2024decoherencewavefunctiondeformationd4,sala2024stabilityloopmodelsdecohering,  sohal2025noisy, ellison2025towards, hauser2024informationdynamicsdecoheredquantum,
lu2024disentanglingtransitionstopologicalorder,sun2025holographicviewmixedstatesymmetryprotected,zhang2025strongtoweakspontaneousbreaking1form}).

In light of our results, it would be interesting to understand under in what circumstances our results generalize to other classes of deformed wavefunctions and further explore the implications our results for the various use cases for these wavefunctions.
As an example, in the context of decoherence, we highlight that in a companion work \cite{companion}, we explore how loop correlation functions indicate that the toric code under decoherence has a rather subtle status between short- and long-range entanglement beyond its error threshold.

Relatedly, it would be interesting to understand the stability of such wavefunctions and their gapless parent Hamiltonians to perturbations, and where they lie in more general phase diagrams. The indistinguishability conditions (\ref{eq:ind1}) and (\ref{eq:ind2}) only hold with $\mathcal O(1/L)$ error terms, but this does not by itself preclude stability to perturbations. For example, in $(3+1)D$, the gapless $U(1)$ spin liquid is known to be stable to perturbations\cite{Hermele04}, even though (\ref{eq:ind1}) and (\ref{eq:ind2}) have error terms that are $\mathcal{O}(1/L^2)$. Previous numerical work has shown that the $(2+0)D$ critical point at $\beta=\beta_c$ immediately flows to conventional $(2+1)D$ quantum Ising criticality upon perturbing the Hamiltonian\cite{Isakov11}, suggesting that the Hamiltonian likely immediately gaps out with such perturbations. 
On the other hand, perturbations related to wavefunction deformations with both Pauli $X$ and $Z$ terms, i.e. $|\psi(g_x,g_z)\rangle\propto\prod_e(1+g_xX_e+g_zZ_e)|\mathsf{TC}\rangle$, have been shown to preserve a single exact (dressed) 1-form symmetry in regions of the phase diagram with respect to $(g_x,g_y)$ \cite{haegeman_2015_gauging,haegeman2015shadows,zhu2019gapless,liu2025}. $|\psi(g_x,g_z)\rangle$ in these regions seems to satisfy the conditions required for gaplessness in Theorem 1, indicating that the gapless properties presented here may persist not only along fine-tuned lines but also in entire regions in parameter space.

Lastly, it would be interesting to investigate these novel types of gap-forbidding correlations from a tensor network point of view. 
In fact, the states discussed in this work admit an exact projected entangled pair state (PEPS) description with bond dimension $D=2$ \cite{verstraete2004renormalizationalgorithmsquantummanybody,nishio2004tensorproductvariationalformulation, Verstraete_2004, Verstraete_2006}. 
The PEPS for the toric code state is known to have a `\emph{virtual}' 1-form symmetry, whose presence is important\footnote{Breaking this virtual symmetry at the PEPS level has provided a mechanism for constructing gapless parent Hamiltonians called `uncle Hamiltonians' \cite{uncle,uncle2,pivot}. In contrast, here we are exploring the scenario where no gapped parent Hamiltonian exists.} to its topological stability \cite{chen20101formpeps,schuch2013topological,Sahinoglu2021-zq, shukla_2018_boson}.
The deformed state [Eq.~\eqref{eq-deformed_TC}] preserves this property, even in the confined regime $\beta > \beta_c$. In addition our state has a \emph{physical} 1-form symmetry, whose disorder operator has long-range order. It would be interesting to investigate what these ingredients imply about the local tensor properties (in particular its symmetries), as this can perhaps provide an alternative route to proving that a parent Hamiltonian must be gapless. As a by-product, this would shed light on the local tensor properties best suited for describing gapped phases of matter.

\section{Acknowledgements}
We would like to thank Tarun Grover, Sarang Gopalakrishnan, Tim Hsieh, David Huse, Vedika Khemani, Michael Levin, Zhu-Xi Luo, Ruochen Ma, Nandgopal Manoj, Lokeshwar Prasad, Shengqi Sang, Raman Sohal, Chong Wang,  Ashvin Vishwanath, Cenke Xu, Yizhi You, and Guo-Yi Zhu for helpful conversations.
We especially thank Haruki Watanabe for pointing out an error in App.~\ref{appendixvariational}, relating to the distinction between locally annihilating and frustration free Hamiltonians. We also thank Pablo Sala for detailed comments on the manuscript and Tibor Rakovszky for pointing out subtleties regarding Glauber dynamics in higher dimensions, and the roughening transition in 3D.
R.S. acknowledges support from the U.S. Department of Energy, Office of Science, Office of Advanced Scientific Computing Research, Department of Energy Computational Science Graduate Fellowship under Award Number
DESC0022158.
C.Z. thanks the audience at the KITP 2024 conference on Correlated Gapless Quantum Matter for thought-provoking feedback, and Ben McDonough, Chao Yin, and Andrew Lucas for collaboration on related work~\cite{mcdonough2025}.
C.Z. is supported by the Harvard Society of Fellows and the Simons Collaboration on Ultra Quantum Matter. 
C.v.K. was supported by a UKRI FLF through MR/Z000297/1 and MR/T040947/2. This work was performed in part at the Aspen Center for Physics, which is supported by National Science Foundation grant PHY-2210452 and a grant from the Alfred P. Sloan Foundation (G-2024-22395).
DMRG simulations in the appendix were performed on the Harvard FASRC facility using the TeNPy Library \cite{Hauschild18}, which was
inspired by a previous library \cite{Kjaell13}.

\bibliography{refs}

\begin{thebibliography}{129}%
\makeatletter
\providecommand \@ifxundefined [1]{%
 \@ifx{#1\undefined}
}%
\providecommand \@ifnum [1]{%
 \ifnum #1\expandafter \@firstoftwo
 \else \expandafter \@secondoftwo
 \fi
}%
\providecommand \@ifx [1]{%
 \ifx #1\expandafter \@firstoftwo
 \else \expandafter \@secondoftwo
 \fi
}%
\providecommand \natexlab [1]{#1}%
\providecommand \enquote  [1]{``#1''}%
\providecommand \bibnamefont  [1]{#1}%
\providecommand \bibfnamefont [1]{#1}%
\providecommand \citenamefont [1]{#1}%
\providecommand \href@noop [0]{\@secondoftwo}%
\providecommand \href [0]{\begingroup \@sanitize@url \@href}%
\providecommand \@href[1]{\@@startlink{#1}\@@href}%
\providecommand \@@href[1]{\endgroup#1\@@endlink}%
\providecommand \@sanitize@url [0]{\catcode `\\12\catcode `\$12\catcode
  `\&12\catcode `\#12\catcode `\^12\catcode `\_12\catcode `\%12\relax}%
\providecommand \@@startlink[1]{}%
\providecommand \@@endlink[0]{}%
\providecommand \url  [0]{\begingroup\@sanitize@url \@url }%
\providecommand \@url [1]{\endgroup\@href {#1}{\urlprefix }}%
\providecommand \urlprefix  [0]{URL }%
\providecommand \Eprint [0]{\href }%
\providecommand \doibase [0]{https://doi.org/}%
\providecommand \selectlanguage [0]{\@gobble}%
\providecommand \bibinfo  [0]{\@secondoftwo}%
\providecommand \bibfield  [0]{\@secondoftwo}%
\providecommand \translation [1]{[#1]}%
\providecommand \BibitemOpen [0]{}%
\providecommand \bibitemStop [0]{}%
\providecommand \bibitemNoStop [0]{.\EOS\space}%
\providecommand \EOS [0]{\spacefactor3000\relax}%
\providecommand \BibitemShut  [1]{\csname bibitem#1\endcsname}%
\let\auto@bib@innerbib\@empty
\bibitem [{\citenamefont {Hastings}(2004{\natexlab{a}})}]{LSM2004Hastings}%
  \BibitemOpen
  \bibfield  {author} {\bibinfo {author} {\bibfnamefont {M.~B.}\ \bibnamefont
  {Hastings}},\ }\bibfield  {title} {\bibinfo {title} {Lieb-schultz-mattis in
  higher dimensions},\ }\href {https://doi.org/10.1103/PhysRevB.69.104431}
  {\bibfield  {journal} {\bibinfo  {journal} {Phys. Rev. B}\ }\textbf {\bibinfo
  {volume} {69}},\ \bibinfo {pages} {104431} (\bibinfo {year}
  {2004}{\natexlab{a}})}\BibitemShut {NoStop}%
\bibitem [{\citenamefont
  {Hastings}(2004{\natexlab{b}})}]{hastings2004locality}%
  \BibitemOpen
  \bibfield  {author} {\bibinfo {author} {\bibfnamefont {M.~B.}\ \bibnamefont
  {Hastings}},\ }\bibfield  {title} {\bibinfo {title} {Locality in quantum and
  markov dynamics on lattices and networks},\ }\href
  {https://doi.org/10.1103/PhysRevLett.93.140402} {\bibfield  {journal}
  {\bibinfo  {journal} {Phys. Rev. Lett.}\ }\textbf {\bibinfo {volume} {93}},\
  \bibinfo {pages} {140402} (\bibinfo {year} {2004}{\natexlab{b}})}\BibitemShut
  {NoStop}%
\bibitem [{\citenamefont {Nachtergaele}\ and\ \citenamefont
  {Sims}(2006)}]{nachtergaele2006}%
  \BibitemOpen
  \bibfield  {author} {\bibinfo {author} {\bibfnamefont {B.}~\bibnamefont
  {Nachtergaele}}\ and\ \bibinfo {author} {\bibfnamefont {R.}~\bibnamefont
  {Sims}},\ }\bibfield  {title} {\bibinfo {title} {Lieb-robinson bounds and the
  exponential clustering theorem},\ }\href
  {https://link.springer.com/article/10.1007/s00220-006-1556-1} {\bibfield
  {journal} {\bibinfo  {journal} {Communications in mathematical physics}\
  }\textbf {\bibinfo {volume} {265}},\ \bibinfo {pages} {119} (\bibinfo {year}
  {2006})}\BibitemShut {NoStop}%
\bibitem [{\citenamefont {Hastings}\ and\ \citenamefont
  {Koma}(2006)}]{hastings2006}%
  \BibitemOpen
  \bibfield  {author} {\bibinfo {author} {\bibfnamefont {M.~B.}\ \bibnamefont
  {Hastings}}\ and\ \bibinfo {author} {\bibfnamefont {T.}~\bibnamefont
  {Koma}},\ }\bibfield  {title} {\bibinfo {title} {Spectral gap and exponential
  decay of correlations},\ }\href
  {https://link.springer.com/article/10.1007/s00220-006-0030-4} {\bibfield
  {journal} {\bibinfo  {journal} {Communications in mathematical physics}\
  }\textbf {\bibinfo {volume} {265}},\ \bibinfo {pages} {781} (\bibinfo {year}
  {2006})}\BibitemShut {NoStop}%
\bibitem [{\citenamefont {Hastings}(2007)}]{hastings2007}%
  \BibitemOpen
  \bibfield  {author} {\bibinfo {author} {\bibfnamefont {M.~B.}\ \bibnamefont
  {Hastings}},\ }\bibfield  {title} {\bibinfo {title} {An area law for
  one-dimensional quantum systems},\ }\href
  {https://iopscience.iop.org/article/10.1088/1742-5468/2007/08/P08024}
  {\bibfield  {journal} {\bibinfo  {journal} {Journal of statistical mechanics:
  theory and experiment}\ }\textbf {\bibinfo {volume} {2007}},\ \bibinfo
  {pages} {P08024} (\bibinfo {year} {2007})}\BibitemShut {NoStop}%
\bibitem [{\citenamefont {Arad}\ \emph {et~al.}(2013)\citenamefont {Arad},
  \citenamefont {Kitaev}, \citenamefont {Landau},\ and\ \citenamefont
  {Vazirani}}]{arad2013}%
  \BibitemOpen
  \bibfield  {author} {\bibinfo {author} {\bibfnamefont {I.}~\bibnamefont
  {Arad}}, \bibinfo {author} {\bibfnamefont {A.}~\bibnamefont {Kitaev}},
  \bibinfo {author} {\bibfnamefont {Z.}~\bibnamefont {Landau}},\ and\ \bibinfo
  {author} {\bibfnamefont {U.}~\bibnamefont {Vazirani}},\ }\bibfield  {title}
  {\bibinfo {title} {An area law and sub-exponential algorithm for 1d
  systems},\ }\href {https://arxiv.org/abs/1301.1162} {\bibfield  {journal}
  {\bibinfo  {journal} {arXiv preprint arXiv:1301.1162}\ } (\bibinfo {year}
  {2013})}\BibitemShut {NoStop}%
\bibitem [{\citenamefont {Anshu}\ \emph {et~al.}(2022)\citenamefont {Anshu},
  \citenamefont {Arad},\ and\ \citenamefont {Gosset}}]{anshu2022}%
  \BibitemOpen
  \bibfield  {author} {\bibinfo {author} {\bibfnamefont {A.}~\bibnamefont
  {Anshu}}, \bibinfo {author} {\bibfnamefont {I.}~\bibnamefont {Arad}},\ and\
  \bibinfo {author} {\bibfnamefont {D.}~\bibnamefont {Gosset}},\ }\bibfield
  {title} {\bibinfo {title} {An area law for 2d frustration-free spin
  systems},\ }in\ \href@noop {} {\emph {\bibinfo {booktitle} {Proceedings of
  the 54th Annual ACM SIGACT Symposium on Theory of Computing}}}\ (\bibinfo
  {year} {2022})\ pp.\ \bibinfo {pages} {12--18}\BibitemShut {NoStop}%
\bibitem [{\citenamefont {Verstraete}\ and\ \citenamefont
  {Cirac}(2004{\natexlab{a}})}]{verstraete2004renormalizationalgorithmsquantummanybody}%
  \BibitemOpen
  \bibfield  {author} {\bibinfo {author} {\bibfnamefont {F.}~\bibnamefont
  {Verstraete}}\ and\ \bibinfo {author} {\bibfnamefont {J.~I.}\ \bibnamefont
  {Cirac}},\ }\href {https://arxiv.org/abs/cond-mat/0407066} {\bibinfo {title}
  {Renormalization algorithms for quantum-many body systems in two and higher
  dimensions}} (\bibinfo {year} {2004}{\natexlab{a}}),\ \Eprint
  {https://arxiv.org/abs/cond-mat/0407066} {arXiv:cond-mat/0407066
  [cond-mat.str-el]} \BibitemShut {NoStop}%
\bibitem [{\citenamefont {Verstraete}\ and\ \citenamefont
  {Cirac}(2004{\natexlab{b}})}]{Verstraete_2004}%
  \BibitemOpen
  \bibfield  {author} {\bibinfo {author} {\bibfnamefont {F.}~\bibnamefont
  {Verstraete}}\ and\ \bibinfo {author} {\bibfnamefont {J.~I.}\ \bibnamefont
  {Cirac}},\ }\bibfield  {title} {\bibinfo {title} {Valence-bond states for
  quantum computation},\ }\bibfield  {journal} {\bibinfo  {journal} {Physical
  Review A}\ }\textbf {\bibinfo {volume} {70}},\ \href
  {https://doi.org/10.1103/physreva.70.060302} {10.1103/physreva.70.060302}
  (\bibinfo {year} {2004}{\natexlab{b}})\BibitemShut {NoStop}%
\bibitem [{\citenamefont {Verstraete}\ \emph {et~al.}(2006)\citenamefont
  {Verstraete}, \citenamefont {Wolf}, \citenamefont {Perez-Garcia},\ and\
  \citenamefont {Cirac}}]{Verstraete_2006}%
  \BibitemOpen
  \bibfield  {author} {\bibinfo {author} {\bibfnamefont {F.}~\bibnamefont
  {Verstraete}}, \bibinfo {author} {\bibfnamefont {M.~M.}\ \bibnamefont
  {Wolf}}, \bibinfo {author} {\bibfnamefont {D.}~\bibnamefont {Perez-Garcia}},\
  and\ \bibinfo {author} {\bibfnamefont {J.~I.}\ \bibnamefont {Cirac}},\
  }\bibfield  {title} {\bibinfo {title} {Criticality, the area law, and the
  computational power of projected entangled pair states},\ }\bibfield
  {journal} {\bibinfo  {journal} {Physical Review Letters}\ }\textbf {\bibinfo
  {volume} {96}},\ \href {https://doi.org/10.1103/physrevlett.96.220601}
  {10.1103/physrevlett.96.220601} (\bibinfo {year} {2006})\BibitemShut
  {NoStop}%
\bibitem [{\citenamefont {Cirac}\ \emph {et~al.}(2021)\citenamefont {Cirac},
  \citenamefont {P\'erez-Garc\'{\i}a}, \citenamefont {Schuch},\ and\
  \citenamefont {Verstraete}}]{Cirac2021MPSReview}%
  \BibitemOpen
  \bibfield  {author} {\bibinfo {author} {\bibfnamefont {J.~I.}\ \bibnamefont
  {Cirac}}, \bibinfo {author} {\bibfnamefont {D.}~\bibnamefont
  {P\'erez-Garc\'{\i}a}}, \bibinfo {author} {\bibfnamefont {N.}~\bibnamefont
  {Schuch}},\ and\ \bibinfo {author} {\bibfnamefont {F.}~\bibnamefont
  {Verstraete}},\ }\bibfield  {title} {\bibinfo {title} {Matrix product states
  and projected entangled pair states: Concepts, symmetries, theorems},\ }\href
  {https://doi.org/10.1103/RevModPhys.93.045003} {\bibfield  {journal}
  {\bibinfo  {journal} {Rev. Mod. Phys.}\ }\textbf {\bibinfo {volume} {93}},\
  \bibinfo {pages} {045003} (\bibinfo {year} {2021})}\BibitemShut {NoStop}%
\bibitem [{\citenamefont {Papanikolaou}\ \emph {et~al.}(2007)\citenamefont
  {Papanikolaou}, \citenamefont {Raman},\ and\ \citenamefont
  {Fradkin}}]{Papanikolaou07}%
  \BibitemOpen
  \bibfield  {author} {\bibinfo {author} {\bibfnamefont {S.}~\bibnamefont
  {Papanikolaou}}, \bibinfo {author} {\bibfnamefont {K.~S.}\ \bibnamefont
  {Raman}},\ and\ \bibinfo {author} {\bibfnamefont {E.}~\bibnamefont
  {Fradkin}},\ }\bibfield  {title} {\bibinfo {title} {Topological phases and
  topological entropy of two-dimensional systems with finite correlation
  length},\ }\href {https://doi.org/10.1103/PhysRevB.76.224421} {\bibfield
  {journal} {\bibinfo  {journal} {Phys. Rev. B}\ }\textbf {\bibinfo {volume}
  {76}},\ \bibinfo {pages} {224421} (\bibinfo {year} {2007})}\BibitemShut
  {NoStop}%
\bibitem [{\citenamefont {Castelnovo}\ and\ \citenamefont
  {Chamon}(2008)}]{castelnovo2008}%
  \BibitemOpen
  \bibfield  {author} {\bibinfo {author} {\bibfnamefont {C.}~\bibnamefont
  {Castelnovo}}\ and\ \bibinfo {author} {\bibfnamefont {C.}~\bibnamefont
  {Chamon}},\ }\bibfield  {title} {\bibinfo {title} {Quantum topological phase
  transition at the microscopic level},\ }\href
  {https://doi.org/10.1103/PhysRevB.77.054433} {\bibfield  {journal} {\bibinfo
  {journal} {Phys. Rev. B}\ }\textbf {\bibinfo {volume} {77}},\ \bibinfo
  {pages} {054433} (\bibinfo {year} {2008})}\BibitemShut {NoStop}%
\bibitem [{\citenamefont {Rokhsar}\ and\ \citenamefont
  {Kivelson}(1988)}]{rokhsar1988}%
  \BibitemOpen
  \bibfield  {author} {\bibinfo {author} {\bibfnamefont {D.~S.}\ \bibnamefont
  {Rokhsar}}\ and\ \bibinfo {author} {\bibfnamefont {S.~A.}\ \bibnamefont
  {Kivelson}},\ }\bibfield  {title} {\bibinfo {title} {Superconductivity and
  the quantum hard-core dimer gas},\ }\href
  {https://doi.org/10.1103/PhysRevLett.61.2376} {\bibfield  {journal} {\bibinfo
   {journal} {Phys. Rev. Lett.}\ }\textbf {\bibinfo {volume} {61}},\ \bibinfo
  {pages} {2376} (\bibinfo {year} {1988})}\BibitemShut {NoStop}%
\bibitem [{\citenamefont {Castelnovo}\ \emph {et~al.}(2005)\citenamefont
  {Castelnovo}, \citenamefont {Chamon}, \citenamefont {Mudry},\ and\
  \citenamefont {Pujol}}]{Castelnovo_2005}%
  \BibitemOpen
  \bibfield  {author} {\bibinfo {author} {\bibfnamefont {C.}~\bibnamefont
  {Castelnovo}}, \bibinfo {author} {\bibfnamefont {C.}~\bibnamefont {Chamon}},
  \bibinfo {author} {\bibfnamefont {C.}~\bibnamefont {Mudry}},\ and\ \bibinfo
  {author} {\bibfnamefont {P.}~\bibnamefont {Pujol}},\ }\bibfield  {title}
  {\bibinfo {title} {From quantum mechanics to classical statistical physics:
  Generalized rokhsar–kivelson hamiltonians and the “stochastic matrix
  form” decomposition},\ }\href {https://doi.org/10.1016/j.aop.2005.01.006}
  {\bibfield  {journal} {\bibinfo  {journal} {Annals of Physics}\ }\textbf
  {\bibinfo {volume} {318}},\ \bibinfo {pages} {316–344} (\bibinfo {year}
  {2005})}\BibitemShut {NoStop}%
\bibitem [{\citenamefont {Henley}(2004)}]{henley2004classical}%
  \BibitemOpen
  \bibfield  {author} {\bibinfo {author} {\bibfnamefont {C.~L.}\ \bibnamefont
  {Henley}},\ }\bibfield  {title} {\bibinfo {title} {From classical to quantum
  dynamics at rokhsar--kivelson points},\ }\href@noop {} {\bibfield  {journal}
  {\bibinfo  {journal} {Journal of Physics: Condensed Matter}\ }\textbf
  {\bibinfo {volume} {16}},\ \bibinfo {pages} {S891} (\bibinfo {year}
  {2004})}\BibitemShut {NoStop}%
\bibitem [{\citenamefont {Kitaev}(2003)}]{Kitaev_2003}%
  \BibitemOpen
  \bibfield  {author} {\bibinfo {author} {\bibfnamefont {A.}~\bibnamefont
  {Kitaev}},\ }\bibfield  {title} {\bibinfo {title} {Fault-tolerant quantum
  computation by anyons},\ }\href
  {https://doi.org/10.1016/s0003-4916(02)00018-0} {\bibfield  {journal}
  {\bibinfo  {journal} {Annals of Physics}\ }\textbf {\bibinfo {volume}
  {303}},\ \bibinfo {pages} {2–30} (\bibinfo {year} {2003})}\BibitemShut
  {NoStop}%
\bibitem [{\citenamefont {Perez-Garcia}\ \emph {et~al.}(2008)\citenamefont
  {Perez-Garcia}, \citenamefont {Verstraete}, \citenamefont {Wolf},\ and\
  \citenamefont {Cirac}}]{gappedHam}%
  \BibitemOpen
  \bibfield  {author} {\bibinfo {author} {\bibfnamefont {D.}~\bibnamefont
  {Perez-Garcia}}, \bibinfo {author} {\bibfnamefont {F.}~\bibnamefont
  {Verstraete}}, \bibinfo {author} {\bibfnamefont {M.~M.}\ \bibnamefont
  {Wolf}},\ and\ \bibinfo {author} {\bibfnamefont {J.~I.}\ \bibnamefont
  {Cirac}},\ }\bibfield  {title} {\bibinfo {title} {Peps as unique ground
  states of local hamiltonians},\ }\href@noop {} {\bibfield  {journal}
  {\bibinfo  {journal} {Quantum Info. Comput.}\ }\textbf {\bibinfo {volume}
  {8}},\ \bibinfo {pages} {650–663} (\bibinfo {year} {2008})}\BibitemShut
  {NoStop}%
\bibitem [{\citenamefont {Ardonne}\ \emph {et~al.}(2004)\citenamefont
  {Ardonne}, \citenamefont {Fendley},\ and\ \citenamefont
  {Fradkin}}]{ardonne2004}%
  \BibitemOpen
  \bibfield  {author} {\bibinfo {author} {\bibfnamefont {E.}~\bibnamefont
  {Ardonne}}, \bibinfo {author} {\bibfnamefont {P.}~\bibnamefont {Fendley}},\
  and\ \bibinfo {author} {\bibfnamefont {E.}~\bibnamefont {Fradkin}},\
  }\bibfield  {title} {\bibinfo {title} {Topological order and conformal
  quantum critical points},\ }\href
  {https://www.sciencedirect.com/science/article/abs/pii/S0003491604000247?via%3Dihub}
  {\bibfield  {journal} {\bibinfo  {journal} {Annals of Physics}\ }\textbf
  {\bibinfo {volume} {310}},\ \bibinfo {pages} {493} (\bibinfo {year}
  {2004})}\BibitemShut {NoStop}%
\bibitem [{\citenamefont {Hastings}(2010)}]{hastings2010locality}%
  \BibitemOpen
  \bibfield  {author} {\bibinfo {author} {\bibfnamefont {M.~B.}\ \bibnamefont
  {Hastings}},\ }\bibfield  {title} {\bibinfo {title} {Locality in quantum
  systems},\ }\href@noop {} {\bibfield  {journal} {\bibinfo  {journal} {Quantum
  Theory from Small to Large Scales}\ }\textbf {\bibinfo {volume} {95}},\
  \bibinfo {pages} {171} (\bibinfo {year} {2010})}\BibitemShut {NoStop}%
\bibitem [{\citenamefont {Isakov}\ \emph {et~al.}(2011)\citenamefont {Isakov},
  \citenamefont {Fendley}, \citenamefont {Ludwig}, \citenamefont {Trebst},\
  and\ \citenamefont {Troyer}}]{Isakov11}%
  \BibitemOpen
  \bibfield  {author} {\bibinfo {author} {\bibfnamefont {S.~V.}\ \bibnamefont
  {Isakov}}, \bibinfo {author} {\bibfnamefont {P.}~\bibnamefont {Fendley}},
  \bibinfo {author} {\bibfnamefont {A.~W.~W.}\ \bibnamefont {Ludwig}}, \bibinfo
  {author} {\bibfnamefont {S.}~\bibnamefont {Trebst}},\ and\ \bibinfo {author}
  {\bibfnamefont {M.}~\bibnamefont {Troyer}},\ }\bibfield  {title} {\bibinfo
  {title} {Dynamics at and near conformal quantum critical points},\ }\href
  {https://doi.org/10.1103/PhysRevB.83.125114} {\bibfield  {journal} {\bibinfo
  {journal} {Phys. Rev. B}\ }\textbf {\bibinfo {volume} {83}},\ \bibinfo
  {pages} {125114} (\bibinfo {year} {2011})}\BibitemShut {NoStop}%
\bibitem [{\citenamefont {{Sahay et al.}}(2024)}]{companion}%
  \BibitemOpen
  \bibfield  {author} {\bibinfo {author} {\bibnamefont {{Sahay et al.}}},\
  }\href@noop {} {\bibinfo {title} {Is the decohered toric code long-range
  entangled beyond its error threshold?}} (\bibinfo {year} {2024}),\ \bibinfo
  {note} {to appear}\BibitemShut {NoStop}%
\bibitem [{\citenamefont {Wegner}(1971)}]{Wegner71}%
  \BibitemOpen
  \bibfield  {author} {\bibinfo {author} {\bibfnamefont {F.~J.}\ \bibnamefont
  {Wegner}},\ }\bibfield  {title} {\bibinfo {title} {Duality in generalized
  ising models and phase transitions without local order parameters},\ }\href
  {https://doi.org/10.1063/1.1665530} {\bibfield  {journal} {\bibinfo
  {journal} {Journal of Mathematical Physics}\ }\textbf {\bibinfo {volume}
  {12}},\ \bibinfo {pages} {2259} (\bibinfo {year} {1971})}\BibitemShut
  {NoStop}%
\bibitem [{\citenamefont {Kogut}(1979)}]{Kogut_Review}%
  \BibitemOpen
  \bibfield  {author} {\bibinfo {author} {\bibfnamefont {J.~B.}\ \bibnamefont
  {Kogut}},\ }\bibfield  {title} {\bibinfo {title} {An introduction to lattice
  gauge theory and spin systems},\ }\href
  {https://doi.org/10.1103/RevModPhys.51.659} {\bibfield  {journal} {\bibinfo
  {journal} {Rev. Mod. Phys.}\ }\textbf {\bibinfo {volume} {51}},\ \bibinfo
  {pages} {659} (\bibinfo {year} {1979})}\BibitemShut {NoStop}%
\bibitem [{\citenamefont {Kogut}\ and\ \citenamefont
  {Susskind}(1975)}]{Kogut_Susskind}%
  \BibitemOpen
  \bibfield  {author} {\bibinfo {author} {\bibfnamefont {J.}~\bibnamefont
  {Kogut}}\ and\ \bibinfo {author} {\bibfnamefont {L.}~\bibnamefont
  {Susskind}},\ }\bibfield  {title} {\bibinfo {title} {Hamiltonian formulation
  of wilson's lattice gauge theories},\ }\href
  {https://doi.org/10.1103/PhysRevD.11.395} {\bibfield  {journal} {\bibinfo
  {journal} {Phys. Rev. D}\ }\textbf {\bibinfo {volume} {11}},\ \bibinfo
  {pages} {395} (\bibinfo {year} {1975})}\BibitemShut {NoStop}%
\bibitem [{\citenamefont {Fradkin}\ and\ \citenamefont
  {Shenker}(1979)}]{FradkinShenker}%
  \BibitemOpen
  \bibfield  {author} {\bibinfo {author} {\bibfnamefont {E.}~\bibnamefont
  {Fradkin}}\ and\ \bibinfo {author} {\bibfnamefont {S.~H.}\ \bibnamefont
  {Shenker}},\ }\bibfield  {title} {\bibinfo {title} {Phase diagrams of lattice
  gauge theories with higgs fields},\ }\href
  {https://doi.org/10.1103/PhysRevD.19.3682} {\bibfield  {journal} {\bibinfo
  {journal} {Phys. Rev. D}\ }\textbf {\bibinfo {volume} {19}},\ \bibinfo
  {pages} {3682} (\bibinfo {year} {1979})}\BibitemShut {NoStop}%
\bibitem [{\citenamefont {Wilson}(1974)}]{Wilson74}%
  \BibitemOpen
  \bibfield  {author} {\bibinfo {author} {\bibfnamefont {K.~G.}\ \bibnamefont
  {Wilson}},\ }\bibfield  {title} {\bibinfo {title} {Confinement of quarks},\
  }\href {https://doi.org/10.1103/PhysRevD.10.2445} {\bibfield  {journal}
  {\bibinfo  {journal} {Phys. Rev. D}\ }\textbf {\bibinfo {volume} {10}},\
  \bibinfo {pages} {2445} (\bibinfo {year} {1974})}\BibitemShut {NoStop}%
\bibitem [{\citenamefont {Gaiotto}\ \emph {et~al.}(2015)\citenamefont
  {Gaiotto}, \citenamefont {Kapustin}, \citenamefont {Seiberg},\ and\
  \citenamefont {Willett}}]{Gaiotto2015}%
  \BibitemOpen
  \bibfield  {author} {\bibinfo {author} {\bibfnamefont {D.}~\bibnamefont
  {Gaiotto}}, \bibinfo {author} {\bibfnamefont {A.}~\bibnamefont {Kapustin}},
  \bibinfo {author} {\bibfnamefont {N.}~\bibnamefont {Seiberg}},\ and\ \bibinfo
  {author} {\bibfnamefont {B.}~\bibnamefont {Willett}},\ }\bibfield  {title}
  {\bibinfo {title} {Generalized global symmetries},\ }\href
  {https://doi.org/10.1007/JHEP02(2015)172} {\bibfield  {journal} {\bibinfo
  {journal} {Journal of High Energy Physics}\ }\textbf {\bibinfo {volume}
  {2015}},\ \bibinfo {pages} {172} (\bibinfo {year} {2015})}\BibitemShut
  {NoStop}%
\bibitem [{\citenamefont {McGreevy}(2023)}]{McGreevy23}%
  \BibitemOpen
  \bibfield  {author} {\bibinfo {author} {\bibfnamefont {J.}~\bibnamefont
  {McGreevy}},\ }\bibfield  {title} {\bibinfo {title} {Generalized symmetries
  in condensed matter},\ }\href
  {https://doi.org/https://doi.org/10.1146/annurev-conmatphys-040721-021029}
  {\bibfield  {journal} {\bibinfo  {journal} {Annual Review of Condensed Matter
  Physics}\ }\textbf {\bibinfo {volume} {14}},\ \bibinfo {pages} {57} (\bibinfo
  {year} {2023})}\BibitemShut {NoStop}%
\bibitem [{\citenamefont {Huxford}\ \emph {et~al.}(2023)\citenamefont
  {Huxford}, \citenamefont {Nguyen},\ and\ \citenamefont {Kim}}]{huxford2023}%
  \BibitemOpen
  \bibfield  {author} {\bibinfo {author} {\bibfnamefont {J.}~\bibnamefont
  {Huxford}}, \bibinfo {author} {\bibfnamefont {D.}~\bibnamefont {Nguyen}},\
  and\ \bibinfo {author} {\bibfnamefont {Y.~B.}\ \bibnamefont {Kim}},\
  }\bibfield  {title} {\bibinfo {title} {Gaining insights on anyon condensation
  and 1-form symmetry breaking across a topological phase transition in a
  deformed toric code model},\ }\href
  {https://www.scipost.org/SciPostPhys.15.6.253?acad_field_slug=astronomy}
  {\bibfield  {journal} {\bibinfo  {journal} {SciPost Physics}\ }\textbf
  {\bibinfo {volume} {15}},\ \bibinfo {pages} {253} (\bibinfo {year}
  {2023})}\BibitemShut {NoStop}%
\bibitem [{\citenamefont {Fan}\ \emph {et~al.}(2024)\citenamefont {Fan},
  \citenamefont {Bao}, \citenamefont {Altman},\ and\ \citenamefont
  {Vishwanath}}]{fan2024diagnostics}%
  \BibitemOpen
  \bibfield  {author} {\bibinfo {author} {\bibfnamefont {R.}~\bibnamefont
  {Fan}}, \bibinfo {author} {\bibfnamefont {Y.}~\bibnamefont {Bao}}, \bibinfo
  {author} {\bibfnamefont {E.}~\bibnamefont {Altman}},\ and\ \bibinfo {author}
  {\bibfnamefont {A.}~\bibnamefont {Vishwanath}},\ }\bibfield  {title}
  {\bibinfo {title} {Diagnostics of mixed-state topological order and breakdown
  of quantum memory},\ }\href {https://doi.org/10.1103/PRXQuantum.5.020343}
  {\bibfield  {journal} {\bibinfo  {journal} {PRX Quantum}\ }\textbf {\bibinfo
  {volume} {5}},\ \bibinfo {pages} {020343} (\bibinfo {year}
  {2024})}\BibitemShut {NoStop}%
\bibitem [{\citenamefont {Bao}\ \emph {et~al.}(2023)\citenamefont {Bao},
  \citenamefont {Fan}, \citenamefont {Vishwanath},\ and\ \citenamefont
  {Altman}}]{bao2023mixed}%
  \BibitemOpen
  \bibfield  {author} {\bibinfo {author} {\bibfnamefont {Y.}~\bibnamefont
  {Bao}}, \bibinfo {author} {\bibfnamefont {R.}~\bibnamefont {Fan}}, \bibinfo
  {author} {\bibfnamefont {A.}~\bibnamefont {Vishwanath}},\ and\ \bibinfo
  {author} {\bibfnamefont {E.}~\bibnamefont {Altman}},\ }\bibfield  {title}
  {\bibinfo {title} {Mixed-state topological order and the errorfield double
  formulation of decoherence-induced transitions},\ }\href
  {https://arxiv.org/abs/2301.05687} {\bibfield  {journal} {\bibinfo  {journal}
  {arXiv preprint arXiv:2301.05687}\ } (\bibinfo {year} {2023})}\BibitemShut
  {NoStop}%
\bibitem [{\citenamefont {McDonough}\ \emph {et~al.}(2025)\citenamefont
  {McDonough}, \citenamefont {Yin}, \citenamefont {Lucas},\ and\ \citenamefont
  {Zhang}}]{mcdonough2025}%
  \BibitemOpen
  \bibfield  {author} {\bibinfo {author} {\bibfnamefont {B.~T.}\ \bibnamefont
  {McDonough}}, \bibinfo {author} {\bibfnamefont {C.}~\bibnamefont {Yin}},
  \bibinfo {author} {\bibfnamefont {A.}~\bibnamefont {Lucas}},\ and\ \bibinfo
  {author} {\bibfnamefont {C.}~\bibnamefont {Zhang}},\ }\bibfield  {title}
  {\bibinfo {title} {Lieb-robinson bounds with exponential-in-volume tails},\
  }\href {https://arxiv.org/abs/2502.02652} {\bibfield  {journal} {\bibinfo
  {journal} {arXiv preprint arXiv:2502.02652}\ } (\bibinfo {year}
  {2025})}\BibitemShut {NoStop}%
\bibitem [{\citenamefont {O'Brien}\ and\ \citenamefont
  {Fendley}(2018)}]{OBrien18}%
  \BibitemOpen
  \bibfield  {author} {\bibinfo {author} {\bibfnamefont {E.}~\bibnamefont
  {O'Brien}}\ and\ \bibinfo {author} {\bibfnamefont {P.}~\bibnamefont
  {Fendley}},\ }\bibfield  {title} {\bibinfo {title} {Lattice supersymmetry and
  order-disorder coexistence in the tricritical ising model},\ }\href
  {https://doi.org/10.1103/PhysRevLett.120.206403} {\bibfield  {journal}
  {\bibinfo  {journal} {Phys. Rev. Lett.}\ }\textbf {\bibinfo {volume} {120}},\
  \bibinfo {pages} {206403} (\bibinfo {year} {2018})}\BibitemShut {NoStop}%
\bibitem [{\citenamefont {Swingle}\ and\ \citenamefont
  {McGreevy}(2016)}]{swingle2016ssource}%
  \BibitemOpen
  \bibfield  {author} {\bibinfo {author} {\bibfnamefont {B.}~\bibnamefont
  {Swingle}}\ and\ \bibinfo {author} {\bibfnamefont {J.}~\bibnamefont
  {McGreevy}},\ }\bibfield  {title} {\bibinfo {title} {Renormalization group
  constructions of topological quantum liquids and beyond},\ }\href
  {https://doi.org/10.1103/PhysRevB.93.045127} {\bibfield  {journal} {\bibinfo
  {journal} {Phys. Rev. B}\ }\textbf {\bibinfo {volume} {93}},\ \bibinfo
  {pages} {045127} (\bibinfo {year} {2016})}\BibitemShut {NoStop}%
\bibitem [{\citenamefont {Huse}\ and\ \citenamefont
  {Fisher}(1987)}]{HuseFisher}%
  \BibitemOpen
  \bibfield  {author} {\bibinfo {author} {\bibfnamefont {D.~A.}\ \bibnamefont
  {Huse}}\ and\ \bibinfo {author} {\bibfnamefont {D.~S.}\ \bibnamefont
  {Fisher}},\ }\bibfield  {title} {\bibinfo {title} {Dynamics of droplet
  fluctuations in pure and random ising systems},\ }\href
  {https://doi.org/10.1103/PhysRevB.35.6841} {\bibfield  {journal} {\bibinfo
  {journal} {Phys. Rev. B}\ }\textbf {\bibinfo {volume} {35}},\ \bibinfo
  {pages} {6841} (\bibinfo {year} {1987})}\BibitemShut {NoStop}%
\bibitem [{\citenamefont {Miyashita}\ and\ \citenamefont
  {Takano}(1985)}]{miyashita1985}%
  \BibitemOpen
  \bibfield  {author} {\bibinfo {author} {\bibfnamefont {S.}~\bibnamefont
  {Miyashita}}\ and\ \bibinfo {author} {\bibfnamefont {H.}~\bibnamefont
  {Takano}},\ }\bibfield  {title} {\bibinfo {title} {Dynamical nature of the
  phase transition of the two-dimensional kinetic ising model},\ }\href
  {https://academic.oup.com/ptp/article/73/5/1122/1846639} {\bibfield
  {journal} {\bibinfo  {journal} {Progress of theoretical physics}\ }\textbf
  {\bibinfo {volume} {73}},\ \bibinfo {pages} {1122} (\bibinfo {year}
  {1985})}\BibitemShut {NoStop}%
\bibitem [{\citenamefont {Henkel}\ and\ \citenamefont
  {Pleimling}(2011)}]{henkel2011}%
  \BibitemOpen
  \bibfield  {author} {\bibinfo {author} {\bibfnamefont {M.}~\bibnamefont
  {Henkel}}\ and\ \bibinfo {author} {\bibfnamefont {M.}~\bibnamefont
  {Pleimling}},\ }\href@noop {} {\emph {\bibinfo {title} {Non-Equilibrium Phase
  Transitions: Volume 2: Ageing and Dynamical Scaling Far from Equilibrium}}}\
  (\bibinfo  {publisher} {Springer Science \& Business Media},\ \bibinfo {year}
  {2011})\BibitemShut {NoStop}%
\bibitem [{\citenamefont {Hastings}(2011)}]{hastings2011finiteT}%
  \BibitemOpen
  \bibfield  {author} {\bibinfo {author} {\bibfnamefont {M.~B.}\ \bibnamefont
  {Hastings}},\ }\bibfield  {title} {\bibinfo {title} {Topological order at
  nonzero temperature},\ }\href
  {https://doi.org/10.1103/PhysRevLett.107.210501} {\bibfield  {journal}
  {\bibinfo  {journal} {Phys. Rev. Lett.}\ }\textbf {\bibinfo {volume} {107}},\
  \bibinfo {pages} {210501} (\bibinfo {year} {2011})}\BibitemShut {NoStop}%
\bibitem [{\citenamefont {Aharonov}\ and\ \citenamefont
  {Touati}(2018)}]{Aharonov18}%
  \BibitemOpen
  \bibfield  {author} {\bibinfo {author} {\bibfnamefont {D.}~\bibnamefont
  {Aharonov}}\ and\ \bibinfo {author} {\bibfnamefont {Y.}~\bibnamefont
  {Touati}},\ }\href {https://arxiv.org/abs/1810.03912} {\bibinfo {title}
  {Quantum circuit depth lower bounds for homological codes}} (\bibinfo {year}
  {2018}),\ \Eprint {https://arxiv.org/abs/1810.03912} {arXiv:1810.03912
  [quant-ph]} \BibitemShut {NoStop}%
\bibitem [{\citenamefont {Li}\ \emph {et~al.}(2024)\citenamefont {Li},
  \citenamefont {Lee},\ and\ \citenamefont {Yoshida}}]{Li2024}%
  \BibitemOpen
  \bibfield  {author} {\bibinfo {author} {\bibfnamefont {Z.}~\bibnamefont
  {Li}}, \bibinfo {author} {\bibfnamefont {D.}~\bibnamefont {Lee}},\ and\
  \bibinfo {author} {\bibfnamefont {B.}~\bibnamefont {Yoshida}},\ }\href
  {https://arxiv.org/abs/2405.07970} {\bibinfo {title} {How much entanglement
  is needed for emergent anyons and fermions?}} (\bibinfo {year} {2024}),\
  \Eprint {https://arxiv.org/abs/2405.07970} {arXiv:2405.07970 [quant-ph]}
  \BibitemShut {NoStop}%
\bibitem [{\citenamefont {Kitaev}(2006)}]{kitaev2006}%
  \BibitemOpen
  \bibfield  {author} {\bibinfo {author} {\bibfnamefont {A.}~\bibnamefont
  {Kitaev}},\ }\bibfield  {title} {\bibinfo {title} {Anyons in an exactly
  solved model and beyond},\ }\href
  {https://www.sciencedirect.com/science/article/abs/pii/S0003491605002381?via%3Dihub}
  {\bibfield  {journal} {\bibinfo  {journal} {Annals of Physics}\ }\textbf
  {\bibinfo {volume} {321}},\ \bibinfo {pages} {2} (\bibinfo {year}
  {2006})}\BibitemShut {NoStop}%
\bibitem [{\citenamefont {Freedman}\ \emph {et~al.}(2003)\citenamefont
  {Freedman}, \citenamefont {Nayak},\ and\ \citenamefont
  {Shtengel}}]{freedman2003non}%
  \BibitemOpen
  \bibfield  {author} {\bibinfo {author} {\bibfnamefont {M.}~\bibnamefont
  {Freedman}}, \bibinfo {author} {\bibfnamefont {C.}~\bibnamefont {Nayak}},\
  and\ \bibinfo {author} {\bibfnamefont {K.}~\bibnamefont {Shtengel}},\
  }\bibfield  {title} {\bibinfo {title} {Non-abelian topological phases in an
  extended hubbard model},\ }\href@noop {} {\bibfield  {journal} {\bibinfo
  {journal} {arXiv preprint cond-mat/0309120}\ } (\bibinfo {year}
  {2003})}\BibitemShut {NoStop}%
\bibitem [{\citenamefont {Freedman}\ \emph {et~al.}(2004)\citenamefont
  {Freedman}, \citenamefont {Nayak}, \citenamefont {Shtengel}, \citenamefont
  {Walker},\ and\ \citenamefont {Wang}}]{Freedman04}%
  \BibitemOpen
  \bibfield  {author} {\bibinfo {author} {\bibfnamefont {M.}~\bibnamefont
  {Freedman}}, \bibinfo {author} {\bibfnamefont {C.}~\bibnamefont {Nayak}},
  \bibinfo {author} {\bibfnamefont {K.}~\bibnamefont {Shtengel}}, \bibinfo
  {author} {\bibfnamefont {K.}~\bibnamefont {Walker}},\ and\ \bibinfo {author}
  {\bibfnamefont {Z.}~\bibnamefont {Wang}},\ }\bibfield  {title} {\bibinfo
  {title} {A class of p,t-invariant topological phases of interacting
  electrons},\ }\href
  {https://doi.org/https://doi.org/10.1016/j.aop.2004.01.006} {\bibfield
  {journal} {\bibinfo  {journal} {Annals of Physics}\ }\textbf {\bibinfo
  {volume} {310}},\ \bibinfo {pages} {428} (\bibinfo {year}
  {2004})}\BibitemShut {NoStop}%
\bibitem [{\citenamefont {Freedman}\ \emph
  {et~al.}(2005{\natexlab{a}})\citenamefont {Freedman}, \citenamefont {Nayak},\
  and\ \citenamefont {Shtengel}}]{Freedman05a}%
  \BibitemOpen
  \bibfield  {author} {\bibinfo {author} {\bibfnamefont {M.}~\bibnamefont
  {Freedman}}, \bibinfo {author} {\bibfnamefont {C.}~\bibnamefont {Nayak}},\
  and\ \bibinfo {author} {\bibfnamefont {K.}~\bibnamefont {Shtengel}},\
  }\bibfield  {title} {\bibinfo {title} {Line of critical points in $2+1$
  dimensions: Quantum critical loop gases and non-abelian gauge theory},\
  }\href {https://doi.org/10.1103/PhysRevLett.94.147205} {\bibfield  {journal}
  {\bibinfo  {journal} {Phys. Rev. Lett.}\ }\textbf {\bibinfo {volume} {94}},\
  \bibinfo {pages} {147205} (\bibinfo {year} {2005}{\natexlab{a}})}\BibitemShut
  {NoStop}%
\bibitem [{\citenamefont {Freedman}\ \emph
  {et~al.}(2005{\natexlab{b}})\citenamefont {Freedman}, \citenamefont {Nayak},\
  and\ \citenamefont {Shtengel}}]{Freedman05b}%
  \BibitemOpen
  \bibfield  {author} {\bibinfo {author} {\bibfnamefont {M.}~\bibnamefont
  {Freedman}}, \bibinfo {author} {\bibfnamefont {C.}~\bibnamefont {Nayak}},\
  and\ \bibinfo {author} {\bibfnamefont {K.}~\bibnamefont {Shtengel}},\
  }\bibfield  {title} {\bibinfo {title} {Extended hubbard model with ring
  exchange: A route to a non-abelian topological phase},\ }\href
  {https://doi.org/10.1103/PhysRevLett.94.066401} {\bibfield  {journal}
  {\bibinfo  {journal} {Phys. Rev. Lett.}\ }\textbf {\bibinfo {volume} {94}},\
  \bibinfo {pages} {066401} (\bibinfo {year} {2005}{\natexlab{b}})}\BibitemShut
  {NoStop}%
\bibitem [{\citenamefont {Freedman}\ \emph {et~al.}(2008)\citenamefont
  {Freedman}, \citenamefont {Nayak},\ and\ \citenamefont
  {Shtengel}}]{freedman2008}%
  \BibitemOpen
  \bibfield  {author} {\bibinfo {author} {\bibfnamefont {M.}~\bibnamefont
  {Freedman}}, \bibinfo {author} {\bibfnamefont {C.}~\bibnamefont {Nayak}},\
  and\ \bibinfo {author} {\bibfnamefont {K.}~\bibnamefont {Shtengel}},\
  }\bibfield  {title} {\bibinfo {title} {Lieb-schultz-mattis theorem for
  quasitopological systems},\ }\href
  {https://doi.org/10.1103/PhysRevB.78.174411} {\bibfield  {journal} {\bibinfo
  {journal} {Phys. Rev. B}\ }\textbf {\bibinfo {volume} {78}},\ \bibinfo
  {pages} {174411} (\bibinfo {year} {2008})}\BibitemShut {NoStop}%
\bibitem [{\citenamefont {Dai}\ and\ \citenamefont {Nahum}(2020)}]{dai2020}%
  \BibitemOpen
  \bibfield  {author} {\bibinfo {author} {\bibfnamefont {Z.}~\bibnamefont
  {Dai}}\ and\ \bibinfo {author} {\bibfnamefont {A.}~\bibnamefont {Nahum}},\
  }\bibfield  {title} {\bibinfo {title} {Quantum criticality of loops with
  topologically constrained dynamics},\ }\href
  {https://doi.org/10.1103/PhysRevResearch.2.033051} {\bibfield  {journal}
  {\bibinfo  {journal} {Phys. Rev. Res.}\ }\textbf {\bibinfo {volume} {2}},\
  \bibinfo {pages} {033051} (\bibinfo {year} {2020})}\BibitemShut {NoStop}%
\bibitem [{\citenamefont {Martinelli}(1992)}]{HF_extended_discussion}%
  \BibitemOpen
  \bibfield  {author} {\bibinfo {author} {\bibfnamefont {F.}~\bibnamefont
  {Martinelli}},\ }\bibfield  {title} {\bibinfo {title} {Low temperature
  stochastic spin dynamics: Metastability, convergence to equilibrium and phase
  segregation},\ }in\ \href@noop {} {\emph {\bibinfo {booktitle} {Mathematical
  Physics X}}},\ \bibinfo {editor} {edited by\ \bibinfo {editor} {\bibfnamefont
  {K.}~\bibnamefont {Schm{\"u}dgen}}}\ (\bibinfo  {publisher} {Springer Berlin
  Heidelberg},\ \bibinfo {address} {Berlin, Heidelberg},\ \bibinfo {year}
  {1992})\ pp.\ \bibinfo {pages} {87--102}\BibitemShut {NoStop}%
\bibitem [{\citenamefont {Ogielski}(1987)}]{Ogielski_HF}%
  \BibitemOpen
  \bibfield  {author} {\bibinfo {author} {\bibfnamefont {A.~T.}\ \bibnamefont
  {Ogielski}},\ }\bibfield  {title} {\bibinfo {title} {Dynamics of fluctuations
  in the ordered phase of kinetic ising models},\ }\href
  {https://doi.org/10.1103/PhysRevB.36.7315} {\bibfield  {journal} {\bibinfo
  {journal} {Phys. Rev. B}\ }\textbf {\bibinfo {volume} {36}},\ \bibinfo
  {pages} {7315} (\bibinfo {year} {1987})}\BibitemShut {NoStop}%
\bibitem [{\citenamefont {Martinelli}(1994)}]{martinelli1994two}%
  \BibitemOpen
  \bibfield  {author} {\bibinfo {author} {\bibfnamefont {F.}~\bibnamefont
  {Martinelli}},\ }\bibfield  {title} {\bibinfo {title} {On the two-dimensional
  dynamical ising model in the phase coexistence region},\ }\href@noop {}
  {\bibfield  {journal} {\bibinfo  {journal} {Journal of Statistical Physics}\
  }\textbf {\bibinfo {volume} {76}},\ \bibinfo {pages} {1179} (\bibinfo {year}
  {1994})}\BibitemShut {NoStop}%
\bibitem [{\citenamefont {Lifshitz}(1962)}]{lifshitz1962}%
  \BibitemOpen
  \bibfield  {author} {\bibinfo {author} {\bibfnamefont {I.}~\bibnamefont
  {Lifshitz}},\ }\bibfield  {title} {\bibinfo {title} {Kinetics of ordering
  during second-order phase transitions},\ }\href
  {http://jetp.ras.ru/cgi-bin/dn/e_015_05_0939.pdf} {\bibfield  {journal}
  {\bibinfo  {journal} {Sov. Phys. JETP}\ }\textbf {\bibinfo {volume} {15}},\
  \bibinfo {pages} {939} (\bibinfo {year} {1962})}\BibitemShut {NoStop}%
\bibitem [{\citenamefont {Allen}\ and\ \citenamefont {Cahn}(1979)}]{allen1979}%
  \BibitemOpen
  \bibfield  {author} {\bibinfo {author} {\bibfnamefont {S.~M.}\ \bibnamefont
  {Allen}}\ and\ \bibinfo {author} {\bibfnamefont {J.~W.}\ \bibnamefont
  {Cahn}},\ }\bibfield  {title} {\bibinfo {title} {A microscopic theory for
  antiphase boundary motion and its application to antiphase domain
  coarsening},\ }\href
  {https://www.sciencedirect.com/science/article/abs/pii/0001616079901962}
  {\bibfield  {journal} {\bibinfo  {journal} {Acta metallurgica}\ }\textbf
  {\bibinfo {volume} {27}},\ \bibinfo {pages} {1085} (\bibinfo {year}
  {1979})}\BibitemShut {NoStop}%
\bibitem [{\citenamefont {Sahni}\ \emph {et~al.}(1981)\citenamefont {Sahni},
  \citenamefont {Dee}, \citenamefont {Gunton}, \citenamefont {Phani},
  \citenamefont {Lebowitz},\ and\ \citenamefont {Kalos}}]{sahni1981}%
  \BibitemOpen
  \bibfield  {author} {\bibinfo {author} {\bibfnamefont {P.~S.}\ \bibnamefont
  {Sahni}}, \bibinfo {author} {\bibfnamefont {G.}~\bibnamefont {Dee}}, \bibinfo
  {author} {\bibfnamefont {J.~D.}\ \bibnamefont {Gunton}}, \bibinfo {author}
  {\bibfnamefont {M.}~\bibnamefont {Phani}}, \bibinfo {author} {\bibfnamefont
  {J.~L.}\ \bibnamefont {Lebowitz}},\ and\ \bibinfo {author} {\bibfnamefont
  {M.}~\bibnamefont {Kalos}},\ }\bibfield  {title} {\bibinfo {title} {Dynamics
  of a two-dimensional order-disorder transition},\ }\href
  {https://doi.org/10.1103/PhysRevB.24.410} {\bibfield  {journal} {\bibinfo
  {journal} {Phys. Rev. B}\ }\textbf {\bibinfo {volume} {24}},\ \bibinfo
  {pages} {410} (\bibinfo {year} {1981})}\BibitemShut {NoStop}%
\bibitem [{\citenamefont {Ohta}\ \emph {et~al.}(1982)\citenamefont {Ohta},
  \citenamefont {Jasnow},\ and\ \citenamefont {Kawasaki}}]{ohta1982}%
  \BibitemOpen
  \bibfield  {author} {\bibinfo {author} {\bibfnamefont {T.}~\bibnamefont
  {Ohta}}, \bibinfo {author} {\bibfnamefont {D.}~\bibnamefont {Jasnow}},\ and\
  \bibinfo {author} {\bibfnamefont {K.}~\bibnamefont {Kawasaki}},\ }\bibfield
  {title} {\bibinfo {title} {Universal scaling in the motion of random
  interfaces},\ }\href {https://doi.org/10.1103/PhysRevLett.49.1223} {\bibfield
   {journal} {\bibinfo  {journal} {Phys. Rev. Lett.}\ }\textbf {\bibinfo
  {volume} {49}},\ \bibinfo {pages} {1223} (\bibinfo {year}
  {1982})}\BibitemShut {NoStop}%
\bibitem [{\citenamefont {Sahni}\ \emph {et~al.}(1983)\citenamefont {Sahni},
  \citenamefont {Grest},\ and\ \citenamefont {Safran}}]{sahni1983}%
  \BibitemOpen
  \bibfield  {author} {\bibinfo {author} {\bibfnamefont {P.~S.}\ \bibnamefont
  {Sahni}}, \bibinfo {author} {\bibfnamefont {G.~S.}\ \bibnamefont {Grest}},\
  and\ \bibinfo {author} {\bibfnamefont {S.~A.}\ \bibnamefont {Safran}},\
  }\bibfield  {title} {\bibinfo {title} {Temperature dependence of domain
  kinetics in two dimensions},\ }\href
  {https://doi.org/10.1103/PhysRevLett.50.60} {\bibfield  {journal} {\bibinfo
  {journal} {Phys. Rev. Lett.}\ }\textbf {\bibinfo {volume} {50}},\ \bibinfo
  {pages} {60} (\bibinfo {year} {1983})}\BibitemShut {NoStop}%
\bibitem [{\citenamefont {Gunton}(1984)}]{gunton1984}%
  \BibitemOpen
  \bibfield  {author} {\bibinfo {author} {\bibfnamefont {J.}~\bibnamefont
  {Gunton}},\ }\bibfield  {title} {\bibinfo {title} {The dynamics of random
  interfaces in phase transitions},\ }\href
  {https://link.springer.com/article/10.1007/BF01009455} {\bibfield  {journal}
  {\bibinfo  {journal} {Journal of Statistical Physics}\ }\textbf {\bibinfo
  {volume} {34}},\ \bibinfo {pages} {1019} (\bibinfo {year}
  {1984})}\BibitemShut {NoStop}%
\bibitem [{\citenamefont {Masaoka}\ \emph {et~al.}(2024)\citenamefont
  {Masaoka}, \citenamefont {Soejima},\ and\ \citenamefont
  {Watanabe}}]{masaoka2024}%
  \BibitemOpen
  \bibfield  {author} {\bibinfo {author} {\bibfnamefont {R.}~\bibnamefont
  {Masaoka}}, \bibinfo {author} {\bibfnamefont {T.}~\bibnamefont {Soejima}},\
  and\ \bibinfo {author} {\bibfnamefont {H.}~\bibnamefont {Watanabe}},\
  }\bibfield  {title} {\bibinfo {title} {Quadratic dispersion relations in
  gapless frustration-free systems},\ }\href
  {https://doi.org/10.1103/PhysRevB.110.195140} {\bibfield  {journal} {\bibinfo
   {journal} {Phys. Rev. B}\ }\textbf {\bibinfo {volume} {110}},\ \bibinfo
  {pages} {195140} (\bibinfo {year} {2024})}\BibitemShut {NoStop}%
\bibitem [{\citenamefont {Haga}\ \emph {et~al.}(2021)\citenamefont {Haga},
  \citenamefont {Nakagawa}, \citenamefont {Hamazaki},\ and\ \citenamefont
  {Ueda}}]{haga2021}%
  \BibitemOpen
  \bibfield  {author} {\bibinfo {author} {\bibfnamefont {T.}~\bibnamefont
  {Haga}}, \bibinfo {author} {\bibfnamefont {M.}~\bibnamefont {Nakagawa}},
  \bibinfo {author} {\bibfnamefont {R.}~\bibnamefont {Hamazaki}},\ and\
  \bibinfo {author} {\bibfnamefont {M.}~\bibnamefont {Ueda}},\ }\bibfield
  {title} {\bibinfo {title} {Liouvillian skin effect: Slowing down of
  relaxation processes without gap closing},\ }\href
  {https://doi.org/10.1103/PhysRevLett.127.070402} {\bibfield  {journal}
  {\bibinfo  {journal} {Phys. Rev. Lett.}\ }\textbf {\bibinfo {volume} {127}},\
  \bibinfo {pages} {070402} (\bibinfo {year} {2021})}\BibitemShut {NoStop}%
\bibitem [{\citenamefont {Rakovszky}\ \emph {et~al.}(2024)\citenamefont
  {Rakovszky}, \citenamefont {Gopalakrishnan},\ and\ \citenamefont {von
  Keyserlingk}}]{rakovszky2024}%
  \BibitemOpen
  \bibfield  {author} {\bibinfo {author} {\bibfnamefont {T.}~\bibnamefont
  {Rakovszky}}, \bibinfo {author} {\bibfnamefont {S.}~\bibnamefont
  {Gopalakrishnan}},\ and\ \bibinfo {author} {\bibfnamefont {C.}~\bibnamefont
  {von Keyserlingk}},\ }\bibfield  {title} {\bibinfo {title} {Defining stable
  phases of open quantum systems},\ }\href
  {https://doi.org/10.1103/PhysRevX.14.041031} {\bibfield  {journal} {\bibinfo
  {journal} {Phys. Rev. X}\ }\textbf {\bibinfo {volume} {14}},\ \bibinfo
  {pages} {041031} (\bibinfo {year} {2024})}\BibitemShut {NoStop}%
\bibitem [{\citenamefont {WEN}(1990)}]{Wen_90}%
  \BibitemOpen
  \bibfield  {author} {\bibinfo {author} {\bibfnamefont {X.~G.}\ \bibnamefont
  {WEN}},\ }\bibfield  {title} {\bibinfo {title} {Topological orders in rigid
  states},\ }\href {https://doi.org/10.1142/S0217979290000139} {\bibfield
  {journal} {\bibinfo  {journal} {International Journal of Modern Physics B}\
  }\textbf {\bibinfo {volume} {04}},\ \bibinfo {pages} {239} (\bibinfo {year}
  {1990})},\ \Eprint
  {https://arxiv.org/abs/https://doi.org/10.1142/S0217979290000139}
  {https://doi.org/10.1142/S0217979290000139} \BibitemShut {NoStop}%
\bibitem [{\citenamefont {Wilczek}(1982)}]{Wilczek82}%
  \BibitemOpen
  \bibfield  {author} {\bibinfo {author} {\bibfnamefont {F.}~\bibnamefont
  {Wilczek}},\ }\bibfield  {title} {\bibinfo {title} {Quantum mechanics of
  fractional-spin particles},\ }\href
  {https://doi.org/10.1103/PhysRevLett.49.957} {\bibfield  {journal} {\bibinfo
  {journal} {Phys. Rev. Lett.}\ }\textbf {\bibinfo {volume} {49}},\ \bibinfo
  {pages} {957} (\bibinfo {year} {1982})}\BibitemShut {NoStop}%
\bibitem [{\citenamefont {Wen}(1989)}]{Wen89}%
  \BibitemOpen
  \bibfield  {author} {\bibinfo {author} {\bibfnamefont {X.~G.}\ \bibnamefont
  {Wen}},\ }\bibfield  {title} {\bibinfo {title} {Vacuum degeneracy of chiral
  spin states in compactified space},\ }\href
  {https://doi.org/10.1103/PhysRevB.40.7387} {\bibfield  {journal} {\bibinfo
  {journal} {Phys. Rev. B}\ }\textbf {\bibinfo {volume} {40}},\ \bibinfo
  {pages} {7387} (\bibinfo {year} {1989})}\BibitemShut {NoStop}%
\bibitem [{\citenamefont {Read}\ and\ \citenamefont
  {Chakraborty}(1989)}]{Read89}%
  \BibitemOpen
  \bibfield  {author} {\bibinfo {author} {\bibfnamefont {N.}~\bibnamefont
  {Read}}\ and\ \bibinfo {author} {\bibfnamefont {B.}~\bibnamefont
  {Chakraborty}},\ }\bibfield  {title} {\bibinfo {title} {Statistics of the
  excitations of the resonating-valence-bond state},\ }\href
  {https://doi.org/10.1103/PhysRevB.40.7133} {\bibfield  {journal} {\bibinfo
  {journal} {Phys. Rev. B}\ }\textbf {\bibinfo {volume} {40}},\ \bibinfo
  {pages} {7133} (\bibinfo {year} {1989})}\BibitemShut {NoStop}%
\bibitem [{\citenamefont {Kivelson}(1989)}]{Kivelson89}%
  \BibitemOpen
  \bibfield  {author} {\bibinfo {author} {\bibfnamefont {S.}~\bibnamefont
  {Kivelson}},\ }\bibfield  {title} {\bibinfo {title} {Statistics of holons in
  the quantum hard-core dimer gas},\ }\href
  {https://doi.org/10.1103/PhysRevB.39.259} {\bibfield  {journal} {\bibinfo
  {journal} {Phys. Rev. B}\ }\textbf {\bibinfo {volume} {39}},\ \bibinfo
  {pages} {259} (\bibinfo {year} {1989})}\BibitemShut {NoStop}%
\bibitem [{\citenamefont {Wen}(2004)}]{Wenbook}%
  \BibitemOpen
  \bibfield  {author} {\bibinfo {author} {\bibfnamefont {X.}~\bibnamefont
  {Wen}},\ }\href
  {https://global.oup.com/academic/product/quantum-field-theory-of-many-body-systems-9780198530947?cc=de&lang=en&}
  {\emph {\bibinfo {title} {Quantum Field Theory of Many-body Systems}}},\
  Oxford graduate texts\ (\bibinfo  {publisher} {Oxford University Press},\
  \bibinfo {year} {2004})\BibitemShut {NoStop}%
\bibitem [{\citenamefont {Einarsson}(1990)}]{einarsson_1990}%
  \BibitemOpen
  \bibfield  {author} {\bibinfo {author} {\bibfnamefont {T.}~\bibnamefont
  {Einarsson}},\ }\bibfield  {title} {\bibinfo {title} {Fractional statistics
  on a torus},\ }\href {https://doi.org/10.1103/PhysRevLett.64.1995} {\bibfield
   {journal} {\bibinfo  {journal} {Phys. Rev. Lett.}\ }\textbf {\bibinfo
  {volume} {64}},\ \bibinfo {pages} {1995} (\bibinfo {year}
  {1990})}\BibitemShut {NoStop}%
\bibitem [{\citenamefont {Sachdev}()}]{sachdev_book_2023}%
  \BibitemOpen
  \bibfield  {author} {\bibinfo {author} {\bibfnamefont {S.}~\bibnamefont
  {Sachdev}},\ }\href {https://doi.org/10.1017/9781009212717} {\emph {\bibinfo
  {title} {Quantum {{Phases}} of {{Matter}}}}}\ (\bibinfo  {publisher}
  {Cambridge University Press})\BibitemShut {NoStop}%
\bibitem [{\citenamefont {Simon}()}]{simon_topological_2023}%
  \BibitemOpen
  \bibfield  {author} {\bibinfo {author} {\bibfnamefont {S.~H.}\ \bibnamefont
  {Simon}},\ }\href@noop {} {\emph {\bibinfo {title} {Topological
  {{Quantum}}}}}\ (\bibinfo  {publisher} {Oxford University Press})\BibitemShut
  {NoStop}%
\bibitem [{\citenamefont {Zhao}\ \emph {et~al.}(2021)\citenamefont {Zhao},
  \citenamefont {Yan}, \citenamefont {Cheng},\ and\ \citenamefont
  {Meng}}]{zhao2021}%
  \BibitemOpen
  \bibfield  {author} {\bibinfo {author} {\bibfnamefont {J.}~\bibnamefont
  {Zhao}}, \bibinfo {author} {\bibfnamefont {Z.}~\bibnamefont {Yan}}, \bibinfo
  {author} {\bibfnamefont {M.}~\bibnamefont {Cheng}},\ and\ \bibinfo {author}
  {\bibfnamefont {Z.~Y.}\ \bibnamefont {Meng}},\ }\bibfield  {title} {\bibinfo
  {title} {Higher-form symmetry breaking at ising transitions},\ }\href
  {https://doi.org/10.1103/PhysRevResearch.3.033024} {\bibfield  {journal}
  {\bibinfo  {journal} {Phys. Rev. Res.}\ }\textbf {\bibinfo {volume} {3}},\
  \bibinfo {pages} {033024} (\bibinfo {year} {2021})}\BibitemShut {NoStop}%
\bibitem [{\citenamefont {Matsumoto}\ \emph {et~al.}(2020)\citenamefont
  {Matsumoto}, \citenamefont {Kawabata}, \citenamefont {Ashida}, \citenamefont
  {Furukawa},\ and\ \citenamefont {Ueda}}]{matsumoto2020}%
  \BibitemOpen
  \bibfield  {author} {\bibinfo {author} {\bibfnamefont {N.}~\bibnamefont
  {Matsumoto}}, \bibinfo {author} {\bibfnamefont {K.}~\bibnamefont {Kawabata}},
  \bibinfo {author} {\bibfnamefont {Y.}~\bibnamefont {Ashida}}, \bibinfo
  {author} {\bibfnamefont {S.}~\bibnamefont {Furukawa}},\ and\ \bibinfo
  {author} {\bibfnamefont {M.}~\bibnamefont {Ueda}},\ }\bibfield  {title}
  {\bibinfo {title} {Continuous phase transition without gap closing in
  non-hermitian quantum many-body systems},\ }\href
  {https://doi.org/10.1103/PhysRevLett.125.260601} {\bibfield  {journal}
  {\bibinfo  {journal} {Phys. Rev. Lett.}\ }\textbf {\bibinfo {volume} {125}},\
  \bibinfo {pages} {260601} (\bibinfo {year} {2020})}\BibitemShut {NoStop}%
\bibitem [{\citenamefont {Kim}\ \emph {et~al.}(2023)\citenamefont {Kim},
  \citenamefont {Levin}, \citenamefont {Lin}, \citenamefont {Ranard},\ and\
  \citenamefont {Shi}}]{kim2023}%
  \BibitemOpen
  \bibfield  {author} {\bibinfo {author} {\bibfnamefont {I.~H.}\ \bibnamefont
  {Kim}}, \bibinfo {author} {\bibfnamefont {M.}~\bibnamefont {Levin}}, \bibinfo
  {author} {\bibfnamefont {T.-C.}\ \bibnamefont {Lin}}, \bibinfo {author}
  {\bibfnamefont {D.}~\bibnamefont {Ranard}},\ and\ \bibinfo {author}
  {\bibfnamefont {B.}~\bibnamefont {Shi}},\ }\bibfield  {title} {\bibinfo
  {title} {Universal lower bound on topological entanglement entropy},\ }\href
  {https://doi.org/10.1103/PhysRevLett.131.166601} {\bibfield  {journal}
  {\bibinfo  {journal} {Phys. Rev. Lett.}\ }\textbf {\bibinfo {volume} {131}},\
  \bibinfo {pages} {166601} (\bibinfo {year} {2023})}\BibitemShut {NoStop}%
\bibitem [{\citenamefont {Bravyi}\ \emph {et~al.}(2006)\citenamefont {Bravyi},
  \citenamefont {Hastings},\ and\ \citenamefont
  {Verstraete}}]{bravyi_2006_topological}%
  \BibitemOpen
  \bibfield  {author} {\bibinfo {author} {\bibfnamefont {S.}~\bibnamefont
  {Bravyi}}, \bibinfo {author} {\bibfnamefont {M.~B.}\ \bibnamefont
  {Hastings}},\ and\ \bibinfo {author} {\bibfnamefont {F.}~\bibnamefont
  {Verstraete}},\ }\bibfield  {title} {\bibinfo {title} {Lieb-robinson bounds
  and the generation of correlations and topological quantum order},\ }\href
  {https://doi.org/10.1103/PhysRevLett.97.050401} {\bibfield  {journal}
  {\bibinfo  {journal} {Phys. Rev. Lett.}\ }\textbf {\bibinfo {volume} {97}},\
  \bibinfo {pages} {050401} (\bibinfo {year} {2006})}\BibitemShut {NoStop}%
\bibitem [{\citenamefont {Bravyi}\ \emph {et~al.}(2010)\citenamefont {Bravyi},
  \citenamefont {Hastings},\ and\ \citenamefont {Michalakis}}]{Bravyi_2010}%
  \BibitemOpen
  \bibfield  {author} {\bibinfo {author} {\bibfnamefont {S.}~\bibnamefont
  {Bravyi}}, \bibinfo {author} {\bibfnamefont {M.~B.}\ \bibnamefont
  {Hastings}},\ and\ \bibinfo {author} {\bibfnamefont {S.}~\bibnamefont
  {Michalakis}},\ }\bibfield  {title} {\bibinfo {title} {Topological quantum
  order: Stability under local perturbations},\ }\bibfield  {journal} {\bibinfo
   {journal} {Journal of Mathematical Physics}\ }\textbf {\bibinfo {volume}
  {51}},\ \href {https://doi.org/10.1063/1.3490195} {10.1063/1.3490195}
  (\bibinfo {year} {2010})\BibitemShut {NoStop}%
\bibitem [{\citenamefont {Bonderson}\ and\ \citenamefont
  {Nayak}(2013)}]{Bonderson13}%
  \BibitemOpen
  \bibfield  {author} {\bibinfo {author} {\bibfnamefont {P.}~\bibnamefont
  {Bonderson}}\ and\ \bibinfo {author} {\bibfnamefont {C.}~\bibnamefont
  {Nayak}},\ }\bibfield  {title} {\bibinfo {title} {Quasi-topological phases of
  matter and topological protection},\ }\href
  {https://doi.org/10.1103/PhysRevB.87.195451} {\bibfield  {journal} {\bibinfo
  {journal} {Phys. Rev. B}\ }\textbf {\bibinfo {volume} {87}},\ \bibinfo
  {pages} {195451} (\bibinfo {year} {2013})}\BibitemShut {NoStop}%
\bibitem [{\citenamefont {Rasmussen}\ and\ \citenamefont
  {Jermyn}(2018)}]{Rasmussen18}%
  \BibitemOpen
  \bibfield  {author} {\bibinfo {author} {\bibfnamefont {A.}~\bibnamefont
  {Rasmussen}}\ and\ \bibinfo {author} {\bibfnamefont {A.~S.}\ \bibnamefont
  {Jermyn}},\ }\bibfield  {title} {\bibinfo {title} {Gapless topological order,
  gravity, and black holes},\ }\href
  {https://doi.org/10.1103/PhysRevB.97.165141} {\bibfield  {journal} {\bibinfo
  {journal} {Phys. Rev. B}\ }\textbf {\bibinfo {volume} {97}},\ \bibinfo
  {pages} {165141} (\bibinfo {year} {2018})}\BibitemShut {NoStop}%
\bibitem [{\citenamefont {Hermele}\ \emph {et~al.}(2004)\citenamefont
  {Hermele}, \citenamefont {Fisher},\ and\ \citenamefont
  {Balents}}]{Hermele04}%
  \BibitemOpen
  \bibfield  {author} {\bibinfo {author} {\bibfnamefont {M.}~\bibnamefont
  {Hermele}}, \bibinfo {author} {\bibfnamefont {M.~P.~A.}\ \bibnamefont
  {Fisher}},\ and\ \bibinfo {author} {\bibfnamefont {L.}~\bibnamefont
  {Balents}},\ }\bibfield  {title} {\bibinfo {title} {Pyrochlore photons: The
  $u(1)$ spin liquid in a $s=\frac{1}{2}$ three-dimensional frustrated
  magnet},\ }\href {https://doi.org/10.1103/PhysRevB.69.064404} {\bibfield
  {journal} {\bibinfo  {journal} {Phys. Rev. B}\ }\textbf {\bibinfo {volume}
  {69}},\ \bibinfo {pages} {064404} (\bibinfo {year} {2004})}\BibitemShut
  {NoStop}%
\bibitem [{\citenamefont {Freedman}(2003)}]{Freedman2003}%
  \BibitemOpen
  \bibfield  {author} {\bibinfo {author} {\bibfnamefont {M.~H.}\ \bibnamefont
  {Freedman}},\ }\bibfield  {title} {\bibinfo {title} {A magnetic model with a
  possible chern-simons phase},\ }\href
  {https://doi.org/10.1007/s00220-002-0785-1} {\bibfield  {journal} {\bibinfo
  {journal} {Communications in Mathematical Physics}\ }\textbf {\bibinfo
  {volume} {234}},\ \bibinfo {pages} {129} (\bibinfo {year}
  {2003})}\BibitemShut {NoStop}%
\bibitem [{\citenamefont {Fendley}(2008)}]{Fendley_2008}%
  \BibitemOpen
  \bibfield  {author} {\bibinfo {author} {\bibfnamefont {P.}~\bibnamefont
  {Fendley}},\ }\bibfield  {title} {\bibinfo {title} {Topological order from
  quantum loops and nets},\ }\href {https://doi.org/10.1016/j.aop.2008.04.011}
  {\bibfield  {journal} {\bibinfo  {journal} {Annals of Physics}\ }\textbf
  {\bibinfo {volume} {323}},\ \bibinfo {pages} {3113–3136} (\bibinfo {year}
  {2008})}\BibitemShut {NoStop}%
\bibitem [{\citenamefont {Fidkowski}\ \emph {et~al.}(2009)\citenamefont
  {Fidkowski}, \citenamefont {Freedman}, \citenamefont {Nayak}, \citenamefont
  {Walker},\ and\ \citenamefont {Wang}}]{Fidkowski2009}%
  \BibitemOpen
  \bibfield  {author} {\bibinfo {author} {\bibfnamefont {L.}~\bibnamefont
  {Fidkowski}}, \bibinfo {author} {\bibfnamefont {M.}~\bibnamefont {Freedman}},
  \bibinfo {author} {\bibfnamefont {C.}~\bibnamefont {Nayak}}, \bibinfo
  {author} {\bibfnamefont {K.}~\bibnamefont {Walker}},\ and\ \bibinfo {author}
  {\bibfnamefont {Z.}~\bibnamefont {Wang}},\ }\bibfield  {title} {\bibinfo
  {title} {From string nets to nonabelions},\ }\href@noop {} {\bibfield
  {journal} {\bibinfo  {journal} {Communications in Mathematical Physics}\
  }\textbf {\bibinfo {volume} {287}},\ \bibinfo {pages} {805} (\bibinfo {year}
  {2009})}\BibitemShut {NoStop}%
\bibitem [{\citenamefont {Fendley}\ \emph {et~al.}(2013)\citenamefont
  {Fendley}, \citenamefont {Isakov},\ and\ \citenamefont {Troyer}}]{Fendley13}%
  \BibitemOpen
  \bibfield  {author} {\bibinfo {author} {\bibfnamefont {P.}~\bibnamefont
  {Fendley}}, \bibinfo {author} {\bibfnamefont {S.~V.}\ \bibnamefont
  {Isakov}},\ and\ \bibinfo {author} {\bibfnamefont {M.}~\bibnamefont
  {Troyer}},\ }\bibfield  {title} {\bibinfo {title} {Fibonacci topological
  order from quantum nets},\ }\href
  {https://doi.org/10.1103/PhysRevLett.110.260408} {\bibfield  {journal}
  {\bibinfo  {journal} {Phys. Rev. Lett.}\ }\textbf {\bibinfo {volume} {110}},\
  \bibinfo {pages} {260408} (\bibinfo {year} {2013})}\BibitemShut {NoStop}%
\bibitem [{\citenamefont {Haegeman}\ \emph
  {et~al.}(2015{\natexlab{a}})\citenamefont {Haegeman}, \citenamefont
  {Van~Acoleyen}, \citenamefont {Schuch}, \citenamefont {Cirac},\ and\
  \citenamefont {Verstraete}}]{haegeman_2015_gauging}%
  \BibitemOpen
  \bibfield  {author} {\bibinfo {author} {\bibfnamefont {J.}~\bibnamefont
  {Haegeman}}, \bibinfo {author} {\bibfnamefont {K.}~\bibnamefont
  {Van~Acoleyen}}, \bibinfo {author} {\bibfnamefont {N.}~\bibnamefont
  {Schuch}}, \bibinfo {author} {\bibfnamefont {J.~I.}\ \bibnamefont {Cirac}},\
  and\ \bibinfo {author} {\bibfnamefont {F.}~\bibnamefont {Verstraete}},\
  }\bibfield  {title} {\bibinfo {title} {Gauging quantum states: From global to
  local symmetries in many-body systems},\ }\href
  {https://doi.org/10.1103/PhysRevX.5.011024} {\bibfield  {journal} {\bibinfo
  {journal} {Phys. Rev. X}\ }\textbf {\bibinfo {volume} {5}},\ \bibinfo {pages}
  {011024} (\bibinfo {year} {2015}{\natexlab{a}})}\BibitemShut {NoStop}%
\bibitem [{\citenamefont {Zhu}\ and\ \citenamefont
  {Zhang}(2019)}]{zhu2019gapless}%
  \BibitemOpen
  \bibfield  {author} {\bibinfo {author} {\bibfnamefont {G.-Y.}\ \bibnamefont
  {Zhu}}\ and\ \bibinfo {author} {\bibfnamefont {G.-M.}\ \bibnamefont
  {Zhang}},\ }\bibfield  {title} {\bibinfo {title} {Gapless coulomb state
  emerging from a self-dual topological tensor-network state},\ }\href
  {https://doi.org/10.1103/PhysRevLett.122.176401} {\bibfield  {journal}
  {\bibinfo  {journal} {Phys. Rev. Lett.}\ }\textbf {\bibinfo {volume} {122}},\
  \bibinfo {pages} {176401} (\bibinfo {year} {2019})}\BibitemShut {NoStop}%
\bibitem [{\citenamefont {Schotte}\ \emph {et~al.}(2019)\citenamefont
  {Schotte}, \citenamefont {Carrasco}, \citenamefont {Vanhecke}, \citenamefont
  {Vanderstraeten}, \citenamefont {Haegeman}, \citenamefont {Verstraete},\ and\
  \citenamefont {Vidal}}]{schotte2019tensornetwork}%
  \BibitemOpen
  \bibfield  {author} {\bibinfo {author} {\bibfnamefont {A.}~\bibnamefont
  {Schotte}}, \bibinfo {author} {\bibfnamefont {J.}~\bibnamefont {Carrasco}},
  \bibinfo {author} {\bibfnamefont {B.}~\bibnamefont {Vanhecke}}, \bibinfo
  {author} {\bibfnamefont {L.}~\bibnamefont {Vanderstraeten}}, \bibinfo
  {author} {\bibfnamefont {J.}~\bibnamefont {Haegeman}}, \bibinfo {author}
  {\bibfnamefont {F.}~\bibnamefont {Verstraete}},\ and\ \bibinfo {author}
  {\bibfnamefont {J.}~\bibnamefont {Vidal}},\ }\bibfield  {title} {\bibinfo
  {title} {Tensor-network approach to phase transitions in string-net models},\
  }\href {https://doi.org/10.1103/PhysRevB.100.245125} {\bibfield  {journal}
  {\bibinfo  {journal} {Phys. Rev. B}\ }\textbf {\bibinfo {volume} {100}},\
  \bibinfo {pages} {245125} (\bibinfo {year} {2019})}\BibitemShut {NoStop}%
\bibitem [{\citenamefont {Xu}\ \emph {et~al.}(2020)\citenamefont {Xu},
  \citenamefont {Zhang},\ and\ \citenamefont {Zhang}}]{xu2020tensor}%
  \BibitemOpen
  \bibfield  {author} {\bibinfo {author} {\bibfnamefont {W.-T.}\ \bibnamefont
  {Xu}}, \bibinfo {author} {\bibfnamefont {Q.}~\bibnamefont {Zhang}},\ and\
  \bibinfo {author} {\bibfnamefont {G.-M.}\ \bibnamefont {Zhang}},\ }\bibfield
  {title} {\bibinfo {title} {Tensor network approach to phase transitions of a
  non-abelian topological phase},\ }\href
  {https://doi.org/10.1103/PhysRevLett.124.130603} {\bibfield  {journal}
  {\bibinfo  {journal} {Phys. Rev. Lett.}\ }\textbf {\bibinfo {volume} {124}},\
  \bibinfo {pages} {130603} (\bibinfo {year} {2020})}\BibitemShut {NoStop}%
\bibitem [{\citenamefont {Xu}\ and\ \citenamefont
  {Schuch}(2021)}]{xu2021characterization}%
  \BibitemOpen
  \bibfield  {author} {\bibinfo {author} {\bibfnamefont {W.-T.}\ \bibnamefont
  {Xu}}\ and\ \bibinfo {author} {\bibfnamefont {N.}~\bibnamefont {Schuch}},\
  }\bibfield  {title} {\bibinfo {title} {Characterization of topological phase
  transitions from a non-abelian topological state and its galois conjugate
  through condensation and confinement order parameters},\ }\href
  {https://doi.org/10.1103/PhysRevB.104.155119} {\bibfield  {journal} {\bibinfo
   {journal} {Phys. Rev. B}\ }\textbf {\bibinfo {volume} {104}},\ \bibinfo
  {pages} {155119} (\bibinfo {year} {2021})}\BibitemShut {NoStop}%
\bibitem [{\citenamefont {Sala}\ \emph {et~al.}(2024)\citenamefont {Sala},
  \citenamefont {Alicea},\ and\ \citenamefont
  {Verresen}}]{sala2024decoherencewavefunctiondeformationd4}%
  \BibitemOpen
  \bibfield  {author} {\bibinfo {author} {\bibfnamefont {P.}~\bibnamefont
  {Sala}}, \bibinfo {author} {\bibfnamefont {J.}~\bibnamefont {Alicea}},\ and\
  \bibinfo {author} {\bibfnamefont {R.}~\bibnamefont {Verresen}},\ }\href
  {https://arxiv.org/abs/2409.12948} {\bibinfo {title} {Decoherence and
  wavefunction deformation of $d_4$ non-abelian topological order}} (\bibinfo
  {year} {2024}),\ \Eprint {https://arxiv.org/abs/2409.12948} {arXiv:2409.12948
  [cond-mat.str-el]} \BibitemShut {NoStop}%
\bibitem [{\citenamefont {Sala}\ and\ \citenamefont
  {Verresen}(2024)}]{sala2024stabilityloopmodelsdecohering}%
  \BibitemOpen
  \bibfield  {author} {\bibinfo {author} {\bibfnamefont {P.}~\bibnamefont
  {Sala}}\ and\ \bibinfo {author} {\bibfnamefont {R.}~\bibnamefont
  {Verresen}},\ }\href {https://arxiv.org/abs/2409.12230} {\bibinfo {title}
  {Stability and loop models from decohering non-abelian topological order}}
  (\bibinfo {year} {2024}),\ \Eprint {https://arxiv.org/abs/2409.12230}
  {arXiv:2409.12230 [quant-ph]} \BibitemShut {NoStop}%
\bibitem [{\citenamefont {Sahay}\ \emph {et~al.}(2023)\citenamefont {Sahay},
  \citenamefont {Vishwanath},\ and\ \citenamefont {Verresen}}]{Sahay_2023}%
  \BibitemOpen
  \bibfield  {author} {\bibinfo {author} {\bibfnamefont {R.}~\bibnamefont
  {Sahay}}, \bibinfo {author} {\bibfnamefont {A.}~\bibnamefont {Vishwanath}},\
  and\ \bibinfo {author} {\bibfnamefont {R.}~\bibnamefont {Verresen}},\
  }\href@noop {} {\bibinfo {title} {Quantum spin puddles and lakes: Nisq-era
  spin liquids from non-equilibrium dynamics}} (\bibinfo {year} {2023}),\
  \Eprint {https://arxiv.org/abs/2211.01381} {arXiv:2211.01381
  [cond-mat.str-el]} \BibitemShut {NoStop}%
\bibitem [{\citenamefont {Sahay}\ and\ \citenamefont
  {Verresen}(2024)}]{sahay2024finitedepthpreparationtensornetwork}%
  \BibitemOpen
  \bibfield  {author} {\bibinfo {author} {\bibfnamefont {R.}~\bibnamefont
  {Sahay}}\ and\ \bibinfo {author} {\bibfnamefont {R.}~\bibnamefont
  {Verresen}},\ }\href {https://arxiv.org/abs/2404.17087} {\bibinfo {title}
  {Finite-depth preparation of tensor network states from measurement}}
  (\bibinfo {year} {2024}),\ \Eprint {https://arxiv.org/abs/2404.17087}
  {arXiv:2404.17087 [quant-ph]} \BibitemShut {NoStop}%
\bibitem [{\citenamefont {Sahay}\ and\ \citenamefont
  {Verresen}(2025)}]{sahay2025classifying}%
  \BibitemOpen
  \bibfield  {author} {\bibinfo {author} {\bibfnamefont {R.}~\bibnamefont
  {Sahay}}\ and\ \bibinfo {author} {\bibfnamefont {R.}~\bibnamefont
  {Verresen}},\ }\bibfield  {title} {\bibinfo {title} {Classifying
  one-dimensional quantum states prepared by a single round of measurements},\
  }\href {https://doi.org/10.1103/PRXQuantum.6.010329} {\bibfield  {journal}
  {\bibinfo  {journal} {PRX Quantum}\ }\textbf {\bibinfo {volume} {6}},\
  \bibinfo {pages} {010329} (\bibinfo {year} {2025})}\BibitemShut {NoStop}%
\bibitem [{\citenamefont {Zhu}\ \emph {et~al.}(2023)\citenamefont {Zhu},
  \citenamefont {Tantivasadakarn}, \citenamefont {Vishwanath}, \citenamefont
  {Trebst},\ and\ \citenamefont {Verresen}}]{zhu2023nishimori}%
  \BibitemOpen
  \bibfield  {author} {\bibinfo {author} {\bibfnamefont {G.-Y.}\ \bibnamefont
  {Zhu}}, \bibinfo {author} {\bibfnamefont {N.}~\bibnamefont
  {Tantivasadakarn}}, \bibinfo {author} {\bibfnamefont {A.}~\bibnamefont
  {Vishwanath}}, \bibinfo {author} {\bibfnamefont {S.}~\bibnamefont {Trebst}},\
  and\ \bibinfo {author} {\bibfnamefont {R.}~\bibnamefont {Verresen}},\
  }\bibfield  {title} {\bibinfo {title} {Nishimori's cat: Stable long-range
  entanglement from finite-depth unitaries and weak measurements},\ }\href
  {https://doi.org/10.1103/PhysRevLett.131.200201} {\bibfield  {journal}
  {\bibinfo  {journal} {Phys. Rev. Lett.}\ }\textbf {\bibinfo {volume} {131}},\
  \bibinfo {pages} {200201} (\bibinfo {year} {2023})}\BibitemShut {NoStop}%
\bibitem [{\citenamefont {Lee}\ \emph {et~al.}(2023)\citenamefont {Lee},
  \citenamefont {Jian},\ and\ \citenamefont {Xu}}]{Lee2023quantum}%
  \BibitemOpen
  \bibfield  {author} {\bibinfo {author} {\bibfnamefont {J.~Y.}\ \bibnamefont
  {Lee}}, \bibinfo {author} {\bibfnamefont {C.-M.}\ \bibnamefont {Jian}},\ and\
  \bibinfo {author} {\bibfnamefont {C.}~\bibnamefont {Xu}},\ }\bibfield
  {title} {\bibinfo {title} {Quantum criticality under decoherence or weak
  measurement},\ }\href {https://doi.org/10.1103/PRXQuantum.4.030317}
  {\bibfield  {journal} {\bibinfo  {journal} {PRX Quantum}\ }\textbf {\bibinfo
  {volume} {4}},\ \bibinfo {pages} {030317} (\bibinfo {year}
  {2023})}\BibitemShut {NoStop}%
\bibitem [{\citenamefont
  {Lee}(2024)}]{lee2024exactcalculationscoherentinformation}%
  \BibitemOpen
  \bibfield  {author} {\bibinfo {author} {\bibfnamefont {J.~Y.}\ \bibnamefont
  {Lee}},\ }\href {https://arxiv.org/abs/2402.16937} {\bibinfo {title} {Exact
  calculations of coherent information for toric codes under decoherence:
  Identifying the fundamental error threshold}} (\bibinfo {year} {2024}),\
  \Eprint {https://arxiv.org/abs/2402.16937} {arXiv:2402.16937
  [cond-mat.stat-mech]} \BibitemShut {NoStop}%
\bibitem [{\citenamefont {Su}\ \emph {et~al.}(2024)\citenamefont {Su},
  \citenamefont {Yang},\ and\ \citenamefont {Jian}}]{su2024tapestry}%
  \BibitemOpen
  \bibfield  {author} {\bibinfo {author} {\bibfnamefont {K.}~\bibnamefont
  {Su}}, \bibinfo {author} {\bibfnamefont {Z.}~\bibnamefont {Yang}},\ and\
  \bibinfo {author} {\bibfnamefont {C.-M.}\ \bibnamefont {Jian}},\ }\bibfield
  {title} {\bibinfo {title} {Tapestry of dualities in decohered quantum error
  correction codes},\ }\href {https://doi.org/10.1103/PhysRevB.110.085158}
  {\bibfield  {journal} {\bibinfo  {journal} {Phys. Rev. B}\ }\textbf {\bibinfo
  {volume} {110}},\ \bibinfo {pages} {085158} (\bibinfo {year}
  {2024})}\BibitemShut {NoStop}%
\bibitem [{\citenamefont
  {Lyons}(2024)}]{lyons2024understandingstabilizercodeslocal}%
  \BibitemOpen
  \bibfield  {author} {\bibinfo {author} {\bibfnamefont {A.}~\bibnamefont
  {Lyons}},\ }\href {https://arxiv.org/abs/2403.03955} {\bibinfo {title}
  {Understanding stabilizer codes under local decoherence through a general
  statistical mechanics mapping}} (\bibinfo {year} {2024}),\ \Eprint
  {https://arxiv.org/abs/2403.03955} {arXiv:2403.03955 [quant-ph]} \BibitemShut
  {NoStop}%
\bibitem [{\citenamefont {Li}\ and\ \citenamefont
  {Mong}(2024)}]{li2024replicatopologicalorderquantum}%
  \BibitemOpen
  \bibfield  {author} {\bibinfo {author} {\bibfnamefont {Z.}~\bibnamefont
  {Li}}\ and\ \bibinfo {author} {\bibfnamefont {R.~S.~K.}\ \bibnamefont
  {Mong}},\ }\href {https://arxiv.org/abs/2402.09516} {\bibinfo {title}
  {Replica topological order in quantum mixed states and quantum error
  correction}} (\bibinfo {year} {2024}),\ \Eprint
  {https://arxiv.org/abs/2402.09516} {arXiv:2402.09516 [quant-ph]} \BibitemShut
  {NoStop}%
\bibitem [{\citenamefont {Chen}\ and\ \citenamefont
  {Grover}(2024{\natexlab{a}})}]{chen_2024_unconventional}%
  \BibitemOpen
  \bibfield  {author} {\bibinfo {author} {\bibfnamefont {Y.-H.}\ \bibnamefont
  {Chen}}\ and\ \bibinfo {author} {\bibfnamefont {T.}~\bibnamefont {Grover}},\
  }\bibfield  {title} {\bibinfo {title} {Unconventional topological mixed-state
  transition and critical phase induced by self-dual coherent errors},\ }\href
  {https://doi.org/10.1103/PhysRevB.110.125152} {\bibfield  {journal} {\bibinfo
   {journal} {Phys. Rev. B}\ }\textbf {\bibinfo {volume} {110}},\ \bibinfo
  {pages} {125152} (\bibinfo {year} {2024}{\natexlab{a}})}\BibitemShut
  {NoStop}%
\bibitem [{\citenamefont {Chen}\ and\ \citenamefont
  {Grover}(2024{\natexlab{b}})}]{chen_2024_separability}%
  \BibitemOpen
  \bibfield  {author} {\bibinfo {author} {\bibfnamefont {Y.-H.}\ \bibnamefont
  {Chen}}\ and\ \bibinfo {author} {\bibfnamefont {T.}~\bibnamefont {Grover}},\
  }\bibfield  {title} {\bibinfo {title} {Separability transitions in
  topological states induced by local decoherence},\ }\href
  {https://doi.org/10.1103/PhysRevLett.132.170602} {\bibfield  {journal}
  {\bibinfo  {journal} {Phys. Rev. Lett.}\ }\textbf {\bibinfo {volume} {132}},\
  \bibinfo {pages} {170602} (\bibinfo {year} {2024}{\natexlab{b}})}\BibitemShut
  {NoStop}%
\bibitem [{\citenamefont {Sohal}\ and\ \citenamefont
  {Prem}(2025)}]{sohal2025noisy}%
  \BibitemOpen
  \bibfield  {author} {\bibinfo {author} {\bibfnamefont {R.}~\bibnamefont
  {Sohal}}\ and\ \bibinfo {author} {\bibfnamefont {A.}~\bibnamefont {Prem}},\
  }\bibfield  {title} {\bibinfo {title} {Noisy approach to intrinsically
  mixed-state topological order},\ }\href
  {https://doi.org/10.1103/PRXQuantum.6.010313} {\bibfield  {journal} {\bibinfo
   {journal} {PRX Quantum}\ }\textbf {\bibinfo {volume} {6}},\ \bibinfo {pages}
  {010313} (\bibinfo {year} {2025})}\BibitemShut {NoStop}%
\bibitem [{\citenamefont {Ellison}\ and\ \citenamefont
  {Cheng}(2025)}]{ellison2025towards}%
  \BibitemOpen
  \bibfield  {author} {\bibinfo {author} {\bibfnamefont {T.~D.}\ \bibnamefont
  {Ellison}}\ and\ \bibinfo {author} {\bibfnamefont {M.}~\bibnamefont
  {Cheng}},\ }\bibfield  {title} {\bibinfo {title} {Toward a classification of
  mixed-state topological orders in two dimensions},\ }\href
  {https://doi.org/10.1103/PRXQuantum.6.010315} {\bibfield  {journal} {\bibinfo
   {journal} {PRX Quantum}\ }\textbf {\bibinfo {volume} {6}},\ \bibinfo {pages}
  {010315} (\bibinfo {year} {2025})}\BibitemShut {NoStop}%
\bibitem [{\citenamefont {Hauser}\ \emph {et~al.}(2024)\citenamefont {Hauser},
  \citenamefont {Bao}, \citenamefont {Sang}, \citenamefont {Lavasani},
  \citenamefont {Agrawal},\ and\ \citenamefont
  {Fisher}}]{hauser2024informationdynamicsdecoheredquantum}%
  \BibitemOpen
  \bibfield  {author} {\bibinfo {author} {\bibfnamefont {J.}~\bibnamefont
  {Hauser}}, \bibinfo {author} {\bibfnamefont {Y.}~\bibnamefont {Bao}},
  \bibinfo {author} {\bibfnamefont {S.}~\bibnamefont {Sang}}, \bibinfo {author}
  {\bibfnamefont {A.}~\bibnamefont {Lavasani}}, \bibinfo {author}
  {\bibfnamefont {U.}~\bibnamefont {Agrawal}},\ and\ \bibinfo {author}
  {\bibfnamefont {M.~P.~A.}\ \bibnamefont {Fisher}},\ }\href
  {https://arxiv.org/abs/2407.07882} {\bibinfo {title} {Information dynamics in
  decohered quantum memory with repeated syndrome measurements: a dual
  approach}} (\bibinfo {year} {2024}),\ \Eprint
  {https://arxiv.org/abs/2407.07882} {arXiv:2407.07882 [quant-ph]} \BibitemShut
  {NoStop}%
\bibitem [{\citenamefont
  {Lu}(2024)}]{lu2024disentanglingtransitionstopologicalorder}%
  \BibitemOpen
  \bibfield  {author} {\bibinfo {author} {\bibfnamefont {T.-C.}\ \bibnamefont
  {Lu}},\ }\href {https://arxiv.org/abs/2404.06514} {\bibinfo {title}
  {Disentangling transitions in topological order induced by boundary
  decoherence}} (\bibinfo {year} {2024}),\ \Eprint
  {https://arxiv.org/abs/2404.06514} {arXiv:2404.06514 [quant-ph]} \BibitemShut
  {NoStop}%
\bibitem [{\citenamefont {Sun}\ \emph {et~al.}(2025)\citenamefont {Sun},
  \citenamefont {Zhang}, \citenamefont {Bi},\ and\ \citenamefont
  {You}}]{sun2025holographicviewmixedstatesymmetryprotected}%
  \BibitemOpen
  \bibfield  {author} {\bibinfo {author} {\bibfnamefont {S.}~\bibnamefont
  {Sun}}, \bibinfo {author} {\bibfnamefont {J.-H.}\ \bibnamefont {Zhang}},
  \bibinfo {author} {\bibfnamefont {Z.}~\bibnamefont {Bi}},\ and\ \bibinfo
  {author} {\bibfnamefont {Y.}~\bibnamefont {You}},\ }\href
  {https://arxiv.org/abs/2410.08205} {\bibinfo {title} {Holographic view of
  mixed-state symmetry-protected topological phases in open quantum systems}}
  (\bibinfo {year} {2025}),\ \Eprint {https://arxiv.org/abs/2410.08205}
  {arXiv:2410.08205 [quant-ph]} \BibitemShut {NoStop}%
\bibitem [{\citenamefont {Zhang}\ \emph {et~al.}(2025)\citenamefont {Zhang},
  \citenamefont {Xu}, \citenamefont {Zhang}, \citenamefont {Xu}, \citenamefont
  {Bi},\ and\ \citenamefont
  {Luo}}]{zhang2025strongtoweakspontaneousbreaking1form}%
  \BibitemOpen
  \bibfield  {author} {\bibinfo {author} {\bibfnamefont {C.}~\bibnamefont
  {Zhang}}, \bibinfo {author} {\bibfnamefont {Y.}~\bibnamefont {Xu}}, \bibinfo
  {author} {\bibfnamefont {J.-H.}\ \bibnamefont {Zhang}}, \bibinfo {author}
  {\bibfnamefont {C.}~\bibnamefont {Xu}}, \bibinfo {author} {\bibfnamefont
  {Z.}~\bibnamefont {Bi}},\ and\ \bibinfo {author} {\bibfnamefont {Z.-X.}\
  \bibnamefont {Luo}},\ }\href {https://arxiv.org/abs/2409.17530} {\bibinfo
  {title} {Strong-to-weak spontaneous breaking of 1-form symmetry and
  intrinsically mixed topological order}} (\bibinfo {year} {2025}),\ \Eprint
  {https://arxiv.org/abs/2409.17530} {arXiv:2409.17530 [quant-ph]} \BibitemShut
  {NoStop}%
\bibitem [{\citenamefont {Haegeman}\ \emph
  {et~al.}(2015{\natexlab{b}})\citenamefont {Haegeman}, \citenamefont {Zauner},
  \citenamefont {Schuch},\ and\ \citenamefont
  {Verstraete}}]{haegeman2015shadows}%
  \BibitemOpen
  \bibfield  {author} {\bibinfo {author} {\bibfnamefont {J.}~\bibnamefont
  {Haegeman}}, \bibinfo {author} {\bibfnamefont {V.}~\bibnamefont {Zauner}},
  \bibinfo {author} {\bibfnamefont {N.}~\bibnamefont {Schuch}},\ and\ \bibinfo
  {author} {\bibfnamefont {F.}~\bibnamefont {Verstraete}},\ }\bibfield  {title}
  {\bibinfo {title} {Shadows of anyons and the entanglement structure of
  topological phases},\ }\href {https://www.nature.com/articles/ncomms9284}
  {\bibfield  {journal} {\bibinfo  {journal} {Nature communications}\ }\textbf
  {\bibinfo {volume} {6}},\ \bibinfo {pages} {8284} (\bibinfo {year}
  {2015}{\natexlab{b}})}\BibitemShut {NoStop}%
\bibitem [{\citenamefont {Liu}\ \emph {et~al.}(2025)\citenamefont {Liu},
  \citenamefont {Xu}, \citenamefont {Pollmann},\ and\ \citenamefont
  {Knap}}]{liu2025}%
  \BibitemOpen
  \bibfield  {author} {\bibinfo {author} {\bibfnamefont {Y.-J.}\ \bibnamefont
  {Liu}}, \bibinfo {author} {\bibfnamefont {W.-T.}\ \bibnamefont {Xu}},
  \bibinfo {author} {\bibfnamefont {F.}~\bibnamefont {Pollmann}},\ and\
  \bibinfo {author} {\bibfnamefont {M.}~\bibnamefont {Knap}},\ }\bibfield
  {title} {\bibinfo {title} {Detecting emergent 1-form symmetries with quantum
  error correction},\ }\href {https://arxiv.org/abs/2502.17572} {\bibfield
  {journal} {\bibinfo  {journal} {arXiv preprint arXiv:2502.17572}\ } (\bibinfo
  {year} {2025})}\BibitemShut {NoStop}%
\bibitem [{\citenamefont {Nishio}\ \emph {et~al.}(2004)\citenamefont {Nishio},
  \citenamefont {Maeshima}, \citenamefont {Gendiar},\ and\ \citenamefont
  {Nishino}}]{nishio2004tensorproductvariationalformulation}%
  \BibitemOpen
  \bibfield  {author} {\bibinfo {author} {\bibfnamefont {Y.}~\bibnamefont
  {Nishio}}, \bibinfo {author} {\bibfnamefont {N.}~\bibnamefont {Maeshima}},
  \bibinfo {author} {\bibfnamefont {A.}~\bibnamefont {Gendiar}},\ and\ \bibinfo
  {author} {\bibfnamefont {T.}~\bibnamefont {Nishino}},\ }\href
  {https://arxiv.org/abs/cond-mat/0401115} {\bibinfo {title} {Tensor product
  variational formulation for quantum systems}} (\bibinfo {year} {2004}),\
  \Eprint {https://arxiv.org/abs/cond-mat/0401115} {arXiv:cond-mat/0401115
  [cond-mat.stat-mech]} \BibitemShut {NoStop}%
\bibitem [{\citenamefont {Fern\'andez-Gonz\'alez}\ \emph
  {et~al.}(2012)\citenamefont {Fern\'andez-Gonz\'alez}, \citenamefont {Schuch},
  \citenamefont {Wolf}, \citenamefont {Cirac},\ and\ \citenamefont
  {P\'erez-Garc\'{\i}a}}]{uncle}%
  \BibitemOpen
  \bibfield  {author} {\bibinfo {author} {\bibfnamefont {C.}~\bibnamefont
  {Fern\'andez-Gonz\'alez}}, \bibinfo {author} {\bibfnamefont {N.}~\bibnamefont
  {Schuch}}, \bibinfo {author} {\bibfnamefont {M.~M.}\ \bibnamefont {Wolf}},
  \bibinfo {author} {\bibfnamefont {J.~I.}\ \bibnamefont {Cirac}},\ and\
  \bibinfo {author} {\bibfnamefont {D.}~\bibnamefont {P\'erez-Garc\'{\i}a}},\
  }\bibfield  {title} {\bibinfo {title} {Gapless hamiltonians for the toric
  code using the projected entangled pair state formalism},\ }\href
  {https://doi.org/10.1103/PhysRevLett.109.260401} {\bibfield  {journal}
  {\bibinfo  {journal} {Phys. Rev. Lett.}\ }\textbf {\bibinfo {volume} {109}},\
  \bibinfo {pages} {260401} (\bibinfo {year} {2012})}\BibitemShut {NoStop}%
\bibitem [{\citenamefont {Fern{\'a}ndez-Gonz{\'a}lez}\ \emph
  {et~al.}(2015)\citenamefont {Fern{\'a}ndez-Gonz{\'a}lez}, \citenamefont
  {Schuch}, \citenamefont {Wolf}, \citenamefont {Cirac},\ and\ \citenamefont
  {P{\'e}rez-Garc{\'i}a}}]{uncle2}%
  \BibitemOpen
  \bibfield  {author} {\bibinfo {author} {\bibfnamefont {C.}~\bibnamefont
  {Fern{\'a}ndez-Gonz{\'a}lez}}, \bibinfo {author} {\bibfnamefont
  {N.}~\bibnamefont {Schuch}}, \bibinfo {author} {\bibfnamefont {M.~M.}\
  \bibnamefont {Wolf}}, \bibinfo {author} {\bibfnamefont {J.~I.}\ \bibnamefont
  {Cirac}},\ and\ \bibinfo {author} {\bibfnamefont {D.}~\bibnamefont
  {P{\'e}rez-Garc{\'i}a}},\ }\bibfield  {title} {\bibinfo {title} {Frustration
  free gapless hamiltonians for matrix product states},\ }\href
  {https://doi.org/10.1007/s00220-014-2173-z} {\bibfield  {journal} {\bibinfo
  {journal} {Communications in Mathematical Physics}\ }\textbf {\bibinfo
  {volume} {333}},\ \bibinfo {pages} {299} (\bibinfo {year}
  {2015})}\BibitemShut {NoStop}%
\bibitem [{\citenamefont {Tantivasadakarn}\ \emph {et~al.}(2023)\citenamefont
  {Tantivasadakarn}, \citenamefont {Thorngren}, \citenamefont {Vishwanath},\
  and\ \citenamefont {Verresen}}]{pivot}%
  \BibitemOpen
  \bibfield  {author} {\bibinfo {author} {\bibfnamefont {N.}~\bibnamefont
  {Tantivasadakarn}}, \bibinfo {author} {\bibfnamefont {R.}~\bibnamefont
  {Thorngren}}, \bibinfo {author} {\bibfnamefont {A.}~\bibnamefont
  {Vishwanath}},\ and\ \bibinfo {author} {\bibfnamefont {R.}~\bibnamefont
  {Verresen}},\ }\bibfield  {title} {\bibinfo {title} {{Pivot Hamiltonians as
  generators of symmetry and entanglement}},\ }\href
  {https://doi.org/10.21468/SciPostPhys.14.2.012} {\bibfield  {journal}
  {\bibinfo  {journal} {SciPost Phys.}\ }\textbf {\bibinfo {volume} {14}},\
  \bibinfo {pages} {012} (\bibinfo {year} {2023})}\BibitemShut {NoStop}%
\bibitem [{\citenamefont {Chen}\ \emph {et~al.}(2010)\citenamefont {Chen},
  \citenamefont {Zeng}, \citenamefont {Gu}, \citenamefont {Chuang},\ and\
  \citenamefont {Wen}}]{chen20101formpeps}%
  \BibitemOpen
  \bibfield  {author} {\bibinfo {author} {\bibfnamefont {X.}~\bibnamefont
  {Chen}}, \bibinfo {author} {\bibfnamefont {B.}~\bibnamefont {Zeng}}, \bibinfo
  {author} {\bibfnamefont {Z.-C.}\ \bibnamefont {Gu}}, \bibinfo {author}
  {\bibfnamefont {I.~L.}\ \bibnamefont {Chuang}},\ and\ \bibinfo {author}
  {\bibfnamefont {X.-G.}\ \bibnamefont {Wen}},\ }\bibfield  {title} {\bibinfo
  {title} {Tensor product representation of a topological ordered phase:
  Necessary symmetry conditions},\ }\href
  {https://doi.org/10.1103/PhysRevB.82.165119} {\bibfield  {journal} {\bibinfo
  {journal} {Phys. Rev. B}\ }\textbf {\bibinfo {volume} {82}},\ \bibinfo
  {pages} {165119} (\bibinfo {year} {2010})}\BibitemShut {NoStop}%
\bibitem [{\citenamefont {Schuch}\ \emph {et~al.}(2013)\citenamefont {Schuch},
  \citenamefont {Poilblanc}, \citenamefont {Cirac},\ and\ \citenamefont
  {P\'erez-Garc\'{\i}a}}]{schuch2013topological}%
  \BibitemOpen
  \bibfield  {author} {\bibinfo {author} {\bibfnamefont {N.}~\bibnamefont
  {Schuch}}, \bibinfo {author} {\bibfnamefont {D.}~\bibnamefont {Poilblanc}},
  \bibinfo {author} {\bibfnamefont {J.~I.}\ \bibnamefont {Cirac}},\ and\
  \bibinfo {author} {\bibfnamefont {D.}~\bibnamefont {P\'erez-Garc\'{\i}a}},\
  }\bibfield  {title} {\bibinfo {title} {Topological order in the projected
  entangled-pair states formalism: Transfer operator and boundary
  hamiltonians},\ }\href {https://doi.org/10.1103/PhysRevLett.111.090501}
  {\bibfield  {journal} {\bibinfo  {journal} {Phys. Rev. Lett.}\ }\textbf
  {\bibinfo {volume} {111}},\ \bibinfo {pages} {090501} (\bibinfo {year}
  {2013})}\BibitemShut {NoStop}%
\bibitem [{\citenamefont {{\c S}ahino{\u g}lu}\ \emph
  {et~al.}(2021)\citenamefont {{\c S}ahino{\u g}lu}, \citenamefont
  {Williamson}, \citenamefont {Bultinck}, \citenamefont {Mari{\"e}n},
  \citenamefont {Haegeman}, \citenamefont {Schuch},\ and\ \citenamefont
  {Verstraete}}]{Sahinoglu2021-zq}%
  \BibitemOpen
  \bibfield  {author} {\bibinfo {author} {\bibfnamefont {M.~B.}\ \bibnamefont
  {{\c S}ahino{\u g}lu}}, \bibinfo {author} {\bibfnamefont {D.}~\bibnamefont
  {Williamson}}, \bibinfo {author} {\bibfnamefont {N.}~\bibnamefont
  {Bultinck}}, \bibinfo {author} {\bibfnamefont {M.}~\bibnamefont
  {Mari{\"e}n}}, \bibinfo {author} {\bibfnamefont {J.}~\bibnamefont
  {Haegeman}}, \bibinfo {author} {\bibfnamefont {N.}~\bibnamefont {Schuch}},\
  and\ \bibinfo {author} {\bibfnamefont {F.}~\bibnamefont {Verstraete}},\
  }\bibfield  {title} {\bibinfo {title} {Characterizing topological order with
  matrix product operators},\ }\href@noop {} {\bibfield  {journal} {\bibinfo
  {journal} {Annales Henri Poincar{\'e}}\ }\textbf {\bibinfo {volume} {22}},\
  \bibinfo {pages} {563} (\bibinfo {year} {2021})}\BibitemShut {NoStop}%
\bibitem [{\citenamefont {Shukla}\ \emph {et~al.}(2018)\citenamefont {Shukla},
  \citenamefont {\ifmmode \mbox{\c{S}}\else
  \c{S}\fi{}ahino\ifmmode~\breve{g}\else \u{g}\fi{}lu}, \citenamefont
  {Pollmann},\ and\ \citenamefont {Chen}}]{shukla_2018_boson}%
  \BibitemOpen
  \bibfield  {author} {\bibinfo {author} {\bibfnamefont {S.~K.}\ \bibnamefont
  {Shukla}}, \bibinfo {author} {\bibfnamefont {M.~B.}\ \bibnamefont {\ifmmode
  \mbox{\c{S}}\else \c{S}\fi{}ahino\ifmmode~\breve{g}\else \u{g}\fi{}lu}},
  \bibinfo {author} {\bibfnamefont {F.}~\bibnamefont {Pollmann}},\ and\
  \bibinfo {author} {\bibfnamefont {X.}~\bibnamefont {Chen}},\ }\bibfield
  {title} {\bibinfo {title} {Boson condensation and instability in the tensor
  network representation of string-net states},\ }\href
  {https://doi.org/10.1103/PhysRevB.98.125112} {\bibfield  {journal} {\bibinfo
  {journal} {Phys. Rev. B}\ }\textbf {\bibinfo {volume} {98}},\ \bibinfo
  {pages} {125112} (\bibinfo {year} {2018})}\BibitemShut {NoStop}%
\bibitem [{\citenamefont {Hauschild}\ and\ \citenamefont
  {Pollmann}(2018)}]{Hauschild18}%
  \BibitemOpen
  \bibfield  {author} {\bibinfo {author} {\bibfnamefont {J.}~\bibnamefont
  {Hauschild}}\ and\ \bibinfo {author} {\bibfnamefont {F.}~\bibnamefont
  {Pollmann}},\ }\bibfield  {title} {\bibinfo {title} {{Efficient numerical
  simulations with Tensor Networks: Tensor Network Python (TeNPy)}},\ }\href
  {https://doi.org/10.21468/SciPostPhysLectNotes.5} {\bibfield  {journal}
  {\bibinfo  {journal} {SciPost Phys. Lect. Notes}\ ,\ \bibinfo {pages} {5}}
  (\bibinfo {year} {2018})}\BibitemShut {NoStop}%
\bibitem [{\citenamefont {Kj\"all}\ \emph {et~al.}(2013)\citenamefont
  {Kj\"all}, \citenamefont {Zaletel}, \citenamefont {Mong}, \citenamefont
  {Bardarson},\ and\ \citenamefont {Pollmann}}]{Kjaell13}%
  \BibitemOpen
  \bibfield  {author} {\bibinfo {author} {\bibfnamefont {J.~A.}\ \bibnamefont
  {Kj\"all}}, \bibinfo {author} {\bibfnamefont {M.~P.}\ \bibnamefont
  {Zaletel}}, \bibinfo {author} {\bibfnamefont {R.~S.~K.}\ \bibnamefont
  {Mong}}, \bibinfo {author} {\bibfnamefont {J.~H.}\ \bibnamefont
  {Bardarson}},\ and\ \bibinfo {author} {\bibfnamefont {F.}~\bibnamefont
  {Pollmann}},\ }\bibfield  {title} {\bibinfo {title} {Phase diagram of the
  anisotropic spin-2 xxz model: Infinite-system density matrix renormalization
  group study},\ }\href {https://doi.org/10.1103/PhysRevB.87.235106} {\bibfield
   {journal} {\bibinfo  {journal} {Phys. Rev. B}\ }\textbf {\bibinfo {volume}
  {87}},\ \bibinfo {pages} {235106} (\bibinfo {year} {2013})}\BibitemShut
  {NoStop}%
\bibitem [{\citenamefont {Kramers}\ and\ \citenamefont
  {Wannier}(1941)}]{kramersandwannier}%
  \BibitemOpen
  \bibfield  {author} {\bibinfo {author} {\bibfnamefont {H.~A.}\ \bibnamefont
  {Kramers}}\ and\ \bibinfo {author} {\bibfnamefont {G.~H.}\ \bibnamefont
  {Wannier}},\ }\bibfield  {title} {\bibinfo {title} {Statistics of the
  two-dimensional ferromagnet. part i},\ }\href
  {https://doi.org/10.1103/PhysRev.60.252} {\bibfield  {journal} {\bibinfo
  {journal} {Phys. Rev.}\ }\textbf {\bibinfo {volume} {60}},\ \bibinfo {pages}
  {252} (\bibinfo {year} {1941})}\BibitemShut {NoStop}%
\bibitem [{\citenamefont {Pfeuty}(1970)}]{pfeuty1970one}%
  \BibitemOpen
  \bibfield  {author} {\bibinfo {author} {\bibfnamefont {P.}~\bibnamefont
  {Pfeuty}},\ }\bibfield  {title} {\bibinfo {title} {The one-dimensional ising
  model with a transverse field},\ }\href@noop {} {\bibfield  {journal}
  {\bibinfo  {journal} {ANNALS of Physics}\ }\textbf {\bibinfo {volume} {57}},\
  \bibinfo {pages} {79} (\bibinfo {year} {1970})}\BibitemShut {NoStop}%
\bibitem [{\citenamefont {Abraham}(1971)}]{abraham1971transfer}%
  \BibitemOpen
  \bibfield  {author} {\bibinfo {author} {\bibfnamefont {D.}~\bibnamefont
  {Abraham}},\ }\bibfield  {title} {\bibinfo {title} {On the transfer matrix
  for the two-dimensional ising model},\ }\href@noop {} {\bibfield  {journal}
  {\bibinfo  {journal} {Studies in Applied Mathematics}\ }\textbf {\bibinfo
  {volume} {50}},\ \bibinfo {pages} {71} (\bibinfo {year} {1971})}\BibitemShut
  {NoStop}%
\bibitem [{\citenamefont {Oshikawa}(2019)}]{oshikawa2019universal}%
  \BibitemOpen
  \bibfield  {author} {\bibinfo {author} {\bibfnamefont {M.}~\bibnamefont
  {Oshikawa}},\ }\bibfield  {title} {\bibinfo {title} {Universal finite-size
  gap scaling of the quantum ising chain},\ }\href@noop {} {\bibfield
  {journal} {\bibinfo  {journal} {arXiv preprint arXiv:1910.06353}\ } (\bibinfo
  {year} {2019})}\BibitemShut {NoStop}%
\bibitem [{\citenamefont {He}\ \emph {et~al.}(1990)\citenamefont {He},
  \citenamefont {Hamer},\ and\ \citenamefont {Oitmaa}}]{he1990high}%
  \BibitemOpen
  \bibfield  {author} {\bibinfo {author} {\bibfnamefont {H.-X.}\ \bibnamefont
  {He}}, \bibinfo {author} {\bibfnamefont {C.}~\bibnamefont {Hamer}},\ and\
  \bibinfo {author} {\bibfnamefont {J.}~\bibnamefont {Oitmaa}},\ }\bibfield
  {title} {\bibinfo {title} {High-temperature series expansions for the (2+
  1)-dimensional ising model},\ }\href@noop {} {\bibfield  {journal} {\bibinfo
  {journal} {Journal of Physics A: Mathematical and General}\ }\textbf
  {\bibinfo {volume} {23}},\ \bibinfo {pages} {1775} (\bibinfo {year}
  {1990})}\BibitemShut {NoStop}%
\bibitem [{\citenamefont {Wu}\ \emph {et~al.}(2012)\citenamefont {Wu},
  \citenamefont {Deng},\ and\ \citenamefont {Prokof'ev}}]{wu2012phase}%
  \BibitemOpen
  \bibfield  {author} {\bibinfo {author} {\bibfnamefont {F.}~\bibnamefont
  {Wu}}, \bibinfo {author} {\bibfnamefont {Y.}~\bibnamefont {Deng}},\ and\
  \bibinfo {author} {\bibfnamefont {N.}~\bibnamefont {Prokof'ev}},\ }\bibfield
  {title} {\bibinfo {title} {Phase diagram of the toric code model in a
  parallel magnetic field},\ }\href@noop {} {\bibfield  {journal} {\bibinfo
  {journal} {Physical Review B—Condensed Matter and Materials Physics}\
  }\textbf {\bibinfo {volume} {85}},\ \bibinfo {pages} {195104} (\bibinfo
  {year} {2012})}\BibitemShut {NoStop}%
\bibitem [{\citenamefont {Vidal}\ \emph {et~al.}(2008)\citenamefont {Vidal},
  \citenamefont {Dusuel},\ and\ \citenamefont {Schmidt}}]{vidal2008low}%
  \BibitemOpen
  \bibfield  {author} {\bibinfo {author} {\bibfnamefont {J.}~\bibnamefont
  {Vidal}}, \bibinfo {author} {\bibfnamefont {S.}~\bibnamefont {Dusuel}},\ and\
  \bibinfo {author} {\bibfnamefont {K.~P.}\ \bibnamefont {Schmidt}},\
  }\bibfield  {title} {\bibinfo {title} {Low-energy effective theory of the
  toric code model in a parallel field},\ }\href@noop {} {\bibfield  {journal}
  {\bibinfo  {journal} {arXiv preprint arXiv:0807.0487}\ } (\bibinfo {year}
  {2008})}\BibitemShut {NoStop}%
\bibitem [{\citenamefont {Tupitsyn}\ \emph {et~al.}(2010)\citenamefont
  {Tupitsyn}, \citenamefont {Kitaev}, \citenamefont {Prokof'ev},\ and\
  \citenamefont {Stamp}}]{tupitsyn_2010}%
  \BibitemOpen
  \bibfield  {author} {\bibinfo {author} {\bibfnamefont {I.~S.}\ \bibnamefont
  {Tupitsyn}}, \bibinfo {author} {\bibfnamefont {A.}~\bibnamefont {Kitaev}},
  \bibinfo {author} {\bibfnamefont {N.~V.}\ \bibnamefont {Prokof'ev}},\ and\
  \bibinfo {author} {\bibfnamefont {P.~C.~E.}\ \bibnamefont {Stamp}},\
  }\bibfield  {title} {\bibinfo {title} {Topological multicritical point in the
  phase diagram of the toric code model and three-dimensional lattice gauge
  higgs model},\ }\href {https://doi.org/10.1103/PhysRevB.82.085114} {\bibfield
   {journal} {\bibinfo  {journal} {Phys. Rev. B}\ }\textbf {\bibinfo {volume}
  {82}},\ \bibinfo {pages} {085114} (\bibinfo {year} {2010})}\BibitemShut
  {NoStop}%
\bibitem [{\citenamefont {Dusuel}\ \emph {et~al.}(2011)\citenamefont {Dusuel},
  \citenamefont {Kamfor}, \citenamefont {Or\'us}, \citenamefont {Schmidt},\
  and\ \citenamefont {Vidal}}]{dusuel_2011}%
  \BibitemOpen
  \bibfield  {author} {\bibinfo {author} {\bibfnamefont {S.}~\bibnamefont
  {Dusuel}}, \bibinfo {author} {\bibfnamefont {M.}~\bibnamefont {Kamfor}},
  \bibinfo {author} {\bibfnamefont {R.}~\bibnamefont {Or\'us}}, \bibinfo
  {author} {\bibfnamefont {K.~P.}\ \bibnamefont {Schmidt}},\ and\ \bibinfo
  {author} {\bibfnamefont {J.}~\bibnamefont {Vidal}},\ }\bibfield  {title}
  {\bibinfo {title} {Robustness of a perturbed topological phase},\ }\href
  {https://doi.org/10.1103/PhysRevLett.106.107203} {\bibfield  {journal}
  {\bibinfo  {journal} {Phys. Rev. Lett.}\ }\textbf {\bibinfo {volume} {106}},\
  \bibinfo {pages} {107203} (\bibinfo {year} {2011})}\BibitemShut {NoStop}%
\bibitem [{\citenamefont {Hastings}\ and\ \citenamefont
  {Wen}(2005)}]{hastings2005qa}%
  \BibitemOpen
  \bibfield  {author} {\bibinfo {author} {\bibfnamefont {M.~B.}\ \bibnamefont
  {Hastings}}\ and\ \bibinfo {author} {\bibfnamefont {X.-G.}\ \bibnamefont
  {Wen}},\ }\bibfield  {title} {\bibinfo {title} {Quasiadiabatic continuation
  of quantum states: The stability of topological ground-state degeneracy and
  emergent gauge invariance},\ }\href
  {https://doi.org/10.1103/PhysRevB.72.045141} {\bibfield  {journal} {\bibinfo
  {journal} {Phys. Rev. B}\ }\textbf {\bibinfo {volume} {72}},\ \bibinfo
  {pages} {045141} (\bibinfo {year} {2005})}\BibitemShut {NoStop}%
\bibitem [{\citenamefont {White}(1993)}]{White93}%
  \BibitemOpen
  \bibfield  {author} {\bibinfo {author} {\bibfnamefont {S.~R.}\ \bibnamefont
  {White}},\ }\bibfield  {title} {\bibinfo {title} {Density-matrix algorithms
  for quantum renormalization groups},\ }\href
  {https://doi.org/10.1103/PhysRevB.48.10345} {\bibfield  {journal} {\bibinfo
  {journal} {Phys. Rev. B}\ }\textbf {\bibinfo {volume} {48}},\ \bibinfo
  {pages} {10345} (\bibinfo {year} {1993})}\BibitemShut {NoStop}%
\bibitem [{\citenamefont {Zaletel}\ \emph {et~al.}(2015)\citenamefont
  {Zaletel}, \citenamefont {Mong}, \citenamefont {Karrasch}, \citenamefont
  {Moore},\ and\ \citenamefont {Pollmann}}]{Zaletel15}%
  \BibitemOpen
  \bibfield  {author} {\bibinfo {author} {\bibfnamefont {M.~P.}\ \bibnamefont
  {Zaletel}}, \bibinfo {author} {\bibfnamefont {R.~S.~K.}\ \bibnamefont
  {Mong}}, \bibinfo {author} {\bibfnamefont {C.}~\bibnamefont {Karrasch}},
  \bibinfo {author} {\bibfnamefont {J.~E.}\ \bibnamefont {Moore}},\ and\
  \bibinfo {author} {\bibfnamefont {F.}~\bibnamefont {Pollmann}},\ }\bibfield
  {title} {\bibinfo {title} {Time-evolving a matrix product state with
  long-ranged interactions},\ }\href
  {https://doi.org/10.1103/PhysRevB.91.165112} {\bibfield  {journal} {\bibinfo
  {journal} {Phys. Rev. B}\ }\textbf {\bibinfo {volume} {91}},\ \bibinfo
  {pages} {165112} (\bibinfo {year} {2015})}\BibitemShut {NoStop}%
\end{thebibliography}%

\newpage

\pagebreak
\onecolumngrid
\appendix

\section{Properties of Deformed Toric Code} \label{app-propertiesofdtc}

In this section of the appendix, we review the properties of the deformed toric code wavefunction $\ket{\psi(\beta)} \propto e^{\beta/2\sum_{\ell} Z_{\ell}}\ket{\mathsf{TC}}$ [Eq.~\eqref{eq-deformed_TC}] that are frequently reported in the main text.
First, we demonstrate ways to see that two-point functions of this model decay exponentially and further show that the `t Hooft string has long-range order.
Subsequently, we review a calculation by Huxford et al~\cite{huxford2023} that demonstrates that the deformed toric code has a perimeter law for the Wilson loop operator.

\subsection{Exponentially Decaying Correlations and Long-Range Order}

To show that the correlation functions of the deformed toric code and  Ising wavefunctions of Eq.~\eqref{eq-deformed_TC}~and~\eqref{eq-deformedIsing} are exponentially decaying and that the 't Hooft string/Ising order parameter exhibit long-range order, it is convenient to work with the Ising wavefunctions.
It is worth noting that the two are related by an exact lattice duality \cite{kramersandwannier}:
\begin{equation} \label{eq-KWem}
    \begin{tikzpicture}[scale = 0.5, baseline={([yshift=-.5ex]current bounding box.center)}]
    \draw[gray] (0, -1) -- (0, 1);
    \draw[gray] (-1.5, 1) -- (1.5, 1);
    \draw[gray] (-1.5, -1) -- (1.5, -1);
    \node at (0,0){\normalsize $Z$};
    \end{tikzpicture} \Longleftrightarrow \begin{tikzpicture}[scale = 0.5, baseline={([yshift=-.5ex]current bounding box.center)}]
    \draw[gray] (0, -1) -- (0, 1);
    \draw[gray] (-1.5, 1) -- (1.5, 1);
    \draw[gray] (-1.5, -1) -- (1.5, -1);
    \node at (-1,0){\normalsize $Z$};
    \node at (1,0){\normalsize $Z$};
    \end{tikzpicture} \qquad 
    \begin{tikzpicture}[scale = 0.5, baseline={([yshift=-.5ex]current bounding box.center)}]
\draw[gray] (-1, -1) -- (-1, 1) -- (1, 1) -- (1, -1) -- cycle;
\node at (0.0, -1) {\normalsize $X$};
\node at (0.0, 1) {\normalsize $X$};
\node at (-1, 0.0) {\normalsize $X$};
\node at (1, 0.0) {\normalsize $X$};
\end{tikzpicture} \Longleftrightarrow \begin{tikzpicture}[scale = 0.5, baseline={([yshift=-.5ex]current bounding box.center)}]
\draw[gray] (-1, -1) -- (-1, 1) -- (1, 1) -- (1, -1) -- cycle;
\node at (0.0, 0.0) {\normalsize $X$};
\end{tikzpicture} ,
\end{equation}
which maps the deformed toric code, defined on the lines of the square lattice to the deformed Ising model, which is now defined on the plaquettes of the square lattice (which we will view as the vertices of a slightly translated square lattice).
We start by showing that all correlation functions of the deformed Ising wavefunction---and hence all $1$-form symmetric correlation functions of the deformed toric code exponentially decay.
To do so, we note that
\begin{equation}
    \ket{\phii(\beta)} = \frac{1}{\sqrt{Z(\beta)}} e^{\beta/2 \sum_{\langle v,w \rangle} Z_v Z_{w}} \ket{+}^{\otimes N} = \frac{1}{\sqrt{Z(\beta)}} \sum_{\{\sigma\}} e^{\beta/2 \sum_{\langle v, w\rangle} \sigma_v \sigma_w} \ket{\{\sigma\}} \ ,
\end{equation}
where $Z(\beta)$ is a normalization factor expressed as
\begin{equation} \label{eq-partitionfunction}
    Z(\beta) = \frac{1}{2^{N}} \sum_{\{\sigma\}}e^{\beta \sum_{\langle v, w\rangle} \sigma_v \sigma_w},
\end{equation}
which takes the form of the partition function of the 2D classical Ising model.
From this form of the wavefunction, we can relate the $\langle Z_v Z_w \rangle$ correlation function of the deformed Ising wavefunction to the the spin-spin correlation function of the 2D classical Ising model at inverse temperature $\beta$ \cite{kramersandwannier}:
\begin{equation}
    \langle Z_v Z_w \rangle_{\phii(\beta)} = \frac{1}{Z(\beta)}\sum_{\{\sigma\}} e^{\beta \sum_{\langle v, w\rangle} \sigma_v \sigma_w}  \sigma_v \sigma_w = \langle \sigma_v \sigma_w \rangle_{\beta}^{\text{classical}}.
\end{equation}
It is known that all correlation functions of the classical Ising model are exponentially decaying away from its critical temperature implying that all correlation functions diagonal in $Z$-basis decay exponentially in the deformed Ising wavefunction.
For off-diagonal operators, let us remark that:
\begin{equation}
    X_{v'} \ket{\phii(\beta)} = \frac{1}{\sqrt{Z(\beta)}} X_{v'}e^{\beta/2 \sum_{\langle v,w \rangle} Z_v Z_{w}} \ket{+}^{\otimes N} =  e^{-\beta \sum_{w |\langle w, v' \rangle} Z_v Z_{w}} \ket{+}^{\otimes N}, 
\end{equation}
where in the last equation, the sum in the exponent is over sites $w$ neighboring $v'$.
Consequently, any expectation value of an off-diagonal operator can be re-written as one of a diagonal operator.
Hence, all correlation functions of the deformed Ising wavefunction and all $1$-form symmetric correlation functions of the deformed toric code wavefunctions exponentially decay.
Note that in the deformed toric code, the correlation function of any operator that is charged under the $1$-form is zero.

The equivalence between correlation functions in the deformed Ising wavefunction and correlation functions in the classical Ising model can be used to argue for the long-range order.
Specifically, the classical Ising model is known to undergo a phase transition at inverse temperature $\beta_c = \log(1 + \sqrt{2})/2$, with a trivial paramagnetic phase at $\beta < \beta_c$ and a ferromagnetic phase for $\beta > \beta_c$.
In the ferromagnetic phase, the classical correlation function $\langle \sigma_v \sigma_w \rangle_{\beta > \beta_c}^{\text{classical}}$ develops long-range order and as a consequence $\langle Z_v Z_w \rangle_{\phii(\beta > \beta_c)}$ develops long-range order.
Under the dualities of Eq.~\eqref{eq-KWem}, this translates to long-range order of the 't Hooft string in the deformed toric code.

\subsection{Perimeter Law}

For the reader's convenience, we now review how to derive the perimeter law scaling of the deformed toric code wavefunction provided by Huxford et al in Ref.~\onlinecite{huxford2023}.
Before doing so, we remark upon a physical perspective for seeing the perimeter law scaling in the Ising language based on statistical mechanics.
In particular, let us suppose that $\mu(\mathcal{R}) = \prod_{v \in \mathcal{R}} X_v$ is the disorder operator in region $\mathcal{R}$.
Moreover, let us divide our lattice into the region $\mathcal{R}$ and its complement $\mathcal{R}^c$.
Then,
\begin{align}
\begin{split}
    \langle \mu(\mathcal{R}) \rangle_{\phii(\beta)} &= \frac{1}{Z(\beta)} \bra{+}^{\otimes N} e^{\beta/2 \sum_{\langle v,w \rangle} Z_v Z_{w}} \mu(\mathcal{R})  e^{\beta/2 \sum_{\langle v,w \rangle} Z_v Z_{w}} \ket{+}^{\otimes N}  \\
    &= \frac{1}{Z(\beta)} \bra{+}^{\otimes N} e^{\beta/2 \sum_{\langle v,w \rangle} Z_v Z_{w}}  \left(e^{-\beta/2 
 \sum_{\langle v \in \mathcal{R}, w \in \mathcal{R}^c\rangle} Z_v Z_{w}} e^{\beta/2\sum_{\langle v,w \rangle \in \mathcal{R}, \mathcal{R}_c} Z_v Z_{w}} \right) \mu(\mathcal{R}) \ket{+}^{\otimes N}\\
 &= \frac{Z_{\mathcal{R}}(\beta) Z_{\mathcal{R}^c}(\beta)}{Z(\beta)} \qquad \text{ where } \qquad  Z_{A}(\beta) = \bra{+}^{\otimes |\mathcal{A}|} e^{\beta \sum_{\langle v,w \rangle \in A} Z_v Z_{w}} \ket{+}^{\otimes |\mathcal{A}|},
 \end{split}
\end{align}
where in the second line, we used the fact that commuting $\mu(\mathcal{R})$ through the exponentials will flip the bonds in the exponent at the boundary of $\mathcal{R}$, $\partial \mathcal{R}$.
Futhermore, the notation $\langle v,w \rangle \in \mathcal{R}, \mathcal{R}_c$ indicates that $v,w$ are either both in $\mathcal{R}$ or both in $\mathcal{R}_c$.
Note that $Z_A(\beta)$ appearing in the above expression looks like the partition function of a classical 2D Ising model in region $A$ (c.f. Eq.~\eqref{eq-partitionfunction}).
To estimate the ratio of partition functions appearing above, recall that the partition function of a statistical mechanics model is related to the free energy $F$ via $Z = e^{\beta F}$ where $F > 0$ since $Z>1$ in our case.
Since the free energy is extensive in system size, $F \sim f (|\mathcal{R}| + |\mathcal{R}_c| + |\partial \mathcal{R}|)$ has contributions from $\mathcal{R}$, its complement, and the boundary between the two:
\begin{equation}
    \langle \mu(\mathcal{R}) \rangle_{\phii(\beta)} = \frac{Z_{\mathcal{R}}(\beta) Z_{\mathcal{R}^c}(\beta)}{Z(\beta)} \sim \frac{e^{\beta f |\mathcal{R}|} e^{\beta f |\mathcal{R}_c|}}{e^{\beta f(|\mathcal{R}| + |\mathcal{R}_c| + |\partial \mathcal{R}|)}} = e^{-\beta f|\partial \mathcal{R}|},
\end{equation}
motivating the perimeter law appearing in these wavefunctions.

This perimeter law is derived more rigorously in the work by Huxford et al \cite{huxford2023}.
The derivation of the perimeter law proceeds as follows.
Let us remark that the deformed Ising wavefunction can be written in a domain wall representation similar to discussion of Sec.~\ref{sgaplessstates}:
\begin{equation}
    \ket{\phii(\beta)} \propto  \sum_{\mathcal{C}} e^{-\beta|\mathcal{C}|/2} \ket{\mathcal{C}}. 
\end{equation}
Then, we have 
\begin{equation}
    \langle \mu(\mathcal{R}) \rangle_{\phii(\beta)}=  \cfrac{ \sum_{\mathcal{C}} e^{-\beta/2 (|\mathcal{C}| + |\partial\mathcal{R} + \mathcal{C}|)}}{\sum_{\mathcal{C}} e^{-\beta |\mathcal{C}|}} ,
\end{equation}
where $\partial \mathcal{R} + \mathcal{C}$ is the domain wall configuration obtained by applying $\mu(\mathcal{R})$ to $\ket{\mathcal{C}}$.
Since the action of $\mu(\mathcal{R})$ on the space of domain wall configurations is bijective, we can re-write the above as: 
\begin{align}
\left\langle \mu(\mathcal{R}) \right\rangle_{\phii(\beta)}  &=  \cfrac{ \sum_{\mathcal{C}} 2e^{-\beta/2 (|\mathcal{C}| + |\partial\mathcal{R} + \mathcal{C}|)}}{\sum_{\mathcal{C}} e^{-\beta |\mathcal{C}|} + e^{-\beta |\partial\mathcal{R} + \mathcal{C}|}} = \cfrac{ \sum_{\mathcal{C}} 2e^{-\beta \bar{C}_{\mathcal{R}}}}{\sum_{\mathcal{C}} e^{-\beta |\mathcal{C}|} + e^{-\beta |\partial\mathcal{R} + \mathcal{C}|}} =  \cfrac{ \sum_{\mathcal{C}} e^{-\beta \bar{C}_{\mathcal{R}}}}{\sum_{\mathcal{C}} e^{-\beta \bar{C}_{\mathcal{R}}}\cosh\left(\beta \Delta C_{\mathcal{R}}/2\right)},
\end{align}
where $\bar{C}_{\mathcal{R}} = (|\mathcal{C}| + |\mathcal{C} + \partial \mathcal{R}|)/2$ is the average between the total length of domain walls before and after applying $\mu(\mathcal{R})$ and $\Delta \mathcal{C}_{\mathcal{R}} = |\mathcal{C}| - |\mathcal{C} + \partial \mathcal{R}|$ is the change in the length of the domain walls.
Note that the change in the length of the domain walls is upper bounded as $\Delta \mathcal{C}_{\mathcal{R}}  \leq |\partial \mathcal{R}|$.
Therefore, 
\begin{equation}
    \left\langle \mu(\mathcal{R}) \right\rangle_{\phii(\beta)} \geq \cfrac{ \sum_{\mathcal{C}} e^{-\beta \bar{C}_{\mathcal{R}}}}{\sum_{\mathcal{C}} e^{-\beta \bar{C}_{\mathcal{R}}}\cosh\left(\beta|\partial \mathcal{R}|/2\right)} = \frac{1}{\cosh\left(\beta|\partial \mathcal{R}|/2\right)} \geq w e^{\alpha |\partial \mathcal{R}|},
\end{equation}
for some $w$ and $\alpha$.
This proves the desired perimeter law.

\section{Supporting Results for the Proof of Theorem 1}

In this section of the appendix, we provide some supporting results that appear in the proof of Theorem 1 of the main text.
For the readers convenience, we restate the main theorem below.

\begin{shaded}
    \textbf{Theorem 1.} (\textit{Gaplessness from Correlations)} Let $\ket{\psi}$ be a wavefunction defined on a 2D lattice of finite dimensional qudits with the global topology of a torus\footnotemark.
    Suppose further that there exist $\mathbb{Z}_n$ unitary string operators $W_{\gamma}$ and $V_{\hat{\gamma}}$ that braid non-trivially (i.e. $W_{\gamma} V_{\hat{\gamma}} = e^{i \theta} V_{\hat{\gamma}} W_{\gamma}$ for $\theta \neq 0$ and $\gamma$ and $\hat{\gamma}$ intersecting once).
    If the state $\ket{\psi}$ exhibits the following properties:
    \begin{enumerate}    
        \item[(i)] a $\mathbb{Z}_n$ $1$-form symmetry given by closed $V_{\hat{\gamma}}$ loops,

        \item[(ii)] a perimeter law for closed $W_{\gamma}$ loops [Eq.~\eqref{eq-perimeterscaling}] and long-range order of open $V_{\hat \alpha}$ strings [Eq.~\eqref{eq-fluxprolif}], 

        \item[(iii)] cluster decomposition of open $V_{\hat \alpha}$ strings [Eq.~\eqref{eq-notcat}], 
        \end{enumerate}
        \noindent
        then $\ket{\psi}$ cannot be the  ground state of any gapped local Hamiltonian with a finite ground state degeneracy.
    
\end{shaded}
\footnotetext[\value{footnote}]{As specified in the main text, this theorem holds for lattices whose global topology corresponds to a closed manifold with a genus $g \geq 1$.}
\noindent
The subsections below are organized into three parts.
\begin{description}
    \item[Subsection 1] In this subsection, we will prove a result on the clustering of correlations in gapped quantum systems that we will need for our proof.
    The result closely follows and slightly extends results in existing work \cite{hastings2004locality,nachtergaele2006}.

    \item[Subsection 2] In this subsection, we expound upon two assumptions about gapped phases of matter expressed in Sec.~\ref{subsec-summ} that enter into our proofs.

    \item[Subsection 3] In this subsection, we prove two lemmas that are essential intermediate steps in the full proof of the main text.

\end{description}

\subsection{Review of Clustering in Gapped Quantum Systems} \label{app-locality}

To begin, we prove a result on the clustering of correlations that is essential for the proof of Theorem~1.
The result is very close to one provided in a seminal work by Hastings~\cite{hastings2004locality}, which demonstrated that gapped local Hamiltonians 
satisfy the following clustering property.
Namely, Hastings showed that  if $A_X$ and $B_Y$ are operators supported on sets $X$ and $Y$ respectively and $\Omega$ is the exactly degenerate ground state of a gapped Hamiltonian, then any state $\ket{\psi} \in \Omega$ must obey
\begin{equation} \label{eq-HastingsSeminalCluster}
        |\langle A_X B_Y \rangle_{\psi} -\langle A_X \mathbb{P}_{\Omega} B_Y\rangle_{\psi}| \leq C_{X, Y} \|A_X\| \|B_Y\| e^{-\mu d(X, Y)},
\end{equation}
where $\mathbb{P}_{\Omega}$ is the projector onto $\Omega$, $\| \mathcal{O} \| = \text{max}_{\ket{\psi}, \braket{\psi} = 1} \sqrt{\langle \mathcal{O}^{\dagger} \mathcal{O} \rangle_{\psi}}$ is the operator norm, $d(X, Y)$ is the minimum distance between sets $X$ and $Y$, $\mu$ is an order one constant, and $C_{X, Y}$ is proportional to thet 
size of $X$ and $Y$. 
If there is a nonzero splitting between the ground states, there is typically also a term on the right hand side proportional to this splitting.
However, we remark that, if one makes our our assumptions about gapped phases of matter, however, this splitting can be removed while maintaining the locality of the Hamiltonian (see discussion in Appendix~\ref{app-assumption-discussion}).
%
%

We aim to show a slightly more general result, which allows a similar bound for three widely separated operators.
More precisely, suppose that we have operators $A_X, B_Y,$ and $C_Z$ supported on sets $X, Y$ and $Z$ respectively that are all far away from one another. We want to show that
\begin{equation} \label{eq-seqclusteringguy}
    \langle A_X B_Y C_Z\rangle \approx \langle A_X \mathbb{P}_{\Omega} B_Y C_Z \rangle \approx \langle A_X \mathbb{P}_{\Omega} B_Y \mathbb{P}_{\Omega} C_Z \rangle.
\end{equation}
To do so requires applying the same proof techniques that Hastings uses for the result of Eq.~\eqref{eq-HastingsSeminalCluster} but applied to bounding the operator
\begin{equation}
        \|\mathbb{P}_{\Omega} A_X B_Y \mathbb{P}_{\Omega} - \mathbb{P}_{\Omega} A_X \mathbb{P}_{\Omega} B_Y \mathbb{P}_{\Omega}\|.
\end{equation}
Indeed, by bounding the above, the sequential clustering of Eq.~\eqref{eq-seqclusteringguy} follows.
Namely, a bound for the above would imply that $\langle A_X B_Y C_Z \rangle \approx \langle A_X \mathbb{P}_{\Omega}(B_Y C_Z) \rangle$.
Furthermore, since $\mathbb{P}_{\Omega}(B_Y C_Z) \mathbb{P}_{\Omega} \approx \mathbb{P}_{\Omega}B_Y \mathbb{P}_{\Omega} C_Z \mathbb{P}_{\Omega}$, we have that $\langle A_X B_Y C_Z \rangle \approx \langle A_X \mathbb{P}_{\Omega} B_Y \mathbb{P}_{\Omega} C_Z \rangle$ as desired.

Before proceeding, we make precise what we mean by ``local Hamiltonians''.
Namely, we say a Hamiltonian is local if $H = \sum_X H_X$ where $H_X$ is supported on sets $X$, such that for any lattice site $v$, there exists constants $\mu$ and $s$ independent of system size such that
\begin{equation} \label{eq-localdef}
    \sum_{X\ni v}\|H_X\||X|e^{\mu\mathrm{diam}(X)}\leq s.
\end{equation}
where, for our purposes, by independent of system size, we mean independent of $L_x$ and $L_y$.
Note that the above demands that the operator weight of any Hamiltonian term decays at least as quickly as exponentially in the diameter of the operator's support. 
For such local Hamiltonians, we prove the following result.

\begin{shaded}
    \textbf{Theorem S.1.} Suppose that we have a local Hamiltonain $H$ defined on a quantum lattice system with a ground state subspace containing exactly degenerate states spanning the space $\Omega$ and an $\mathcal{O}(1)$ spectral gap $\Delta E$ above this space.
    Then, for any operators $A_X$, $B_Y$ supported on sets $X, Y$:
    \begin{equation}
        \|\mathbb{P}_{\Omega} A_X B_Y \mathbb{P}_{\Omega} - \mathbb{P}_{\Omega} A_X \mathbb{P}_{\Omega} B_Y \mathbb{P}_{\Omega}\| \leq C_{X, Y}\|A_X\| \|B_Y\| e^{-\mu \ell} ,
    \end{equation}
    where $\mathbb{P}_{\Omega}$ is the operator that projects onto $\Omega$, $\ell=\min_{x\in X,y\in Y}\text{dist}(x,y)$ is the minimum distance between the sets $X$ and $Y$, $C_{X,Y} \sim \mathsf{poly}(|X|, |Y|)$, and $\mu$ is a constant that is independent of system size.
\end{shaded}
\noindent
The subsequent subsubsections will provide a proof of this theorem.
\subsubsection{Useful Notation}
We start by establishing some useful notation.
In particular, we define the positive and negative energy components of an operator $\mathcal{O}$ as  
\begin{equation}
    \bra{n} \mathcal{O}^{\pm} \ket{m} = \bra{n} \mathcal{O} \ket{m} \begin{cases}
        \Theta(E_n - E_m) & + \text{ case} \\  \Theta(E_m - E_n) & - \text{ case}
    \end{cases},
\end{equation}
where $\Theta$ is the Heaviside step function.
These components can be conveniently expressed using the operator in Heisenberg picture as\footnote{Note that when we write $\frac{1}{\pm it+\varepsilon}$, we are implicitly taking $\varepsilon\to 0^+$.}
\begin{equation}
     \mathcal{O}^{\pm} = \frac{1}{2\pi} \int_{-\infty}^{\infty} dt\ \mathcal{O}(t) \frac{1}{\pm it + \varepsilon}.
\end{equation}
Moreover, we will often isolate the ``low-energy'' and ``high energy'' part of the operator as

\begin{equation}
    \mathcal{O}_{\text{low}}  = \mathbb{P}_{\Omega}\mathcal{O} \mathbb{P}_{\Omega} \qquad \mathcal{O}_{\text{high}} = \mathcal{O} - \mathcal{O}_{\text{low}}.
\end{equation}
Notation in hand, we now provide the high level intuition for the proof.

\subsubsection{Overview of the Proof}
Let's first remark that we can write (suppressing $X, Y$)
\begin{align}
  \mathbb{P}_{\Omega} A B \mathbb{P}_{\Omega} &= A_{\text{low}} B_{\text{low}} + \mathbb{P}_{\Omega} A_{\text{high}} B_{\text{high}}\mathbb{P}_{\Omega}\\
  &=A_{\text{low}} B_{\text{low}} + \mathbb{P}_{\Omega}A^-_{\text{high}} B_{\text{high}}\mathbb{P}_{\Omega} = \mathbb{P}_{\Omega} A \mathbb{P}_{\Omega} B \mathbb{P}_{\Omega} + \mathbb{P}_{\Omega} [A^-_{\text{high}}, B] \mathbb{P}_{\Omega},
\end{align}
where in the last line we used the fact that $A_{\text{high}}^- \mathbb{P}_{\Omega} = 0$, which follows from the definition of $A_{\text{high}}^-$.
Consequently, we have that 
\begin{equation}
    \|\mathbb{P}_{\Omega} A B \mathbb{P}_{\Omega} - \mathbb{P}_{\Omega} A \mathbb{P}_{\Omega} B \mathbb{P}_{\Omega}\| = \|\mathbb{P}_{\Omega} [A_{\text{high}}^-, B] \mathbb{P}_{\Omega}\|.
\end{equation}
The goal is to bound the left hand side.
The basic idea of the proof will be to find an operator $\widetilde{A}_{\text{high}}^-$ that approximates the action of $A_{\text{high}}^-$ in the low energy subspace $\Omega$ but is also localized around the support of $A$, such that its commutator with $B$ is small.
This approximation is provided by
\begin{equation}
    \mathcal{O}^{\pm} \longrightarrow \widetilde{\mathcal{O}}^{\pm} \equiv \frac{1}{2\pi}\int_{-\infty}^{\infty} \widetilde{\mathcal{O}}(t) \frac{1}{\pm it + \varepsilon} \qquad \text{ where } \widetilde{O}(t) =  \mathcal{O}(t) \exp\left[- \frac{(t \Delta E)^2}{2q} \right],
\end{equation}
where $\mathcal{O}(t) = e^{i Ht} \mathcal{O} e^{-i H t}$, and locality is optimized by choosing $q$ as: 
\begin{equation}
    q = \ell \Delta E/v
\end{equation}
where $v$ is the Lieb-Robinson velocity of the Hamiltonian.
This approximation has a nice intuitive explanation in the energy eigenbasis.
In particular, 
\begin{equation}
    \widetilde{\mathcal{O}}^+_{nm} = \mathcal{O}_{nm} \Theta_q(E_n - E_m) \qquad \Theta_q(\omega) = \int_0^{\infty} \frac{d \omega}{2\pi} \frac{\sqrt{2\pi q}}{\Delta E} \exp\left( - \frac{q (\omega - \omega')^2}{2 \Delta E^2} \right) ,
\end{equation}
where $\Theta_q$ better approximates $\Theta$ as $q \to \infty$.
With this approximation in mind, we will re-write
\begin{equation}\label{eq-terms2bound}
     \mathbb{P}_{\Omega}[A_{\text{high}}^-, B]\mathbb{P}_{\Omega} = \mathbb{P}_{\Omega}[\widetilde{A}^-, B]\mathbb{P}_{\Omega} + \mathbb{P}_{\Omega}\left( [A_{\text{high}}^-, B] - [\widetilde{A}_{\text{high}}^-, B]\right)\mathbb{P}_{\Omega} - \mathbb{P}_{\Omega}[\widetilde{A}_{\text{low}}^-, B]   \mathbb{P}_{\Omega} .
\end{equation}
The rest of the proof proceeds by showing that each of the terms on the right hand side above is bounded by $C_{X, Y}\|A_X\| \|B_Y\| e^{-\mu \ell}$ where $C_{X, Y} \sim \mathsf{poly}(|X|, |Y|)$ for some constant $\mu$ that is independent of system size.

\subsubsection{Bounding Commutators}
\noindent
We start by bounding the first term $\mathbb{P}_{\Omega}[\widetilde{A}^-, B]\mathbb{P}_{\Omega}$.
Doing so is straightforward and relies on the Lieb-Robinson bound, which says that for time evolution generated by local Hamiltonians,\footnote{While the factor $\frac{v|t|}{\ell}$ does not appear in the usual statement of Lieb-Robinson bounds, it is not hard to show that this form follows immediately from the usual form \cite{hastings2004locality}. See also Eq.~(30) of the arXiv version of Ref.~\onlinecite{hastings2010locality}.}
\begin{equation}
    \|[A_X(t), B_Y]\| \leq \frac{v|t|}{\ell} C_{X, Y} \|A \| \|B\| e^{-\nu (\ell - v |t|)},
\end{equation}
where once again $v$ is the Lieb-Robinson velocity of the Hamiltonian and $C_{X, Y} \sim \mathsf{poly}(|X|, |Y|)$.
With this in mind, we have that 
\begin{align} \label{eq-term1bound}
     \|\mathbb{P}_{\Omega} [\widetilde{A}^-, B] \mathbb{P}_{\Omega}\| \leq \|[\widetilde{A}^{-}, B] \| &= \left|\frac{1}{2\pi} \int_{0}^{\infty} dt e^{- \frac{\Delta E vt^2}{2\ell}} \|[A(t), B] \| \frac{1}{-it + \varepsilon} \right| \\
    &\leq \frac{1}{2\pi} \left| \int_{|t| < \ell/v} dt  e^{- \frac{\Delta E vt^2}{2\ell}} \|[A(t), B] \| \frac{1}{-it + \varepsilon} \right| + \frac{1}{2\pi} \left| \int_{|t| \geq \ell/v} dt  e^{- \frac{\Delta E vt^2}{2\ell}} \|[A(t), B] \| \frac{1}{-it + \varepsilon} \right| \\
    & \leq \frac{1}{\pi} \|A\| \|B \| C_{X, Y} \left( e^{-\nu \ell} + \sqrt{\frac{2 \pi v}{\ell \Delta E}} e^{- \Delta E \ell/2v} \right)  ,
\end{align}
where the first term in the second line is bounded by Lieb Robinson bounds and the second term in that line is bounded by fast decay of the Gaussian for large $t$.
Now, we bound the second term: $\mathbb{P}_{\Omega}\left( [A_{\text{high}}^-, B] - [\widetilde{A}_{\text{high}}^-, B]\right)\mathbb{P}_{\Omega}$. We will see that this term is small because $\tilde{A}_{\text{high}}^{-}$ well approximates $A_{\text{high}}^{-}$ on the ground state subspace.
For the readers convenience, we recall that
\begin{equation}
    A_{\text{high}}^- = \int_{-\infty}^{\infty} A_{\text{high}}(t) \frac{1}{-it + \varepsilon} \qquad \widetilde{A}_{\text{high}}^- = \int A_{\text{high}}(t) e^{- \frac{(t \Delta E)^2}{2q}} \frac{1}{-it  + \varepsilon},
\end{equation}
which we remark decays exponentially in distance.
The difference between these two quantities is made clear by working in the energy eigenbasis: 
\begin{equation}
    (A_{\text{high}}^-)_{nm} = (A_{\text{high}})_{nm} \Theta(E_n - E_m) \qquad     (\widetilde{A}_{\text{high}}^-)_{nm} = (A_{\text{high}})_{nm} \Theta_q(E_n - E_m).
\end{equation}
Now, since $\Theta_q$ is an approximation to the the $\Theta$ function, satisfying $|\Theta_q(\omega) - \Theta(\omega)| \leq e^{-q/2}/\sqrt{2 \pi q}$ for $\omega \geq \Delta E$, we can arrive at
\begin{equation}
    \|A_{\text{high}}^{-}\mathbb{P}_{\Omega} - \widetilde{A}_{\text{high}}^{-} \mathbb{P}_{\Omega} \| \leq  \|A \| \frac{e^{-q/2}}{\sqrt{2\pi q}} \qquad  \|\mathbb{P}_{\Omega}A_{\text{high}}^{-} - \mathbb{P}_{\Omega}\widetilde{A}_{\text{high}}^{-}  \| \leq  \|A \| \frac{e^{-q/2}}{\sqrt{2\pi q}}.
\end{equation}
Consequently, the second term is bounded as:
\begin{equation} \label{eq-term2bound}
    \| \mathbb{P}_{\Omega} [A^{-}_{\text{high}}, B] \mathbb{P}_{\Omega}  - 
 \mathbb{P}_{\Omega}[\widetilde{A}^{-}_{\text{high}}, B] \mathbb{P}_{\Omega}\| \leq 2 \|A \| \|B\| \frac{e^{-q/2}}{\sqrt{2\pi q}} = 2 \|A \| \|B \| \frac{e^{-\ell \Delta E/(2v)}}{\sqrt{2\pi \ell \Delta E/v}}.
\end{equation}
We now bound the final commutator, $\mathbb{P}_{\Omega} [\widetilde{A}^-_{\text{low}}, B] \mathbb{P}_{\Omega} $ which is the most involved.
To do so, we start by making a few observations.
First, we can immediately see that $\mathbb{P}_{\Omega}[\widetilde{A}^-_{\text{low}}, B] \mathbb{P}_{\Omega} = \mathbb{P}_{\Omega}[\widetilde{A}^-_{\text{low}}, B_{\text{low}}] \mathbb{P}_{\Omega}$.
Second, since $\Omega$ corresponds to the exact ground state manifold of $H$, it is clear in the energy eigenbasis that
\begin{equation}
    (\widetilde{A}_{\text{low}}^{-})_{nm} = A_{nm} \begin{cases}
        \Theta_q(0) & \ket{n}, \ket{m} \in \Omega \\
        0 & \text{otherwise}
    \end{cases}.
\end{equation}
But note that $\Theta_q(0) = 1/2$.
Hence,
\begin{equation}\label{eq-term3-0}
\mathbb{P}_{\Omega}[\widetilde{A}^-_{\text{low}}, B] \mathbb{P}_{\Omega} = \mathbb{P}_{\Omega}[\widetilde{A}^-_{\text{low}}, B_{\text{low}}] \mathbb{P}_{\Omega} = \frac{1}{2} [A_{\text{low}}, B_{\text{low}}].
\end{equation}
Thus, bounding the final commutator amounts to bounding $[A_{\text{low}}, B_{\text{low}}]$.
The key insight for doing so was introduced in Ref.~\onlinecite{hastings2004locality} and involves considering the following operator:
\begin{equation}
    \widetilde{A}^0 = \frac{\Delta E}{\sqrt{2\pi q}} \int_{-\infty}^{\infty} dt \widetilde{A}(t).
\end{equation}
Let us remark that since $\widetilde{A}(t) = A(t) \exp\left(- \frac{(t \Delta E)^2}{2q}\right)$, the above operator is quasi-local.
Namely, at small times, when $A(t)$ operator is localized near $X$, the Gaussian factor is order one.
In contrast, at later times, when $A(t)$ is delocalized, its weight is suppressed by the Gaussian factor.
With this in mind, our goal will be to show that $\mathbb{P}_{\Omega} [\widetilde{A}^0, B]\mathbb{P}_{\Omega}$ is close to $[A_{\text{low}}, B_{\text{low}}]$ and then finally that $\mathbb{P}_{\Omega} [\widetilde{A}^0, B]\mathbb{P}_{\Omega}$ is small.
We start by first showing the latter.
This follows from the fact that  $\widetilde{A}^0$ is a smeared out version of $A$ (which in turn follows from the standard Lieb-Robinson bound and long-time attenuation of $\widetilde{A}(t)$).
Namely,
\begin{align}
    \|\mathbb{P}_{\Omega}[\widetilde{A}^0, B] \mathbb{P}_{\Omega}  \| \leq \|[\widetilde{A}^0, B] \| &\leq  \frac{\Delta E}{\sqrt{2\pi q}} \left| \int_{|t| < \ell/v} dt\, \| [\widetilde{A}(t), B] \| \right| + \frac{\Delta E}{\sqrt{2\pi q}} \left| \int_{|t| > \ell/v} dt\, \| [\widetilde{A}(t), B] \| \right| \\
    &\leq \frac{\Delta E}{\sqrt{2\pi q}} C_{X, Y}  \|A \| \|B \|  \left(\int_{|t| < \ell/v} dt\, \frac{v|t|}{\ell}
e^{-\nu (\ell - v |t|)} +  \underbrace{\int_{|t| > \ell/v} dt\, e^{-\frac{\Delta E v t^2}{2\ell}}}_{\mathsf{erfc}}\right) \\
    &\leq \frac{\Delta E}{\sqrt{2\pi q}} C_{X, Y} \|A \| \|B \|\left( e^{- \nu \ell} \frac{\ell}{v} +  2\sqrt{\frac{2\pi \ell}{\Delta E v}} e^{-\Delta E \ell/2v}  \right) \\
    &\leq  C_{X, Y}\|A \| \|B \| \left( \sqrt{ \frac{\ell \Delta}{2 \pi v}} e^{-\nu \ell} +  2e^{-\Delta E \ell/2v}  \right) ,\label{eq-term3-1}
\end{align}
where we used a known bound for the $\mathsf{erfc}$ in the second to last step.
Now, we will show that $\mathbb{P}_{\Omega}[\widetilde{A}^0, B] \mathbb{P}_{\Omega}$ is close to $[A_{\text{low}}, B_{\text{low}}]$.
To do so, we break the problem up into two parts:
\begin{equation}
    \mathbb{P}_{\Omega} [\widetilde{A}^0, B] \mathbb{P}_{\Omega} -  [A_{\text{low}}, B_{\text{low}}]   = ( [\widetilde{A}^0_{\text{low}}, B_{\text{low}}] -  [A_{\text{low}}, B_{\text{low}}]  ) +  \mathbb{P}_{\Omega}[\widetilde{A}^0_{\text{high}}, B] \mathbb{P}_{\Omega} .
\end{equation}
It is easy to see that the first term has to be zero by using the spectral representation. 
In particular, assuming that all low-energy states are exactly degenerate and have energy zero without loss of generality, we get that
\begin{align}\label{eq-term3-2}
    &\| [\widetilde{A}^0_{\text{low}}, B_{\text{low}}] - [A_{\text{low}}, B_{\text{low}}] \| \\
    &=\left\| \sum_{n, m\in \Omega} \left( \sqrt{\frac{\Delta E v}{2\pi\ell}}\int_{-\infty}^{\infty} dt e^{- \Delta E v t^2/2\ell}  \left(\sum_{k \in \Omega} A_{nk} B_{km} - B_{nk} A_{km} \right) - \left(\sum_{k \in \Omega} A_{nk} B_{km} - B_{nk} A_{km} \right) \right) \ket{n} \bra{m}\right\| \\
    &= 0.
\end{align}
The second expression can be bound by the spectral representation.
In particular,
\begin{align} \label{eq-term3-3}
    \| \mathbb{P}_{\Omega} [\widetilde{A}^0_{\text{high}}, B] \mathbb{P}_{\Omega} \| &= \left\| \sum_{n, m \in \Omega}\sqrt{\frac{\Delta E v}{2\pi\ell}}\int_{-\infty}^{\infty} dt e^{- \Delta E v t^2/2\ell} \sum_{k \notin \Omega} \left(A_{nk} B_{km} e^{-iE_k t} - B_{nk} A_{km} e^{iE_kt} \right) \ket{n} \bra{m}  \right\| \\
    &= \left\|\sum_{n, m \in \Omega}  \sum_{k \notin \Omega} \left[A_{nk} B_{km} -  B_{nk} A_{km}\right]e^{- (\ell E_k^2)/(2\Delta E v)} \right\| 
    \leq 2 \|A \| \|B\| e^{-\ell \Delta E/(2v)}.
\end{align}

\subsubsection{Putting the bounds together}
Overall, we considered the following object in Eq.~\eqref{eq-terms2bound}:
\begin{align}
    \|\mathbb{P}_{\Omega} A B \mathbb{P}_{\Omega} - \mathbb{P}_{\Omega} A \mathbb{P}_{\Omega} B \mathbb{P}_{\Omega}\| &= \|\mathbb{P}_{\Omega} [A_{\text{high}}^-, B] \mathbb{P}_{\Omega}\|\\ 
    &=\|\mathbb{P}_{\Omega}[\widetilde{A}^-, B]\mathbb{P}_{\Omega} + \mathbb{P}_{\Omega}\left( [A_{\text{high}}^-, B] - [\widetilde{B}_{\text{high}}^-, B]\right)\mathbb{P}_{\Omega} - \mathbb{P}_{\Omega}[\widetilde{A}_{\text{low}}^-, B]   \mathbb{P}_{\Omega}\| 
\end{align}
and bounded the operator norm of each term appearing above in Eqs.~\eqref{eq-term1bound}~\eqref{eq-term2bound}~\eqref{eq-term3-0}~\eqref{eq-term3-1}~\eqref{eq-term3-2}~and~\eqref{eq-term3-3} with a function that decays exponentially in distance with a pre-factor that depends polynomially on the support of operators $A$ and $B$.
This means that there exists some constant $\mu$ such that: 
\begin{equation}
        \|\mathbb{P}_{\Omega} A_X B_Y \mathbb{P}_{\Omega} - \mathbb{P}_{\Omega} A_X \mathbb{P}_{\Omega} B_Y \mathbb{P}_{\Omega}\| \leq C_{X, Y}\|A_X\| \|B_Y\| e^{-\mu \ell} 
    \end{equation}
which proves the desired result.

\hspace{0.95\textwidth} $\blacksquare$

\subsection{Extended Discussion on Assumptions about Gapped Phases} \label{app-assumption-discussion}

In the main text, we remarked upon two plausible assumptions that we made regarding gapped phases of matter and Hamiltonians that were necessary for proving our theorem.
Our first assumption posited that if a family of quantum states defined on increasing number of qudits $L_x L_y$ remains gapped in the thermodynamic limit on an open range of finite aspect ratios, it remains gapped for any finite aspect ratio.
This was necessary to extend our proof to all aspect ratios, which was strictly proven for an aspect ratio $a = L_x/L_y$ beyond a finite threshold value which is independent of system size.
Our second assumption was made to extend beyond the case of exact ground state degeneracy.
In particular, we were inspired by the fact that in gapped quantum systems, finite-size splitting is intuitively understood via the necessity of going to extensive order in perturbation theory to generate operators $A_{L_x, L_y}$ that distinguish between different ground states.
Consequently, we assumed that the finite-size splitting between degenerate ground states of gapped local Hamiltonians is exponentially small in the support of such operators $\sim e^{-\kappa \cdot \text{support}(A_{L_x, L_y})}$ and that such operators can be added to the Hamiltonian to then exactly cancel this degeneracy.

As stated in the main text, we remark that the first assumption is closely related to the generalized $s$-source conjecture of Ref.~\cite{swingle2016ssource} and we refer the reader to this reference for more intuition about the assumption.
In this subsection, we provide two paradigmatic examples where our second assumption is borne out exactly.

\subsubsection{Example 1: Ising Ferromagnet}

Our first example is the transverse field Ising model in 1D: 
\begin{equation}
    H = -\sum_{v = 0}^{L - 2} Z_v Z_{v + 1} - g \sum_{v = 0}^{L -1} X_v,
\end{equation}
which is symmetric under a $\mathbb{Z}_2$ symmetry generated by $\prod_{v = 0}^{L-1} X_v$.
Let us recall that this model has two phases as a function of $g$: a trivial symmetric phase for $g > 1$ and an ordered $\mathbb{Z}_2$ spontaneous symmetry breaking phase for $g < 1$.
In the ordered phase at $g = 0$, the model is exactly two-fold degenerate with ground states: 
\begin{equation}
    \ket{\mathsf{GHZ}_{\pm}} = \frac{1}{\sqrt{2}} \left( \ket{\uparrow \cdots \uparrow} \pm \ket{\downarrow \cdots \downarrow} \right)
\end{equation}
one of which lives in the $\mathbb{Z}_2$ even sector and the other in the $\mathbb{Z}_2$ odd sector.
When $g \neq 0$ and less than $1$, the system remains in the ferromagnetic phase but its degenerate ground states are now split by an amount that is exponentially small in system size.
This splitting $\mathcal{S}(H)$ scales with system size as $\sim gL e^{- b L}$ \cite{pfeuty1970one, abraham1971transfer, oshikawa2019universal} for some constant $b$, and the the even sector state ends up being lower in energy than the odd sector state.

We now show that a term can be added to the Hamiltonian that keeps the Hamiltonian local [per the definition of Eq.~\eqref{eq-localdef}] but makes the degeneracy exact.
To do so, let us recall that at $L$-th order in perturbation theory in $g$, the operator $-\frac{1}{2}\mathcal{S}(H) \prod_{v = 0}^{L-1} X_v$ is generated that splits the two GHZ states above apart in energy. 
Consequently, we consider the auxiliary Hamiltonian:
\begin{equation}
    H' = H + \frac{1}{2} \mathcal{S}(H) \prod_{v = 0}^{L-1} X_v
\end{equation}
The perturbation to $H$ will leave the ground state subspace invariant but will reduce the ground state splitting to zero.
Crucially, this auxilliary Hamiltonian is still local because $\mathcal{S}(H)$ is exponentially small in the support of $\prod_v X_v$.

\subsubsection{Example 2: Toric Code}

Let us further consider the example of the toric code of $L_x \times L_y$ qubits placed on the links of a square lattice with the global topology of a cylinder (chosen for conveninece in the subsequent example) compactified in the $L_y$ direct.
In this case, consider the following Hamiltonian:
\begin{equation}
    H = - \sum_p \begin{tikzpicture}[scale = 0.5, baseline = {([yshift=-.5ex]current bounding box.center)}]
        \draw[color = gray] (-1, -1) -- (-1, 1) -- (1, 1) -- (1, -1) -- cycle;
        \node at (-1, 0) {\small $X$};
        \node at (1, 0) {\small $X$};
        \node at (0, -1) {\small $X$};
        \node at (0, 1) {\small $X$};
        \node at (0, 0) {\small $\textcolor{gray}{p}$};
    \end{tikzpicture}\ - \sum_v \begin{tikzpicture}[scale = 0.5, baseline = {([yshift=-.5ex]current bounding box.center)}]
    \draw[gray] (1.5,0) -- (-1.5, 0);
    \draw[gray] (0,1.5) -- (0,-1.5);
    \node at (0.75, 0) {\normalsize $Z$};
    \node at (-0.75, 0) {\normalsize $Z$};
    \node at (0, 0.75) {\normalsize $Z$};
    \node at (0, -0.75) {\normalsize $Z$};
    \node at (-0, -0) {\small $\textcolor{gray}{v}$};
\end{tikzpicture} + g \sum_{\ell} Z_\ell
\end{equation}
Let us recall that the above is in a $\mathbb{Z}_2$ topologically ordered phase for $ g < g_c =  0.1642(2)$ \cite{he1990high, wu2012phase, Kitaev_2003, vidal2008low, tupitsyn_2010, dusuel_2011}.
When $g = 0$, the model has two exactly degenerate ground states that can be labeled by the values of any non-contractible loop operator $V_{\hat{\gamma}} = \prod_{v \in \hat{\gamma}_{x, y}} Z_v$:
\begin{equation}
    \ket{\pm 1} = \ket{V_{\hat{\gamma}} = \pm 1}
\end{equation}
When $g \neq 0$ but below the critical threshhold, the ground states are each split   $\mathcal{S}(H) \sim \mathsf{poly}(L_x, L_y) e^{-b L_y}$ \cite{Kitaev_2003}.
Nevertheless, this splitting can be removed adding $\frac{1}{2}\mathcal{S}(H) V_{\hat{\gamma}}$ to the Hamiltonian as: 
\begin{equation}
    H' = H + \frac{1}{2} \mathcal{S}(H) V_{\hat{\gamma}}
\end{equation}
where $\hat{\gamma}$ is any non-contractible loop operator.
Once again, this perturbation to $H$ will leave the ground state subspace invariant but will reduce the ground state splitting to zero.
Crucially, this auxilliary Hamiltonian is still local because $\mathcal{S}(H)$ is exponentially small in the support of $V_{\hat{\gamma}}$.
In the case where a more generic perturbation is added to the toric code, we envision that the dressed logical operators of Ref.~\cite{hastings2005qa} could be used to remove the ground state degeneracy, but we leave a more detailed discussion to future work.

\subsection{Proof of Two Useful Lemmas} \label{app-twolemmas}

\begin{shaded}
    \textbf{Lemma S.1.} Suppose that $\ket{\psi}$ is a wavefunction defined on a two-dimensional lattice with the global topology of the torus and suppose it is invariant under a $\mathbb{Z}_n$ $1$-form symmetry $V_{\hat{\gamma}}$.
    If it obeys the perimeter law scaling of Eq.~\eqref{eq-perimeterscaling}, i.e.
    \begin{equation}
    |\langle W_{\gamma} \rangle_{\psi(\beta)}| = \left|\left \langle \begin{tikzpicture}[scale = 0.9, baseline = {([yshift=-0.5ex]current bounding box.center)}]
    \draw[lightdodgerblue, line width =0.5mm] (0,0) circle (15pt);
    \node at (16 pt, -16 pt) {\small $\gamma$};
    \node at (16 pt, 16 pt) {\small $\ $};
    \node at (-16 pt, -16 pt) {\small $\ $};
    \node at (-16 pt, 16 pt) {\small $\ $};
    \end{tikzpicture}\right \rangle \right|\geq we^{- \alpha |\gamma|},
\end{equation}
for some system size independent constant $w>0$ and any homologically trivial loop $\gamma$, then, any gapped parent Hamiltonian of $\ket{\psi}$  must have a non-trivial ground state degeneracy.
Suppose now that this ground state degeneracy is exact and $\mathbb{P}_{\Omega}$ is the projector onto this degenerate subspace.
Then, if $\gamma_L$ and $\gamma_R$ are non-contractible loops along the $y$-direction of the torus whose length is independent of $L_x$ and whose minimum distance apart is $cL_x$ for some constant $c$ (c.f. Fig.~\ref{fig:loop_configs_on_torus}), we can conclude that
\begin{equation}
    |\langle W^{\dagger}_{\gamma_L} \mathbb{P}_{\Omega} W_{\gamma_R} \rangle|  \geq w' e^{-\alpha' (|\gamma_L| + |\gamma_R|)}  - |\varepsilon_{L_x, L_y}|, 
\end{equation}
where $|\varepsilon_{L_x, L_y}| \leq \mathsf{poly}(L_x, L_y) e^{-\mu L_x}$ for some constants $w',\alpha',\mu >0$. 
\end{shaded}

\textit{Proof.} Let us remark that $\ket{\psi}$ has the property that
\begin{equation}
    |\langle W^{\dagger}_{\gamma_L} W_{\gamma_R} \rangle| \geq  w e^{-\alpha (|\gamma_L| + |\gamma_R|)} \geq w' e^{-2 \alpha' L_y},
\end{equation}
where $w'$ and $\alpha'$ depends on the microscopic structure of the loops (in the simplest case where we choose the loops to be straight, we have $w'=w$ and $\alpha'=\alpha$, which is in fact sufficient for our purposes).
If $\ket{\psi}$ is the unique ground state of a gapped Hamiltonian, then
\begin{equation}
    \langle W^{\dagger}_{\gamma_L} W_{\gamma_R} \rangle = \langle W^{\dagger}_{\gamma_L}\rangle \langle W_{\gamma_R} \rangle  + \varepsilon_{L_x, L_y},
\end{equation}
where $|\varepsilon_{L_x, L_y}| \leq \mathsf{poly}(L_x, L_y)e^{-\mu L_x/2}$ for some $\mathcal{O}(1)$ $\mu$.

$\ $

However, since $V_{\hat{\gamma}}\ket{\psi} = \ket{\psi}$ for all closed loops $\hat{\gamma}$, if we choose $\hat{\gamma}$ that intersects with $\gamma_L$ once, we have $\langle W_{\gamma_L} \rangle = \langle W_{\gamma_L} V_{ \hat{\gamma}} \rangle = e^{i \theta} \langle V_{\hat{\gamma}}W_{\gamma_L} \rangle = e^{i \theta} \langle W_{\gamma_L} \rangle$.
It follows that, $\langle W_{\gamma_L} \rangle = 0$.
Hence,
\begin{equation}
     w' e^{-2\alpha' L_y} \leq |\langle W^{\dagger}_{\gamma_L} W_{\gamma_R} \rangle| =  |\varepsilon_{L_x, L_y}| \leq \mathsf{poly}(L_x, L_y)e^{-\mu L_x/2}.
\end{equation}
Now, by choosing $a = L_x/L_y$ sufficiently large (but still independent of system size), we can obtain a contradiction.
This proves that any gapped Hamiltonian must have a nontrivial ground state degeneracy.

Now, let us assume that $H$ has a exact ground state subspace $\Omega$.
Then, from Theorem S.1. it follows that
\begin{equation} \label{eq-WWcluster}
     \langle W^{\dagger}_{\gamma_L} W_{\gamma_R} \rangle = \langle W^{\dagger}_{\gamma_L} \mathbb{P}_{\Omega} W_{\gamma_R} \rangle  + \varepsilon_{L_x, L_y}.
\end{equation}
If we take the absolute value of both sides and use the triangle inequality, we arrive at 
\begin{equation}
    w e^{-\alpha (|\gamma_L| + |\gamma_R|)} \leq |\langle W^{\dagger}_{\gamma_L} W_{\gamma_R} \rangle|  \leq |\langle W^{\dagger}_{\gamma_L} \mathbb{P}_{\Omega} W_{\gamma_R} \rangle|  + |\varepsilon_{L_x, L_y}|.
\end{equation}
\noindent
Then, as an immediate consequence, we conclude that: 
\begin{equation}
    |\langle W^{\dagger}_{\gamma_L} \mathbb{P}_{\Omega} W_{\gamma_R} \rangle|  \geq w' e^{-\alpha' (|\gamma_L| + |\gamma_R|)}  - |\varepsilon_{L_x, L_y}| ,   
\end{equation}
proving the desired result. 

\hspace{0.95\textwidth} $\blacksquare$

\begin{shaded}
    \textbf{Lemma S.2.}  Suppose that $\Omega$ is the exact ground state subspace of a gapped local Hamiltonian and $\ket{\psi}$ is a quantum state with a $1$-form symmetry generated by operators $V_{\hat{\gamma}}$ for closed curves $\hat{\gamma}$.
    Let us suppose that $\text{dim}(\Omega)$ is bounded from above by a constant $M_{\Omega}$ at all system sizes.
    Further suppose that for all open curves $\hat{\alpha}_1$ and $\hat{\alpha}_2$: 
    \begin{equation}
        \langle V_{\hat{\alpha}_1} V_{\hat{\alpha}_2} \rangle = \langle V_{\hat{\alpha}_1} \rangle \langle V_{\hat{\alpha}_2} \rangle + \varepsilon 
    \end{equation}
    where $|\varepsilon| \leq c e^{-\mu \mathsf{D}(\hat{\alpha}_1, \hat{\alpha}_2)}$ where $\mathsf{D}$ is the minimum distance between the endpoints of $\hat{\alpha}_1$ and $\hat{\alpha}_2$, $\mu$ is some system size independent constant, and $c$ depends polynomially in the distance between the endpoints of $\hat{\alpha}_1$ and the distance between the endpoints of $\hat{\alpha}_2$. 
    Then there exists an open curve $\hat{\alpha}$ whose endpoints are a distance $L_x/2$ apart such that: 
    \begin{equation}\label{projcond}
        \mathbb{P}_{\Omega} V_{\hat{\alpha}}\ket{\psi} = a \ket{\psi} + \varepsilon \ket{\phii}
    \end{equation}
    where $\ket{\phii} \in \Omega$ is a normalized vector orthogonal to $\ket{\psi}$ and $|\varepsilon| \leq \mathsf{poly}(L_x) e^{-b L_x/4}$ for some constant $b$.
\end{shaded}

\begin{figure}[H]
    \centering
    \includegraphics[width=0.3\linewidth]{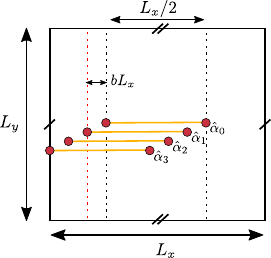}
    \caption{\textbf{String Configurations in Lemma S2.} We draw the family of curves referenced in the proof of Lemma S.2. The curves are all of length $L_x/2$ and are displaced in the $x$-direction by $b L_x$ from one another.
    Above, we displace the curves slightly in the $y$-direction for ease of visualization.}
    \label{fig:opencurves}
\end{figure}

\textit{Proof.} Let us start by writing
\begin{equation} \label{eq-abdecomp}
    \mathbb{P}_{\Omega} V_{\hat{\alpha}} \ket{\psi} = a_{\hat{\alpha}} \ket{\psi}  + b_{\hat{\alpha}} \ket{\phii_{\hat{\alpha}}},
\end{equation}
where $\braket{\psi}{\phii_{\hat{\alpha}}} = 0$ and $\ket{\phii_{\hat{\alpha}}} \in \Omega$.
Moreover, note that by Theorem S.1. (see also Refs.~\onlinecite{hastings2010locality, hastings2004locality}), we have that 
\begin{equation}
    \langle V_{\hat{\alpha}_L} V_{\hat{\alpha}_R} \rangle = \langle V_{\hat{\alpha}_L} \mathbb{P}_{\Omega} V_{\hat{\alpha}_R} \rangle + \varepsilon',
\end{equation}
where $|\varepsilon'| \leq c e^{- \mu \mathsf{D}(\alpha_L, \alpha_R)}$ where $\mu$ is a constant independent of $L_x$ and $L_y$ and  $c$ is depends polynomially on the distance between the endpoints of $\alpha_L$ and the distance between the endpoints of $\alpha_R$.
Consequently, using the assumption of the theorem, we have that
\begin{equation}
     \bar{a}_{\hat{\alpha}_L} a_{\hat{\alpha}_R} =\langle V_{\hat{\alpha}_L}  \rangle \langle V_{\hat{\alpha}_R} \rangle 
 = \langle V_{\hat{\alpha}_L} \mathbb{P}_{\Omega} V_{\hat{\alpha}_R} \rangle + \varepsilon' - \varepsilon = \bar{a}_{\hat{\alpha}_L} a_{\hat{\alpha}_R} + \bar{b}_{\hat{\alpha}_L} b_{\hat{\alpha}_R} \braket{\phii_{\hat{\alpha}_L}}{\phii_{\hat{\alpha}_R}}+ \underbrace{\varepsilon' - \varepsilon}_{- \varepsilon''}.
\end{equation}
As a consequence, we can derive the inequality
\begin{equation}
    |\bar{b}_{\hat{\alpha}_L} b_{\hat{\alpha}_R} \braket{\phii_{\hat{\alpha}_L}}{\phii_{\hat{\alpha}_R}}| = |\varepsilon''| \leq c e^{- \nu \mathsf{D}(\alpha_L, \alpha_R)},
\end{equation}
which holds for all pairs $\hat{\alpha}_L$ and $\hat{\alpha}_R$.
Now, let us consider a family of curves $\{\alpha_j\}_{j = 0}^{M-1}$ of length $L_x/2$ that coincides with $\alpha_L$ of Fig.~\ref{fig:loop_configs_on_torus} for $\alpha_{j = 0}$ and $\alpha_j$ corresponds to translating $\alpha_{j-1}$ by $L_x/{4M} \equiv b L_x$ for an $\mathcal{O}(1)$ integer $M > M_{\Omega}$.
This family of curves is schematically shown in Fig.~\ref{fig:opencurves}.

\noindent
We know that
\begin{equation}
    |\bar{b}_{\alpha_j}| |b_{\alpha_{j + k}}| |\bra{\phii_{\alpha_j}}\ket{\phii_{\alpha_{j + k}}}| \leq \mathsf{poly}(L_x)e^{- \mu b k L_x}.
\end{equation}
Now, suppose for sake of contradiction that for all $j$, $|\bar{b}_{\alpha_j}| > \mathsf{poly}(L_x) e^{-\mu b L_x/4}$.
Then,
\begin{equation}
     |\bra{\phii_{\alpha_j}}\ket{\phii_{\alpha_{j + k}}}| \leq \mathsf{poly}(L_x)e^{- \mu b (k - 1/2) L_x} \leq \mathsf{poly}(L_x)e^{- \mu b L_x/2}.
\end{equation}
This means that $\{\ket{\phii_{\hat{\alpha}_j}}, \ket{\psi}\}$ are $M+1$ quantum states in $\Omega$ whose overlap is bounded from above by $\mathsf{poly}(L_x)e^{- \mu b L_x/2}$, which can be taken to zero in the thermodynamic limit.
Consequently, $\text{dim}(\Omega) \geq M + 1$.
However, this contradicts the assumption that $M > M_{\Omega}> \text{dim}(\Omega)$ and hence there exists a $j$ such that
\begin{equation}
    \mathbb{P}_{\Omega} V_{\hat{\alpha}_j} \ket{\psi} = a_{\hat{\alpha}_j} \ket{\psi}  + b_{\hat{\alpha}_j} \ket{\phii_{\hat{\alpha}}},
\end{equation}
where $|\bar{b}_{\alpha_j}| \leq \mathsf{poly}(L_x) e^{-\mu b L_x/4}$.

\hspace{0.95\textwidth} $\blacksquare$

Assuming that no other states descend to an exponentially small energy at finite $\beta$ in the Castelnovo-Chamon model, it is clear that (\ref{projcond}) holds for all open curves $\hat{\alpha}$. A simple example of a model where (\ref{projcond}) does not hold for all open curves but does hold for \emph{some} open curves is simply the Castelnovo-Chamon Hamiltonian with some local terms dropped. The Hamiltonian is still frustration free, but with a larger ground state subspace, and is still forced to be gapless. Not only does Theorem 1 still apply, but Theorem 2 also still applies, to upper bound the gap.

\section{Detailed Variational Argument}\label{appendixvariational}

In this section of the appendix, we provide a more detailed version of the argument presented that proves the following theorem (repeated from the main text for the convenience of the reader):
\begin{shaded}%
\textbf{Theorem 2.} Let $\ket{\phii(\beta)}$ be the deformed Ising
wavefunction on the $D$-dimensional torus with $\beta>\beta_c$.
Then, the many-body gap $\Delta E$ of any locally annihilating parent
Hamiltonian of $\ket{\phii(\beta)}$ is bounded as $\Delta E\leq c/L^{D+1}$ above its two-fold degenerate ground states,
where $L$ is the linear extent of the system and $c$ is an $\mathcal{O}(1)$
constant.
\end{shaded}
\noindent
The ``locally annihilating'' condition is a weaker condition than frustration-free. We say that $|\varphi(\beta)\rangle$ has a locally annihilating gapped parent Hamiltonian if it has a parent Hamiltonian $H=\sum_vH_v$ with $H_v|\varphi(\beta)\rangle=0$ and $H_v$ quasi-local. 
Note this Hamiltonian is not necessarily frustration free as $H_v$ need not be positive semi-definite. It is, however, locally annihilating in the sense that the ground state is annihilated by each of the local terms $H_v$.
Beyond this theorem, we further prove an important corollary of this theorem that is within the scope of the results of Theorem 1.
In particular,
\begin{shaded}
    \textbf{Corollary S.1.} Any Ising symmetric parent Hamiltonian of $\ket{\phii(\beta)}$ cannot have a unique gapped ground state in its even parity sector.
\end{shaded}
\noindent
The corollary establishes that such a Hamiltonian has gaplessness beyond the exponential spectral degeneracy coming from spontaneous symmetry breaking.
The subsections below are organized as follows:
\begin{description}
    
    \item[Subsection 1] First, we provide a ``warm-up'' to the variation proof of Theorem 2 in a one-dimensional setting, that uses many of the same techniques.

    \item[Subsection 2] Second, we provide the proof of Theorem 2. Specifically, we prove that since $\ket{\varphi(\beta)}$ is the ground state of this locally-annihilating Hamiltonian $H$, we can construct (Ising-symmetric) variational states $\ket{\varphi[\theta]}$ ($\beta$ dependence is suppressed) for $H$ with small overlap with the ground state $\ket{\varphi(\beta)}$ and with a gap that decreases as $\mathcal{O}(1/(L_x^2 L_y)$ energy in system size.
    \item[Subsection 3] We provide numerical evidence for the bound on the gap by investigating the gap of the Hamiltonian of Eq.~\eqref{eq-dualCCHam} using the density matrix renormalization group (DMRG).
    
    \item[Subsection 4]  Finally, we prove an important Lemma, which is necessary for the proof of Corollary S.1.
    Namely, if a Hamiltonian is gapped in the Ising symmetry even sector, then it can be rewritten in a quasi-local locally-annihilating form, extending an argument in \cite{kitaev2006} for Hamiltonians with unique ground states. 
    Using the result of Theorem 2 and this lemma, we can immediately prove Corollary S.1.

\end{description}

\subsection{Warm-Up: Particle Hopping in One Dimension}

The following simplified example demonstrates some of the ideas underlying
the variational argument in the main text. Consider the many-body
Hamiltonian

\begin{align*}
H= & \sum_{j=1}^{L}\underbrace{(c_{j+1}^{\dagger}-c_{j}^{\dagger})(c_{j+1}-c_{j})}_{h_{j}}
\end{align*}
with periodic boundary conditions on $L$ sites, where the $c_{j}^{\dagger},c_{j}$
are Fermion creation/annihilation operators. In the single-particle
sector, $H$ has a unique ground state $|\psi\rangle=c_{k=0}^{\dagger}|0\rangle=L^{-1/2}\sum_{j}|j\rangle$
with zero-energy, where $|j\rangle=c_{j}^{\dagger}|0\rangle$ and $\ket{0}$ is the fermionic vacuum.
Within this sector $H$ has a spectrum $4\sin^{2}(k/2)$, which implies
a gap scaling to zero as $1/L^{2}$ in the thermodynamic limit.

We now use a variational argument to argue in a different way that
the gap obeys $\Delta E\leq O(L^{-2})$. To that end we consider a (un-normalized) trial state $|\psi[\theta]\rangle=L^{-1/2}\sum_{j}\theta_{j}|j\rangle$ where $\theta_j$ is a function we will choose shortly.
For such a state,
\begin{align*}
\langle\psi[\theta]|h_{j}|\psi[\theta]\rangle & =\frac{1}{L}(\theta_{j+1}-\theta_{j})^{2},
\end{align*}
closely mirroring Eq.~\ref{eq:config_laplacian} in the main text.
To tightly bound the gap, we would like this expectation value to be small relative to the normalization of the state $\sum_j |\theta_j|^2$.
One obvious approach to this problem is to make $\varphi$
piecewise constant, however we will show that does not lead to a good estimate of the gap. So, for example, take 
\begin{align}\label{eq:phi_sharp}
\theta_{j} & =\delta_{0<j\leq R},
\end{align}
Then, the expectation value of the Hamiltonian is
\begin{equation}
\langle\psi[\theta]|h_{j}|\psi[\theta]\rangle  =\frac{1}{L}(\delta_{jR}+\delta_{j0}),
\end{equation}
an expression that is small (zero) except precisely when $j$ lies at the boundary of those sites in the support of $\varphi$.
Summing over $j$ then yields $\langle\psi[\theta]|H|\psi[\theta]\rangle=\frac{2}{L}$.
On the other hand, the norm of $|\psi_{\varphi}\rangle$ takes the form $\langle\psi[\theta]|\psi[\theta]\rangle=R/L$
(we will see shortly that the gap is minimized by choosing $R$ such that $R/L=\mathcal{O}(1)$). Putting these results together gives:
\begin{align}
\frac{\langle\psi[\theta]|H|\psi[\theta]\rangle}{\langle\psi[\theta]|\psi[\theta]\rangle} & =2/R\label{eq:exp_energy_sep}
\end{align}
As in the main text, letting $c_{0}$ denote the normalized overlap
of $|\psi\rangle$ with $|\psi[\theta]\rangle$. This is $c_{0}^{2}\equiv\frac{\langle\psi|\psi[\theta]\rangle^{2}}{\langle\psi[\theta]|\psi[\theta]\rangle}=\frac{(R/L)^{2}}{R/L}=R/L$
which gives a bound on the gap of
\begin{align*}
\Delta\leq\frac{2/R}{(1-|c_{0}|^{2})} & .
\end{align*}
This bound implies $\Delta\leq O(1/L)$ for an appropriate choice
of $R=O(L)$, which is a very loose bound on the gap.

We improve the variational argument by making $\theta$ vary more
slowly with $j$, avoiding the sharp features present in Eq.~\ref{eq:phi_sharp}
in order to reduce the expected energy Eq.~\ref{eq:exp_energy_sep},
and in turn reduce the upper bound on the gap. The trick, which
we also use in the main text, is to gradually ramp $\theta$ between
$0$ and $1$ over a buffer region that scales with system size.
The precise details of the ramp appear unimportant; indeed any $\varphi_{j}$
of the form

\begin{equation}
\theta_{j}  =\min\left(\left(j/w\right)^{n},1\right),
\end{equation}
for a fixed $n>0$ and $w=O(L)$ will work. The matrix element for
the energy now goes as

\begin{equation}
\langle\psi[\theta]|H|\psi[\theta]\rangle\sim  \frac{1}{L}\sum_{j}\left(\frac{nj^{n-1}}{w^{n}}\right)^{2}\delta_{j<w} =\mathcal{O}\left(\frac{1}{Lw}\right).
\end{equation}

When we take $w=\mathcal{O}(L)$ this is $\mathcal{O}(1/L^{2})$ which is substantially
smaller than Eq.~\ref{eq:exp_energy_sep}. At the same time, it
is easy to verify that $c_{0},\langle\psi[\theta]|\psi[\theta]\rangle=\mathcal{O}(1)$
so that we can bound the gap as $\Delta\leq \mathcal{O}(1/L^{2})$.

\subsection{Variational States above the Deformed Ising State}
The Hilbert space is spanned by the set of domain
wall (DW) configurations $\left\{ |\mathcal{C}\rangle\right\} $;
this set has a natural notion of Hamming distance encoded by metric, $\mathsf{d}(\mathcal{C},\mathcal{C}')$. The zero-energy ground state is written as
\begin{equation}
\ket{\varphi} =\sum_{\mathcal{C}}\varphi_{\mathcal{C}}|\mathcal{C}\rangle.
\end{equation}
For the $\varphi$ featuring in this work, $\varphi_{\mathcal{C}}\propto e^{-\beta|\mathcal{C}|}$
varies uniformly with respect to $\mathsf{d}(\mathcal{C},\mathcal{C}')$. For
example if $\mathsf{d}(\mathcal{C},\mathcal{C}')\leq1$ then $|\varphi_{\mathcal{C}}|\leq e^{4\beta}|\varphi_{\mathcal{C}'}|$
and visa versa. Form the (un-normalized) variational state
\begin{equation}\label{varwave}
\ket{\varphi[\theta]}  =\sum_{\mathcal{C}}\theta_{\mathcal{C}}\varphi_{\mathcal{C}}|\mathcal{C}\rangle,
\end{equation}
by modulating the ground state amplitudes with a real-valued function
$\theta(\mathcal{C})$ on DW configurations. $\theta$ will change
slowly with respect to the Hamming distance in a sense we make precise later. While
the $\{H_v\}$ need only be quasi-local, we will assume for simplicty they are strictly local in this section. 
Therefore, $H_v$ acts only on a set $A$ containing sites within a ball of $\mathcal{O}(1)$ radius centered at $v$.
Let $B$ denote the complement of this set. 
Write $\ket{\mathcal{C}}= \ket{\mathfrak{a}} \otimes \ket{\mathfrak{b}}$
where $\mathfrak{a},\mathfrak{b}$ denote the part of the configuration
$\mathcal{C}$ in $A$, $B$ respectively. 
We then have
\begin{equation}
H_v\ket{\varphi[\theta]}  =\sum_{\mathfrak{a},\mathfrak{b}}\theta_{\mathfrak{a},\mathfrak{b}}\varphi_{\mathfrak{a},\mathfrak{b}}\left[H_v|\mathfrak{a}\rangle\right]\otimes|\mathfrak{b}\rangle =\sum_{\mathfrak{a},\mathfrak{a}',\mathfrak{b}}\theta_{\mathfrak{a},\mathfrak{b}}\varphi_{\mathfrak{a},\mathfrak{b}}(H_v)_{\mathfrak{a}',\mathfrak{a}}|\mathfrak{a}'\rangle\otimes|\mathfrak{b}\rangle.
\end{equation}
By virtue of local-annihilation, we have $\sum_{\mathfrak{a}}(H_v)_{\mathfrak{a}',\mathfrak{a}}\varphi_{\mathfrak{a},\mathfrak{b}}=0$,
from which it follows that
\begin{align*}
H_v\ket{\varphi[\theta]}  & =\sum_{\mathfrak{a},\mathfrak{a}',\mathfrak{b}}(\theta_{\mathfrak{a},\mathfrak{b}}-\theta_{\mathfrak{a}',\mathfrak{b}})\varphi_{\mathfrak{a},\mathfrak{b}}(H_v)_{\mathfrak{a}'\mathfrak{a}}|\mathfrak{a}'\rangle\otimes|\mathfrak{b}\rangle.
\end{align*}
We now lighten the notation. Denote $\delta_{\mathcal{C},\mathcal{C}'}^{B}$
as the Kronecker delta function ensuring that global configurations
$\mathcal{C},\mathcal{C}'$ agree on $B$. Then
\begin{align*}
H_v\ket{\varphi[\theta]}  & =\sum_{\mathcal{C},\mathcal{C}'}\delta_{\mathcal{C},\mathcal{C}'}^{B}(\theta_{\mathcal{C}}-\theta_{\mathcal{C}'})\varphi_{\mathcal{C}}(H_v)_{\mathcal{C}'\mathcal{C}}|\mathcal{C}'\rangle,
\end{align*}
and
\begin{align}
\begin{split}
\bra{\varphi[\theta]} H_v\ket{\varphi[\theta]}  & =\sum_{\mathcal{C}''}\theta_{\mathcal{C}''}\varphi_{\mathcal{C}''}^{*}\langle\mathcal{C}''|\sum_{\mathcal{C},\mathcal{C}'}\delta_{\mathcal{C},\mathcal{C}'}^{B}(\theta_{\mathcal{C}}-\theta_{\mathcal{C}'})\varphi_{\mathcal{C}}(H_v)_{\mathcal{C}'\mathcal{C}}|\mathcal{C}'\rangle\\
 & =\sum_{\mathcal{C},\mathcal{C}'}\delta_{\mathcal{C},\mathcal{C}'}^{B}\varphi_{\mathcal{C}'}^{*}\varphi_{\mathcal{C}}\times\theta_{\mathcal{C}'}(\theta_{\mathcal{C}}-\theta_{\mathcal{C}'})(H_v)_{\mathcal{C}'\mathcal{C}}.
\end{split}
\end{align}

Using the fact that $H_v$ is hermitian, we obtain
\begin{align}
0\leq |\bra{\varphi[\theta]} H_v\ket{\varphi[\theta]}| & =\left| \frac{1}{2}\sum_{\mathcal{C},\mathcal{C}'}\delta_{\mathcal{C},\mathcal{C}'}^{B}\varphi_{\mathcal{C}'}^{*}\varphi_{\mathcal{C}}\times(\theta_{\mathcal{C}'}-\theta_{\mathcal{C}})^{2}(H_v)_{\mathcal{C}'\mathcal{C}}\nonumber\right| \\
 & \leq\frac{1}{2}\sum_{\mathcal{C},\mathcal{C}'}\delta_{\mathcal{C},\mathcal{C}'}^{B}|\varphi_{\mathcal{C}'}^{*}\varphi_{\mathcal{C}}|(\theta_{\mathcal{C}'}-\theta_{\mathcal{C}})^{2}|(H_v)_{\mathcal{C}'\mathcal{C}}|.\label{eq:initialbound}
\end{align}
Assume $H_v$ can only change the Hamming distance by an $\mathcal{O}(|A|)$ amount. In such cases, the wavefunction amplitudes are close to one
another e.g., $|\varphi_{\mathcal{C}'}|\leq2\alpha|\varphi_{\mathcal{C}}|$
for system size independent $\alpha$. Moreover noting the matrix
element size $|(H_v)_{\mathcal{C}'\mathcal{C}}|$ is uniformly bounded
above, gives
\begin{align}
0\leq|\bra{\varphi[\theta]} H_v\ket{\varphi[\theta]}|  & \leq\sum_{\mathcal{C},\mathcal{C}'}\delta_{\mathcal{C},\mathcal{C}'}^{B}|\varphi_{\mathcal{C}}|^{2}(\theta_{\mathcal{C}'}-\theta_{\mathcal{C}})^{2}\label{eq:config_laplacian},
\end{align}
up to system size independent multiplicative constants \textbf{\emph{which
we routinely ignore here and subsequently in this section. }}

As in the main text, $\mathbb{W}$ is the space of states consisting
of two (non-contractible) domain walls that wrap around a system in
the $y$-direction and any number of contractible domain walls. $\Delta x(\mathcal{C})$
denotes the difference in the center of mass of between these two
domain walls, which can be at most $L_{x}/2$ in magnitude on a torus.
Then we take
\begin{align}
\theta_{\mathcal{C}} & =\begin{cases}
\text{min}\left(4\Delta x(\mathcal{C})/L_{x}-1,1\right) & \mathcal{C}\in\mathbb{W}\text{ and }\Delta x/L_{x}>1/4\\
0 & \text{otherwise}
\end{cases},\label{eq:eq-trialtheta_app}
\end{align}
on configurations where the center of mass of the domain walls $\Delta x(\mathcal{C})$
is greater than a distance $L_{x}/4$. We let $\mathbb{S}$ denote
all those configurations for which $\theta_{\mathcal{C}}\neq0$.

We bound the energy of the state $|\varphi[\theta]\rangle$
and then use the variational principle to bound the gap of $H$. The
variational principle can be used because the state $|\varphi[\theta]\rangle$
becomes orthogonal to $|\varphi(\beta)\rangle$ in the thermodynamic
limit. $|\langle\varphi(\beta)|\varphi[\theta]\rangle|^{2}$ decaying
exponentially in $L_{y}$ as it is bounded from above by the probability
of a configuration being within $\mathbb{S}\subset\mathbb{W}$, which
is equivalent to the probability of the same configuration appearing
within the $D$-dimensional classical Ising model. Since the probability of having a domain wall (anywhere in the system) is suppressed exponentially in their length for domain walls larger than $\mathcal{O}(\log(L))$ for $\beta>\beta_{c}$, this means that $|\langle\varphi(\beta)|\varphi[\theta]\rangle|^{2}$
decays exponentially in $L_{y}$.

With this in mind, we now show that the energy of such states $E=\sum_{v}\langle\varphi[\theta]|H_v|\varphi[\theta]\rangle/\langle\varphi[\theta]|\varphi[\theta]\rangle\leq c/(L^2_{x}L_{y})$\footnote{In general, the result is $\frac{1}{L_x L_y \text{max}(L_x, L_y)}$ but we take $L_x > L_y$.}.
In particular, using Eq.~(\ref{eq:config_laplacian}) we can bound
the local energy by 
\begin{equation}
\frac{|\langle\varphi[\theta]|H_v|\varphi[\theta]\rangle|}{\langle\varphi[\theta]|\varphi[\theta]\rangle}\leq c'\frac{\sum_{\mathcal{C},\mathcal{C}'}(\mathcal{\theta}_{\mathcal{C}}-\mathcal{\theta}_{\mathcal{C}'})^{2}|\varphi_{\mathcal{C}}|^{2}\delta_{\mathcal{C},\mathcal{C}'}^{v}}{\sum_{\mathcal{C}}|\theta_{\mathcal{C}}\varphi_{\mathcal{C}}|^{2}},\label{eq-variationalenergybound}
\end{equation}
where $\delta^{v}=1$ when $\mathcal{C},\mathcal{C}'$ are connected
by $H_v$ and zero otherwise. Here, and henceforth, $c'$ will represent
unimportant constants independent of system size. The above makes
clear that low-energy states must have a $\theta$ that mostly ``varies
slowly'' in $\mathcal{C}$ and where it doesn't, the probability
$|\varphi_{\mathcal{C}}|^{2}$ should be small. We now show that this
is the case for our proposed $\theta$ in Eq.~\eqref{eq:eq-trialtheta_app}.

To see this, we break down the sum over $\mathcal{C},\mathcal{C}'$
in Eq.~\eqref{eq-variationalenergybound} into a few cases. In the
case $\mathcal{C},\mathcal{C}'\notin\mathbb{S}$, note $\theta_{\mathcal{C}}$
is not varying with $\mathcal{C}$ and hence their contribution to
Eq.~\eqref{eq-variationalenergybound} is zero. The only remaining
cases are that both $\mathcal{C},\mathcal{C}'\in\mathbb{S}$, $\mathcal{C}\in\mathbb{S}$
and $\mathcal{C}'\notin\mathbb{S}$, or $\mathcal{C}\notin\mathbb{S}$
and $\mathcal{C}'\in\mathbb{S}$ . Using the symmetry between $\mathcal{C},\mathcal{C}'$
apparent in the initial bound, Eq.~\eqref{eq-variationalenergybound},
these latter two contributions are of the same order, so we will just
consider the second case, so that in all subsequent calculations $\mathcal{C}\in\mathbb{S}$.

Consider first the case $\mathcal{C},\mathcal{C}'\in\mathbb{S}$.
Here we will find that $\theta_{\mathcal{C}}$ is varying slowly leading
to a small energy contribution in Eq.~\eqref{eq-variationalenergybound}.
$\mathcal{C}$ is changed into $\mathcal{C}'$ by the action of a
local Hamiltonian term $H_v$ on the domain walls of $\mathcal{C}$.
This shifts the center of mass difference by $r/L_{y}$; we will later
reason that $r$ will tend to be $\mathcal{O}(1)$ despite the presence of possibly numerous contractible domain walls.
For the moment, we take this as a given, and we find that
\begin{equation} \label{eq-thetac-cp}
(\theta_{\mathcal{C}}-\theta_{\mathcal{C}'})^{2}=\frac{16(\Delta x(\mathcal{C})-\Delta x(\mathcal{C}'))^{2}}{L_{x}^{2}}\leq\frac{16r^{2}}{L_{x}^{2}L_{y}^{2}}.
\end{equation}
Note that $r$ is zero in the cases where $v$ is not next to either
of the domain walls, but that in this case the contribution to the
energy is evidently zero. Therefore we need only consider cases where
$r\geq1$ so that $H_v$ shifts one of the domain walls, which in
turn requres that $v$ must lie along one of the domain walls. The
total energy contribution in the case where $\mathcal{C},\mathcal{C}'\in\mathbb{S}$
is then bounded as:
\begin{align}
E_{\mathcal{C},\mathcal{C}'\in\mathbb{S}} &\leq \left|\sum_{v} \frac{\bra{\phii[\theta]} H_v \ket{\phii[\theta]}}{\braket{\phii[\theta]}} \right| \leq \sum_v \frac{|\langle\varphi[\theta]|H_v|\varphi[\theta]\rangle|}{\langle\varphi[\theta]|\varphi[\theta]\rangle}  \\
&\leq \frac{c'}{L_{x}^{2}L_{y}^{2}}\sum_{v}\sum_{r\geq1}\left(\frac{\sum_{\mathcal{C},\mathcal{C}'\in\mathbb{S}}|\varphi_{\mathcal{C}}|^{2}r^{2}\delta_{\mathcal{C},\mathcal{C}'}^{v}\delta(|\Delta x(\mathcal{C})-\Delta x(\mathcal{C}')|=r/L_{y})}{\sum_{\mathcal{C}}|\theta_{\mathcal{C}}|^{2}|\varphi_{\mathcal{C}}|^{2}}\right),
\end{align}
where the first inequality follows from the fact that $E_{\mathcal{C}, \mathcal{C}' \in \mathbb{S}} \geq 0$ and the second follows from the triangle inequality.
Note that by performing the sum over $v$, we get an extra factor
of $L_{x}L_{y}$ to give
\begin{equation}
E_{\mathcal{C},\mathcal{C}'\in\mathbb{S}}\leq\frac{c'}{L_{x}L_{y}}\underbrace{\frac{1}{\sum_{\mathcal{C}}|\theta_{\mathcal{C}}|^{2}|\varphi_{\mathcal{C}}|^{2}}}_{1/\bigstar}\times\sum_{r\geq1}\left(r^{2}\underbrace{\sum_{\mathcal{C},\mathcal{C}'\in\mathbb{S}}|\varphi_{\mathcal{C}}|^{2}\delta_{\mathcal{C},\mathcal{C}'}^{v}\delta(|\Delta x(\mathcal{C})-\Delta x(\mathcal{C}')|=r/L_{y})}_{\diamond_{r}}\right).
\end{equation}

The $\bigstar$ term is the thermal expectation value of the function
$\langle|\theta_{\mathcal{C}}|^{2}\rangle$. Using $|\theta_{\mathcal{C}}|^{2}\geq\frac{1}{4}\delta(\theta_{\mathcal{C}}>1/2)$
gives $\bigstar\geq\frac{1}{4}\mathbb{P}(\theta_{\mathcal{C}}>1/2)$.
On the other hand, $\mathbb{P}(\theta_{\mathcal{C}}>1/2)=\mathbb{P}(\theta_{\mathcal{C}}>1/2|\mathbb{S})\mathbb{P}(\mathbb{S})$, where $\mathbb{P}(\mathbb{S})$ is the probability of being in the space $\mathbb{S}$.
Note however that $\mathbb{P}(\theta_{\mathcal{C}}>1/2|\mathbb{S})$
is proportional to the probability of having two large loops with
$\Delta x(\mathcal{C})>3L_{x}/8$, given that there are two large
loops that are at least $L_{x}/4$ separated. Such a probability is
$O(1)$ because the Boltzmanm weight of large loop configurations
is largely independent of $\Delta x(\mathcal{C})$. 
In contrast, $\mathbb{P}(\mathbb{S})$ is exponentially small in $L_y$ since all states in $\mathbb{S}$ have a domain wall of length at least $L_y$, and domain walls are suppressed exponentially in their length in the ordered phase.
Thus, we have that
\begin{equation}
1/\bigstar\leq c'/\mathbb{P}(\mathbb{S}).\label{eq:starfinal}
\end{equation}
We now deal with the $\diamond_{r}$ term. This is the probability
that $\mathcal{C\in\mathbb{S}}$ has one of its large domains touching
site $v$, and that acting on site $v$ connects such a large domain
to a smaller domain of linear extent at least $r-1$, so that the
resulting configuration $\Delta x(\mathcal{C}')$ has a center of
mass which differs by at least $r/L_{y}$. Such an event is bounded
above by the probability of having one of the large domains walls
pass through site $v$ (this has probability $\sim O(1/L_{x})$),
and having an nearby domain of linear extent $r$. This we expect
to be bounded by $\diamond_{r}\leq\mathbb{P}(\mathbb{S})\times Ce^{-r/\xi}/L_{x}$,
using the fact that domain walls have a linear free energy cost in
the ordered phase. Putting these contributions together gives
\begin{align}
E_{\mathcal{C},\mathcal{C}'\in\mathbb{S}} & \leq\frac{c'}{L_{x}^{2}L_{y}}\label{eq:ccpin}
\end{align}
Lastly we need to consider the case $\mathcal{C}\in\mathbb{S}$ and
$\mathcal{C}'\notin\mathbb{S}$. Here 
\begin{align*}
(\theta_{\mathcal{C}}-\theta_{\mathcal{C}'})^{2}=(4\Delta x(\mathcal{C})/L_{x}-1)^{2}.
\end{align*}
Plugging this into Eq.~\eqref{eq-variationalenergybound} and summing
over $v$ gives contribution
\begin{equation}
E_{\mathcal{C}\in\mathbb{S},\mathcal{C}'\notin\mathbb{S}}\leq c'\frac{L_{y}}{L_{x}}\frac{1}{\bigstar}\underbrace{\sum_{\mathcal{C}\in\mathbb{S},\mathcal{C}'\notin\mathbb{S}}|\varphi_{\mathcal{C}}|^{2}\left(\Delta x(\mathcal{C})-\frac{L_{x}}{4}\right)^{2}\delta_{\mathcal{C},\mathcal{C}'}^{v}}_{\blacksquare}.\label{eq:ECCpout}
\end{equation}

We now bound $\blacksquare$ above. The events contributing to this
sum are those where acting at site $v$ takes $\mathcal{C}$ out of
the set $\mathbb{S}$. As such, $v$ must lie on one of the long
domains in $\mathcal{C}$, and either i) flipping $v$ must shift
the center of mass difference by at least $\Delta x(\mathcal{C})-\frac{L_{x}}{4}$
or ii) causes the two long domains to merge at a meeting point \footnote{There is also the probability that flipping $v$ introduces a new pair of large
 domain walls, which is the inverse process of the meeting point configuration of (ii). Such a process is suppressed exponentially in system size in the same manner as (ii).}.

The second of these contributions is the easiest to deal with. It
is bounded above by
\begin{equation}
|\blacksquare_{\mathrm{ii}}|\leq L_{x}^{2}\mathbb{P}(\mathcal{C}\in\mathbb{K}_{v}\cap\mathbb{S}),
\end{equation}
where $\mathbb{K}_{v}\cap\mathbb{S}$ is the event where the two long
domains in $\mathcal{C}$ have a meeting point at vertex $v$, but
a center of mass separation of at least $L_{x}/4$. However, modeling the domain wall as a Brownian bridge, we expect that the two
loops deviate in the $x$ direction by a distance of
at most $O(\sqrt{L_{y}})$, which is much smaller than their separation.
Therefore, we expect that $\mathbb{P}(\mathbb{K}_{v}|\mathbb{S})\leq Ce^{-\gamma L_{x}^{2}/L_{y}}$
which is exponentially small in system size. Thus 
\begin{align}
|\blacksquare_{\mathrm{ii}}| & \leq L_{x}^{2}Ce^{-\gamma L_{x}^{2}/L_{y}}\mathbb{P}(\mathbb{S}).\label{eq:block2}
\end{align}

Lastly we deal with the contribution from i). In this case, acting
at $v$ must join one of the long domain walls to another domain wall
$\mathcal{D}$ which shifts the center of mass difference by an amount
$\delta_{v}\mathcal{D}$. $\delta_{v}\mathcal{D}$ must moreover exceed
$\Delta x(\mathcal{C})-L_{x}/4$ if $\mathcal{C}'$ is to exit $\mathbb{S}$.
Defining $s(\mathcal{C})\equiv(\Delta x(\mathcal{C})-L_{x}/4)L_{y}$, and
$r_{v}(\mathcal{D})=\delta_{v}\mathcal{D}L_{y}$, we need only consider
events for which $r_v(\mathcal{D})\geq s(\mathcal{C})$, giving a bound
\begin{equation}
|\blacksquare_{\mathrm{i}}|\leq\frac{1}{L_{y}^{2}}\sum_{\mathcal{C}\in\mathbb{S},\mathcal{C}'\notin\mathbb{S}}|\varphi_{\mathcal{C}}|^{2}s(\mathcal{C})^{2}\delta(r_{v}(\mathcal{D})\geq s(\mathcal{C})).
\end{equation}
Breaking the sum up according to the values of $s(\mathcal{C}),r_{v}(\mathcal{D})$
we may further write 
\begin{equation}
|\blacksquare_{\mathrm{i}}|\leq\frac{\mathbb{P}(\mathbb{S})}{L_{y}^{2}}\sum_{s>0}s^{2}\sum_{r\geq s}\mathbb{P}_{|\mathbb{S}}(\{r_{v}(\mathcal{D})=r\}\cap\{s(\mathcal{C})=s\}).
\end{equation}

We expect the following (loose) bound on the conditional probability
\begin{align*}
\mathbb{P}_{|\mathbb{S}}(\{r_{v}(\mathcal{D}) & =r\}\cap\{s(\mathcal{C})=s\})\leq CL_{x}^{-1}e^{-r/\xi},
\end{align*}
 for some constant $C$. This arises because shifting one of the long
domain COMs by an amount $r/L_{y}$ requires $\mathcal{D}$ to be
an additional domain of linear extent at least $r$, which is exponentially
suppressed in the ordered phase. The factor of $L_{x}^{-1}$ comes
from the fact that $v$ must lie on one of the long domains. Substituting
in these bounds gives $|\blacksquare_{\mathrm{i}}|\leq C\frac{\mathbb{P}(\mathbb{S})}{L_{y}^{2}L_{x}}$.

Returning to Eq.~\eqref{eq:ECCpout} and using Eq.~\eqref{eq:starfinal}
gives
\begin{equation}
E_{\mathcal{C}\in\mathbb{S},\mathcal{C}'\notin\mathbb{S}}\leq\frac{c'}{L_{x}^{2}L_{y}}+e^{-O(L)}.\label{eq:finalEccpout}
\end{equation}
Combining Eq.~\eqref{eq:ccpin} and Eq.~\eqref{eq:finalEccpout}
gives the desired $O(\frac{1}{L_{x}^{2}L_{y}})$ bound on the variational
energy.

\subsection{Numerical Evidence for Gap Bound} \label{app-numerics}
\begin{figure}[t]
    \centering
    \includegraphics[width=0.6\linewidth]{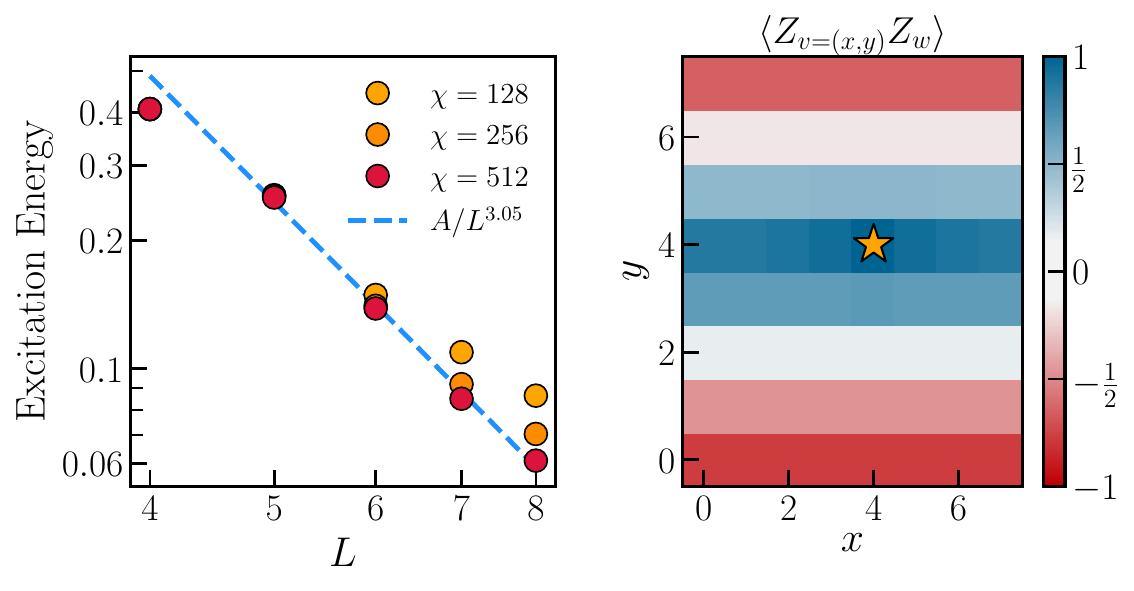}
    \caption{\textbf{Numerical Evidence for Gaplessness Bound.} In the left panel, we evaluate the energy of the first even parity excited state for the Hamiltonian of Eq.~\eqref{eq-dualCCHam} using DMRG for a system of $L \times L$ qubits.
    We plot the energy on a log-log scale as function of system size and find that the gap above the ground state decreases as $A/L^{3.05}$ for the largest bond dimension (fitting to the last four data points), nearly saturating our bound. Here, $A=3.51$ is obtained by numerically fitting the data.
    We remark that since the energies of the largest system sizes remain decreasing with bond dimension, the precise power of $1/L^{\alpha}$ is subject to increase with increased bond dimension.
    Further numerical work is necessary to precisely determine this power.
    In the right panel, we probe the character of these states via $\langle Z_v Z_w \rangle$ as a function of $v = (x, y)$ (with $w$ demarcated with a star) for the first excited state obtained in a $8 \times 8$ system.
    Intriguingly, the state indicates a thick domain wrapping a cycle, resembling our variational state.
    }
    \label{fig:appnumerics}
\end{figure}

We report numerical evidence for the bound of Theorem 2.
In particular, we investigate using the density matrix renormalization group \cite{White93, Hauschild18} the following frustration-free parent Hamiltonian for $\ket{\phii(\beta)}$:
\begin{equation} 
  \widetilde{H}(\beta) = \sum_v \left( e^{-\beta \sum_{\langle v, w \rangle} Z_v Z_w} - X_v \right),   
\end{equation}
placed on a torus of size $L \times L$ for $\beta = 1$.
Specifically, we evaluate the excitation energy of the system by first preparing the ground state:
\begin{equation}
    \ket{\phii(\beta)} \propto e^{\beta/2 \sum_{\langle v, w \rangle} Z_v Z_w}\ket{+}^{\otimes N}
\end{equation}
by imaginary-time evolving the $\ket{+}$ state using the time-evolution algorithm of Ref.~\onlinecite{Zaletel15}.
The result after bond dimension truncation achieves a ground state energy of $\leq 10^{-8}$ for all system size and bond dimensions studied.
Subsequently, excited states are obtained by performing DMRG in the even parity sector and orthogonalizing against the ground state.
The numerical results are shown in Fig.~\ref{fig:appnumerics} for a bond dimensions of $\chi = 128, 256, 512$.
We find in the left panel of Fig.~\ref{fig:appnumerics} that the ground state energy decreases as a function of system size (consistent with a gapless theory) falling as $1/L^{3.05}$ for the largest bond dimension data.
Such an exponent is consistent with and nearly saturates the bound we prove in Theorem 2, suggesting that the bound is tight.
We remark that since the energies of the largest system sizes remain decreasing with bond dimension, the precise power of $1/L^{\alpha}$ will likely increase with increased bond dimension and further numerics are required to determine $\alpha$.
In the right panel of Fig.~\ref{fig:appnumerics}, we examine the character of the excited states by plotting the correlation function $\langle Z_v Z_w \rangle$ as a function of $v = (x, y)$ for fixed $w$.
We find that this function is independent of $x$ and oscillates from $+1$ to $-1$ as a function of $y$, revealing a ``domain wall wave'' similar to the ones predicted by our variational ansatz.

\subsection{Extension to Generic Ising Symmetric Hamiltonians}\label{appendixff}

\begin{shaded}
    \textbf{Lemma S.3.} Suppose $H=\sum_{v}H_v$ is a gapped local Ising symmetric Hamiltonian with a ground state $\ket{\varphi}$ in its Ising even sector and any other exponentially split states in the Ising odd sector such that $\langle H_v\rangle_{\varphi}=0$ (without loss of generality).
    Then, we can write $H=\sum_{v}\tilde{H}_v$ for quasi-local $\tilde{H}_v$ with $\tilde{H}_v|\varphi\rangle=0$.
\end{shaded}

\textit{Proof.} Note that, by assumption, we have an $\mathcal{O}(1)$ spectral gap $\Delta E$ in the Ising even sector. 
We now show that we can write $H=\sum_{v}\tilde{H}_v$
for quasi-local $\tilde{H}_v$ with $\tilde{H}_v|\varphi\rangle=0$.
By Ising symmetry we may, without loss of generality assume $H_v$ are individually Ising
symmetric.
Using the trick in \cite{kitaev2006} we write
\begin{equation}
H=  \int_{-\infty}^{\infty}dtf(t)e^{\mathrm{i}Ht}He^{-\mathrm{i}Ht} =\sum_{v}\underbrace{\int_{-\infty}^{\infty}dtf(t)e^{\mathrm{i}Ht}H_ve^{-\mathrm{i}Ht}}_{\tilde{H}_v}
\end{equation}
with a filter function obeying $\int_{-\infty}^{\infty}dt\:f(t)=1$
and with a Fourier transform having $|\tilde{f}(\omega)|=0$ for $\omega\geq\Delta E$.
Such a filter function can be found that decays (almost) exponentially quickly with $t$. It is easy to check that $\sum_v H_v = \sum_v\tilde{H}_v$, but $\tilde{H}_v$ each individually annihilate the ground state $|\varphi\rangle$: 

\begin{align}
    \begin{split}
        \tilde{H}_v|\varphi\rangle&=\sum_{|n,+\rangle}|n,+\rangle\langle n,+|\tilde{H}_v|\varphi\rangle\\
        &=\langle H_v\rangle_{\varphi}+\sum_{|n,+\rangle\neq|\varphi\rangle}|n,+\rangle\langle n,+|H_v|\varphi\rangle\tilde{f}(E_n)=0,
    \end{split}
\end{align}
where $E_n\geq\Delta E$ is the energy of state $n$ and $|n,+\rangle$ label Ising even eigenstates of $H$. The first term is zero due to the vanishing
ground-state energy, and the second is zero due to the Fourier transform properties of the filter function.

Corollary S.1 follows immediately from the rewriting of $H$ as a locally annihilating Hamiltonian. Once we write $H$ as a locally annihilating Hamiltonian, we can apply our variational argument to show that the Hamiltonian is gapless. This is then inconsistent with the original assumption that $H$ has a $\mathcal{O}(1)$ gap in its Ising even sector, so $H$ cannot be gapped in its Ising even sector.

\end{document}